\RequirePackage{lineno}
\documentclass[superscriptaddress,aps,preprint,amsmath,amssymb,floatfix]{revtex4-1}

\usepackage[T1]{fontenc}
\usepackage{graphicx,morefloats}
\usepackage{subfigure}
\usepackage{color,soul}
\usepackage{bm}
\usepackage{longtable}

\usepackage{amssymb}
\usepackage{amsbsy}
\usepackage{amsmath}
\usepackage{amsthm}
\usepackage{times}
\usepackage{hyperref}
\usepackage{dsfont}
\usepackage{bbm}
\usepackage[normalem]{ulem}
\usepackage{gensymb}

\usepackage{pifont}

\newcommand{\xmark}{\ding{55}}

\newcommand {\R}{\textcolor {black}}

\newcommand{\be}[0]{\begin{equation}}
\newcommand{\ee}[0]{\end{equation}}
\newcommand{\ba}[0]{\begin{eqnarray}}
\newcommand{\ea}[0]{\end{eqnarray}}
\newcommand{\mx}[0]{\begin{pmatrix}}
\newcommand{\ex}[0]{\end{pmatrix}}

\newcommand{\bsub}{\begin{subequations}}
\newcommand{\esub}{\end{subequations}}

\definecolor{darkred}{rgb}{0.8,0,0}
\definecolor{royalblue}{rgb}{0.0, 0.14, 0.4}
\definecolor{magenta}{cmyk}{0,.9,0,0.2}
\definecolor{amethyst}{rgb}{0.6, 0.4, 0.8}
\definecolor{cadmiumgreen}{rgb}{0.0, 0.42, 0.24}
\definecolor{deepcarmine}{rgb}{0.66, 0.13, 0.24}
\definecolor{forestgreen}{rgb}{0.13, 0.55, 0.13}

\definecolor{riceblue}{RGB}{0,32,91}
\definecolor{ricegray}{RGB}{124,126,127}
\definecolor{lightbluegray}{RGB}{173, 199, 220}
\definecolor{lightgray}{RGB}{224, 226, 230}
\definecolor{brightblue}{RGB}{159, 221, 249}
\definecolor{mediumblue}{RGB}{77, 154, 212}
\definecolor{richblue}{RGB}{10, 80, 158}
\definecolor{midnightblue}{RGB}{19, 19, 62}
\definecolor{darkgray}{RGB}{68, 71, 79}
\definecolor{darkbluegray}{RGB}{48, 59, 97}
\definecolor{warmyellow}{RGB}{233, 161, 57}
\definecolor{brickred}{RGB}{192, 72, 41}
\definecolor{burgundy}{RGB}{104, 19, 46}
\definecolor{shadowpurple}{RGB}{54, 46, 82}
\definecolor{leafemerald}{RGB}{0, 91, 80}
\definecolor{treegreen}{RGB}{0, 67, 44}
\definecolor{grassgreen}{RGB}{53, 146, 69}
\definecolor{brightgreen}{RGB}{165, 193, 81}

\begin{document}
\hyphenation{va-ni-sh-ing}
\begin{center}
\thispagestyle{empty}

{\large\bf Chiral Weyl-Kondo semimetals and hexagonal heavy fermion systems} 
\\
[0.3cm]

Kuan-Sen\ Lin$^{1,\dagger}$, 
Yuan\ Fang$^{1,\dagger}$, 
Henrique\ Fabrelli$^{2,\dagger}$,
Runhan\ Li$^{3,4,\dagger}$, 
Andrey\  Prokofiev$^2$, 
Fang\ Xie$^{1}$, 
Jennifer\ Cano$^{5,6}$,
Maia\ G.\ Vergniory$^{3,4,7}$,
Silke Paschen$^{2,1}$,
Qimiao Si$^{1,\ast}$
\\
[0.1cm]

{\em $^1$Department of Physics and Astronomy, Extreme Quantum Materials Alliance, Smalley-Curl Institute, Rice University,
Houston, Texas, 77005, USA \\

$^2$Institute of Solid State Physics, TU Wien, Wiedner Hauptstr.\ 8-10, 1040 Vienna, Austria\\

$^3$D\'epartement de Physique et Institut Quantique,  
Universit\'e de Sherbrooke, Sherbrooke, J1K 2R1, Qu\'ebec, Canada.\\

$^4$Regroupement Qu\'eb\'ecois sur les Mat\'eriaux de Pointe (RQMP), Quebec H3T 3J7, Canada

$^5$Department of Physics and Astronomy, Stony Brook University, Stony Brook, NY 11794, USA\\

$^{6}$Center for Computational Quantum Physics, Flatiron Institute, New York, NY 10010, USA\\

$^7$Donostia International  Physics  Center,  P. Manuel  de Lardizabal 4,  20018 Donostia-San Sebastian,  Spain \\
}

\end{center}

\vspace{0.16cm}
{\bf Strong correlation, in concert with symmetry and topology, engenders novel gapless phases of matter, though only a tip of the iceberg has been seen. An exemplary framework is provided by Weyl-Kondo semimetals, in which Weyl fermions develop through crystalline symmetry constraints on the emergent low-energy heavy-fermion excitations. This paradigm has opened up new opportunities to explore correlated topologies without a noninteracting counterpart, but fully realizing this potential requires a large base of candidate materials. Here we confront the challenge on both fronts by studying heavy fermion systems with hexagonal space groups. This family contains a large number of chiral nonsymmorphic crystal structures that promote Weyl degeneracies and, in addition, feature geometric frustration in the $f$-electron magnetism. Our calculations for the heavy fermion states identify Weyl-Kondo semimetals with chiral or achiral Weyl nodes in the respective structural classes. We also develop the first search strategy of any kind for the difficult case of strongly correlated materials, which is also suitable for automation, using a combination of materials database, symmetry classification and search for desired experimental properties, and propose as candidate topological heavy fermion systems the chiral CePt$_2$B and achiral Ce$_2$NiGe$_3$ and Ce$_6$Co$_{2-\delta}$Si$_3$. Our findings raise the prospect for strongly correlated metallic topology in the unusual setting of exotic quantum magnetism and, moreover, point a way to go beyond serendipity in the search for novel strongly correlated quantum materials.
}

\clearpage
\newpage
\noindent
{\large \bf\R{Main} }
\\
In quantum materials, strong correlations agitate the electrons and, thus, enhance their physical responses \cite{Keimer-Moore_2017,Paschen-Si_2020}, whereas topology protects the robustness in their properties.
Accordingly, strongly correlated topological matter promises to exhibit the best features from both worlds.
Topological materials with Weyl nodes in the noninteracting electron excitation spectrum are broadly referred to as Weyl semimetals~\cite{armitage2018weylrmp,hu2019transportannurev}.
They can be characterized by topological invariants, \textit{e.g.} quantized Berry phases associated with the Weyl nodes.
In weakly interacting cases, Weyl semimetals have a variety of properties, including topologically-protected surface states and anomalous Hall effects that have received relatively comprehensive theoretical understandings and extensive experimental evidences~\cite{armitage2018weylrmp,cano2021band,hu2019transportannurev,son2013chiral,hosur2013recent,zyuzin2012topological,huang2017topological}.
By contrast, the investigation of strongly correlated metallic topology represents an outstanding challenge and is currently at the forefront in quantum materials research.

Weyl-Kondo semimetals (WKSMs) represent a class of strongly correlated gapless topological states.
First advanced in the theoretical study of Kondo lattice models with nonsymmorphic symmetries~\cite{lai2018weylkondo,grefe2020weylkondo,fang2024magneticweylkondosemimetalsinduced} and, concurrently, in the experimental investigations of heavy fermion semimetals~\cite{dzsaber2017kondo,dzsaber2021giant}, WKSMs develop because space group (SG) symmetries constrain the low-energy electronic excitations pinned to the Fermi energy.
Thus far, WKSMs have been studied in cubic and square-net crystals in various models~\cite{lai2018weylkondo,grefe2020weylkondo,chen2022topologicalsemimetal} as well as in candidate materials such as Ce$_3$Bi$_4$Pd$_3$~\cite{dzsaber2017kondo,dzsaber2021giant} and Ce$_2$Au$_3$In$_5$~\cite{chen2022topologicalsemimetal}.
Because they operate in the extreme correlation regime, the landscape of WKSMs is expected to be exceedingly rich; strong correlations per se can produce a variety of phenomena, including exotic quantum magnetism~\cite{Broholm2020}, unconventional superconductivity and unusual charge density wave, to name a few.
Notwithstanding recent advances of topological semimetals lacking any noninteracting counterpart in the case of quantum critical phase~\cite{Hu2021} and near a quantum critical point~\cite{Kirschbaum2026}, this potential remains under-tapped. 
One roadblock is that the material base for strongly correlated metallic topology is still small.
This is to be contrasted with strongly correlated materials per se, where decades of discoveries through serendipity and the clear-cut nature of experimental signatures for strong correlation have led to an enormous materials base, a fact that is intimately connected with the emergence of a rich landscape of correlation physics itself.

Here, we confront both overarching challenges by addressing heavy fermion systems with chiral crystalline structure~\cite{flack2003chiral,chang2018topologicalquantumchiral,fecher2022chirality,wang2024chiral}. In chiral crystals, nonsymmorphic symmetry groups can enforce Weyl nodes~\cite{zhang2018topological}. 
Due to the absence of inversion, mirror, and roto-inversion symmetries, the Weyl and anti-Weyl points that carry opposite chiral charges in a chiral crystal generally are not located at the same energy. This makes a Weyl semimetal with a chiral crystal structure an ideal platform for certain special properties.
For example, it provides one of the rare quantization in measurable properties such as circular photogalvanic effect (CPGE)~\cite{deJuan2017quantized,vazifeh2013electromagnetic,sipe2000second,deyo2009semiclassical,morimoto2016topological}. 
It has been suggested that interactions, even perturbatively, have an important effect on this quantization~\cite{avdoshkin2020interactions,wu2024absence}. It therefore is particularly timely to study chiral Weyl semimetals in the extreme correlation regime, for which the search for material candidates becomes pivotal~\cite{iwasa2023weyl}.

\begin{figure}[ht]
    \centering
    \includegraphics[width=0.9\linewidth]{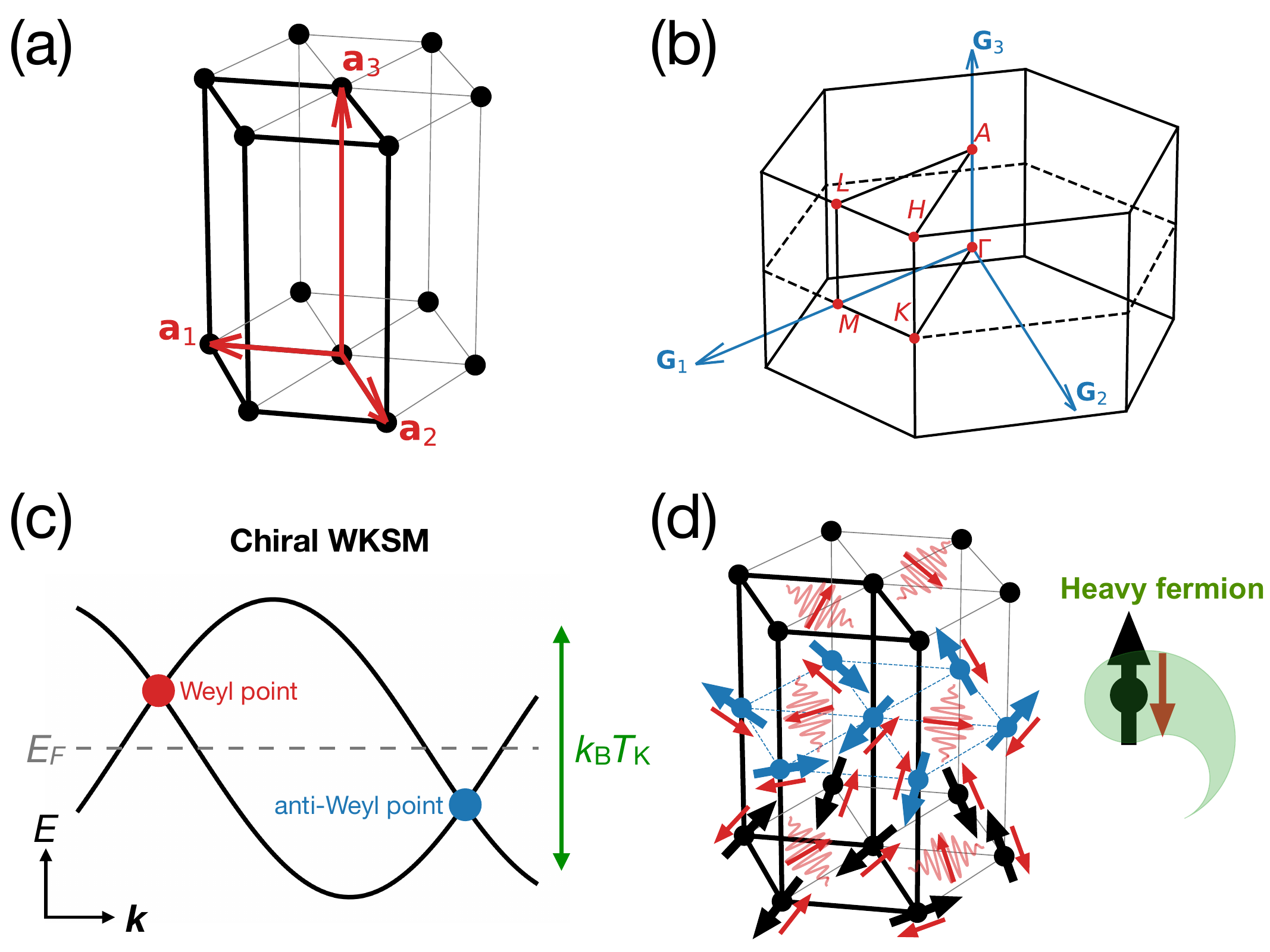}
    \caption{(a)~Hexagonal Bravais lattice in three dimensions in which the primitive unit cell is denoted by thick black lines.
    The three Bravais lattice vectors $\mathbf{a}_i$ ($i=1,2,3$) are also shown.
    (b)~Brillouin zone (BZ) and the high-symmetry points of hexagonal crystal systems. 
    The three reciprocal lattice vectors $\mathbf{G}_i$ ($i=1,2,3$) are also shown.
    Throughout this work, we refer to the top (or bottom) BZ boundary as $\mathbf{k} = k_1 \mathbf{G}_1 + k_2 \mathbf{G}_2 + k_3 \mathbf{G}_3$ with $k_3 = 1/2$ (or $k_3 = - 1/2$).
    (c)~Schematic representation of the chiral WKSM phase, where the Kondo-driven Weyl point and anti-Weyl point that carry opposite chiral charges are located at different energies. 
    The heavy-fermion energy scale is set by the Kondo scale $k_{\rm B} T_{\rm K}$. The Fermi energy is denoted as $E_F$.
    (d)~The crystal structure of a periodic Anderson model on a hexagonal crystal system with two sub-lattices per unit cell. The sub-lattice at $(0,0,0)$ is denoted by black filled circles, and the sub-lattice at $(0,0,1/2)$ is denoted by blue filled circles. 
    The itinerant $c$ electrons are denoted by wave packets with their electron spins represented by arrows.
    The $f$ electrons are localized on the sub-lattices $(0,0,0)$ and $(0,0,1/2)$, and we also use arrows to indicate their fluctuating magnetic moments. 
    The Kondo singlet emerges through the antiferromagnetic Kondo interaction between the localized $f$ electron and the itinerant $c$ electron.
    This leads to the paramagnetic phase of the periodic Anderson model with heavy-fermion excitations.}
    \label{fig:hexagonal_Bravais_lattice_BZ_schematic_model}
\end{figure}

With these considerations in mind, we focus on hexagonal heavy fermion materials. The reason is the following.
Among the hexagonal crystal structures [Fig.~\ref{fig:hexagonal_Bravais_lattice_BZ_schematic_model}(a,b)] with band crossings enforced by nonsymmorphic symmetry in the presence of time-reversal symmetry and spin-orbit coupling, the majority (10 out of 13) are chiral~\cite{zhang2018topological}.
Furthermore, the Ce-ions in these systems often occupy lattices with geometrical frustration (see Supplementary Note 16), which promotes highly entangled quantum magnetism~\cite{Broholm2020}.
Specifically, in Kondo lattice models, we demonstrate chiral WKSM phases [Fig.~\ref{fig:hexagonal_Bravais_lattice_BZ_schematic_model}(c)] in the chiral hexagonal SGs no.~173 ($P 6_3$) and no.~180 ($P 6_2 22$).
As a comparison, topological Kondo semimetal phases are also advanced in several achiral hexagonal SGs.
Our investigation is based on the Anderson/Kondo lattice models solved by parton saddle-point calculations.

Furthermore, we have developed the first suitable-for-automation search procedure of any kind for materials in the presence of strong correlations and illustrate its power by identifying candidate materials to realize the strongly correlated topological phases we advance theoretically.
In sharp contrast with the more familiar cases for weakly correlated compounds, the procedure combines database search, symmetry classification and {\it identification of experimental signatures of desired properties.}
In this way, out of 282 compounds from the database, we identify several candidate nonsymmorphic hexagonal materials for the proposed correlated Kondo semimetal phases.
Throughout this work, we refer to the SGs of hexagonal crystal systems as hexagonal SGs.

\begin{figure}[t]
    \centering
    \includegraphics[width=\linewidth]{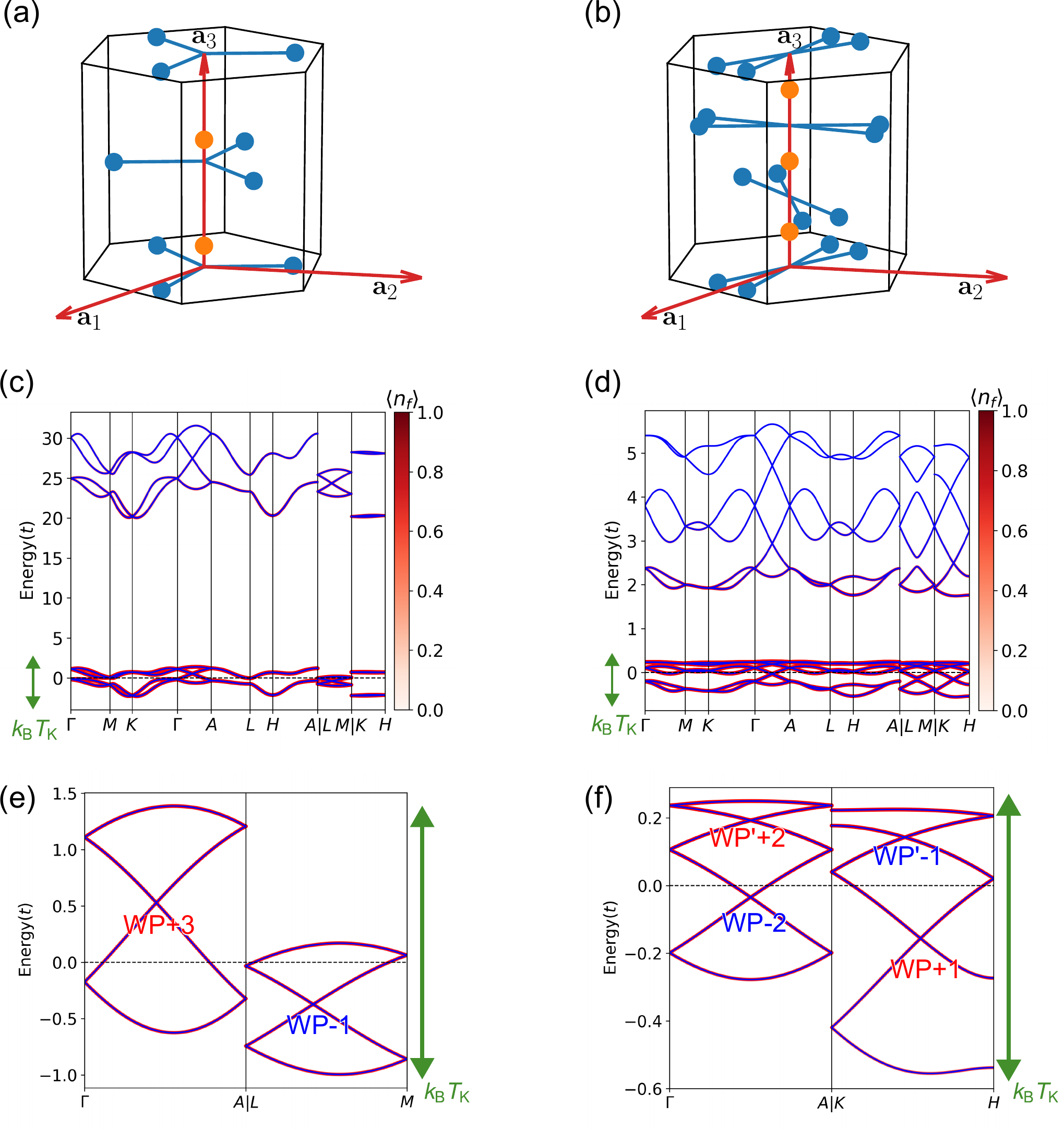}
    \caption{(a)~Unit cell of a chiral crystal in space group no.~173. 
    (b)~Unit cell of a chiral crystal in space group no.~180.
    (c)~The excitation spectrum of the hexagonal periodic Anderson model with chiral space group no.~173 ($P 6_3$) and
    (d)~with chiral space group no.~180 ($P 6_2 22$). 
    The color map indicates the portion of the $f$ electron for a given excitation.
    The energies are in unit of nearest hopping amplitude $t$.
    In (e) and (f), we show the enlarged view of the band crossings demonstrating the low-energy heavy Weyl fermions in (c) and (d), respectively.
    In (e) and (f) we also label the chiral charge of each Weyl point.
    The Kondo energy scale is denoted as $k_{\rm B} T_{\rm K}$.}
    \label{fig:Fig_chiral}
\end{figure}

\noindent
\noindent {\bf{Hexagonal space groups} ~~~}
Among the hexagonal SGs with nonsymmorphic band crossings~\cite{zhang2018topological}, we find that SGs no.~173 ($P 6_3 $), no.~176 ($P 6_3 / m$), no.~180 ($P 6_2 22$), and no.~190 ($P \bar{6} 2c$) have relatively abundant cerium-, uranium-, and ytterbium-based compounds in the Inorganic Crystal Structure Database (ICSD) (see the Supplementary Note 11 for the statistics).
Hence, our study of topological Kondo semimetals will be centered around these SGs.
Specifically, SGs $P 6_3 $ [Fig.~\ref{fig:Fig_chiral}(a)] and $P 6_2 22$ [Fig.~\ref{fig:Fig_chiral}(b)] are chiral, while SGs $P 6_3 / m$ and $P \bar{6} 2c$ are achiral.
We focus on band-crossing features of hexagonal crystal systems in the presence of time-reversal symmetry and spin-orbit coupling.

We now demonstrate the topological Kondo semimetal phases realized in the paramagnetic states of our hexagonal periodic Anderson models for the aforementioned SGs.
The model details are provided in Methods and in the Supplementary Notes 1-6 and 9.
The numerical details on the evaluation of topological chiral charges of Weyl points are provided in the Supplementary Note 7.

\noindent {\bf{Chiral topological Kondo semimetals} ~~~}
Due to the simpler space-group structure,
we begin our demonstration of chiral WKSMs from a model calculation 
in the case of SG no.~173 ($P 6_3$).
The excitation spectrum of the hexagonal periodic Anderson model with this SG is shown in [Fig.~\ref{fig:Fig_chiral}(c)].
In the low-energy heavy-fermion sector, the excitations have a predominant $f$-electron character.
Specifically, there are hourglass-type band crossings along $\Gamma-A$ and $M-L$ high-symmetry lines [Fig.~\ref{fig:Fig_chiral}(e)], indicating the heavy-fermion Weyl points.
In Supplementary Note 3, we show the distribution of the hourglass-type Weyl points in the Brillouin zone (BZ) [Fig.~\ref{fig:hexagonal_Bravais_lattice_BZ_schematic_model}(b)] and their corresponding topological charges.
Interestingly, the Weyl point along $\Gamma - A$ exhibits a high topological charge of $+3$ (Ref.\,\citenum{tsirkin2017composite}), while the Weyl point along $M - L$ has a topological charge of $-1$.
Furthermore, the Weyl points with topological charges $+3$ and $-1$ are not at the same energies as they are not related by symmetries. 
We hence demonstrate the chiral WKSM phase in the paramagnetic state of a system with this hexagonal SG.

In addition, since the SG describes chiral crystal structures, in the presence of time-reversal symmetry and spin-orbit coupling it could host Kramers-Weyl fermions at time-reversal-invariant momenta~\cite{chang2018topologicalquantumchiral}.
For example, in Fig.~\ref{fig:Fig_chiral}(e), at the energetically-lowest Kramers degeneracy at $\Gamma$ and $M$ there are Kramers-Weyl fermions with topological charge $-1$ and $+1$, respectively. 
In other words, the chiral WKSM phase hosts both hourglass-type Weyl points and Kramers-Weyl fermions. 

The excitation spectrum of the periodic Anderson model with SG no.~180 ($P 6_2 22$) is shown in Fig.~\ref{fig:Fig_chiral}(d). 
Along $\Gamma-A$ [Fig.~\ref{fig:Fig_chiral}(f)], we verify that there exist Weyl points with chiral charges of magnitude $2$ (Ref.\,\citenum{tsirkin2017composite,zhang2018topological}).
Considering all the (symmetry-related) Weyl points, the summation of the chiral charges is zero, consistent with the Nielsen-Ninomiya theorem~\cite{nielsen1981absenceI,nielsen1981absenceII}.

\noindent {\bf{Achiral topological Kondo semimetals} ~~~}
We now consider the model for the achiral hexagonal SGs for comparison.
Further details of the model calculations are given in Supplementary Notes 5 and 6.

The achiral hexagonal SG no.~176 ($P 6_3 / m$) contains the screw rotation symmetry $\{ C_{6z} | 0,0,1/2 \}$ and inversion symmetry $\mathcal{I}$.
It is related to $P6_3$ by $P 6_3 / m = (P 6_3) \cup (\mathcal{I} P 6_3)$.
Hence, breaking the inversion symmetry reduces $P 6_3 / m$ to $P 6_3 $.
Symmetry analysis shows that there are symmetry enforced four-fold degenerate Dirac nodal lines in $P 6_3 / m$~\cite{zhang2018topological}.
In addition, $P 6_3 / m$ supports quantized non-Abelian Berry phases, which give rise to non-trivial drumhead surface states bounded by Dirac nodal lines according to bulk-boundary correspondence.
For detailed analysis, see Supplementary Note 8.

The achiral hexagonal SG no.~190 ($P \bar{6} 2c$) contains the roto-inversion symmetry $\{ C_{6z} \mathcal{I} | 0,0,1/2 \}$ and glide symmetry $\{m_{1,-1,0}|0,0,1/2\}$.
There are symmetry enforced Weyl nodal lines in $P \bar{6} 2c$, which is characterized by quantized Abelian Berry phase and drumhead surface states~\cite{chan20163,muechler2020modular}.

\begin{figure}[th]
    \centering
    \includegraphics[width=\linewidth]{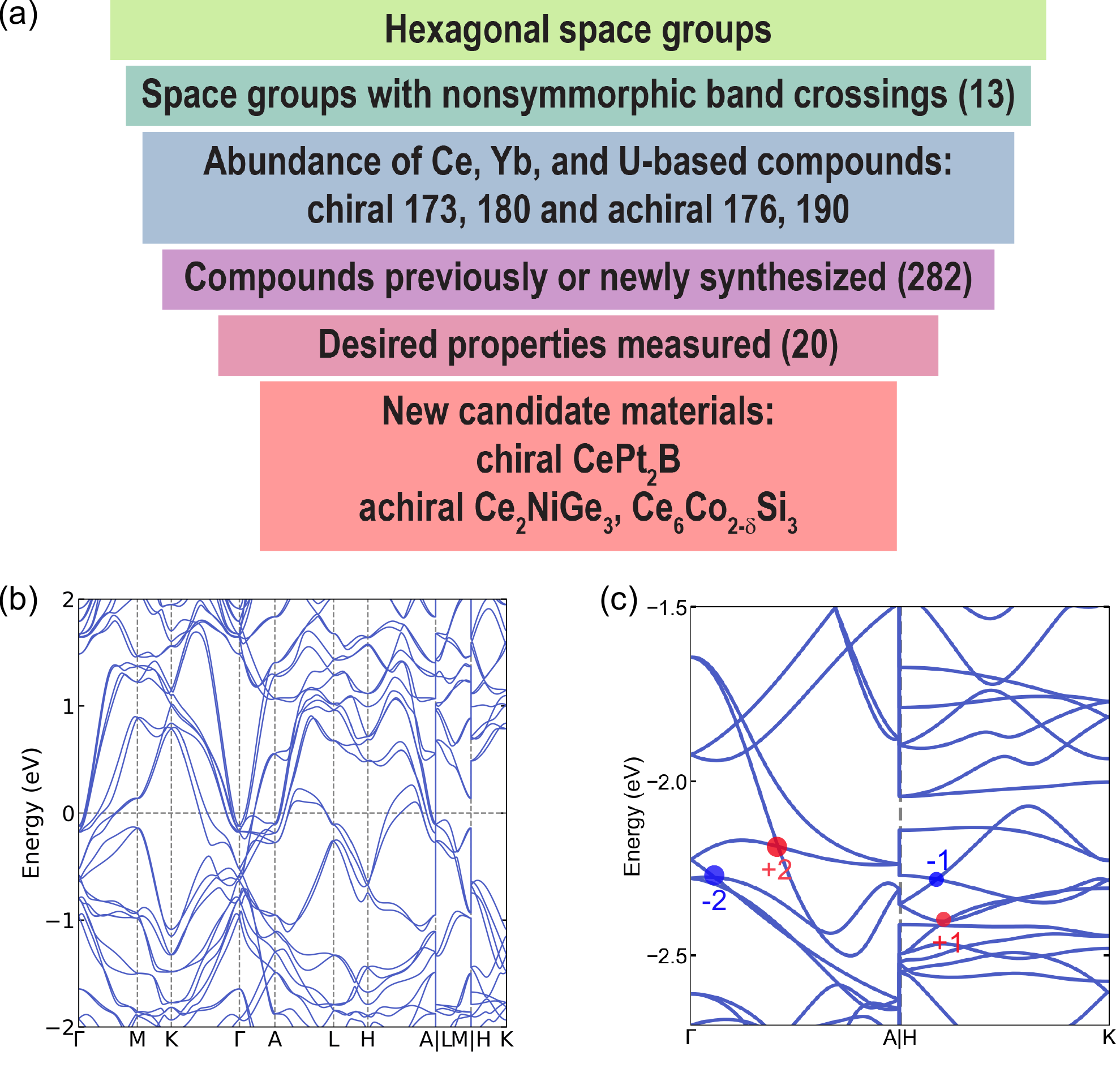}
    \caption{(a)~Schematic illustration of the materials searching procedure for hexagonal crystal systems. Starting from compounds crystallizing in hexagonal structures, we identify space groups hosting nonsymmorphic band crossings. 
    We further focus on Ce-, Yb-, and U-based compounds that have been previously or newly synthesized and experimentally characterized. 
    This screening process leads to the identification of new candidate materials for chiral and achiral WKSMs. 
    (b)~DFT band structures (with spin-orbit couplings) of CePt$_2$B with chiral space group no.~180 ($P 6_2 22$). 
    (c)~The corresponding zoomed-in band structure along $\Gamma-A$ and $K-H$. The symmetry enforced Weyl points are marked with their topological charges.}
    \label{fig:DFT_band_structures_main_text}
\end{figure}

\noindent {\bf{Candidate materials} ~~~}
We next turn to the search for candidate materials.
 For this purpose,
we develop a search procedure as illustrated in Fig.~\ref{fig:DFT_band_structures_main_text}(a).
We start with the compounds in ICSD followed by a literature search and measurements on the physical properties. 
The desired properties include electrical resistivity indicating non-insulating behavior;
an enhanced Sommerfeld coefficient $\gamma$ determined from specific heat measurements indicating strong correlations (related to
Kondo physics); and materials without magnetic transitions or low magnetic transition temperatures.
In Table~\ref{tab:primary_candidate_materials_table_main_text} we summarize 
the physical properties of the primary candidate materials.
The details of the materials properties, as well as other candidate materials, are provided in the Supplementary Notes 12-15.

We succeeded in our search procedure for materials in the chiral SG no.~180 ($P 6_2 22$). The primary candidate material is CePt$_2$B, which is reported to have a magnetic transition with a relatively low transition temperature~\cite{sologub2000newstructuretype,lackner2005lowtemperature}.
Above the magnetic transition temperature and around the characteristic temperature $T_{\rm K}$, the paramagnetic phase of CePt$_2$B could realize a chiral WKSM phase with Kondo-driven Weyl points located along $\Gamma-A$, together with a second-order anomalous Hall response ($j_{\mu} \propto \sigma_{\mu \alpha \beta} E_\alpha E_\beta$) induced by the Berry curvature.
This chiral WKSM phase also serves as a platform to investigate 
CPGE.
If the time-reversal symmetry is broken, 
\textit{e.g.} via the Zeeman effect,
while the space-group symmetries of $P 6_2 22$ are all preserved, CePt$_2$B would still host a second-order spontaneous Hall response induced by the Berry curvature.

For the achiral SG no.~190 ($P \bar{6} 2c$), the primary candidate materials include Ce$_2$NiGe$_3$~\cite{kalsi2014neutron} and Ce$_2$RhSi$_3$~\cite{szytula1993neutron,leciejewicz1995antiferromagnetic}, which are both reported to have a magnetic transition.
Above the magnetic transition and in the paramagnetic phase, these heavy fermion compounds host Kondo-driven Weyl-nodal lines on the BZ boundary, together with a fourth-order anomalous Hall response ($j_\mu \propto \sigma_{\mu\alpha\beta\gamma\delta} E_\alpha E_\beta E_\gamma E_\delta$) as the leading order response induced by the Berry curvature. On the other hand, if the time-reversal symmetry is broken, the Kondo state of Ce$_2$NiGe$_3$ and Ce$_2$RhSi$_3$ with SG no.~190 ($P \bar{6} 2c$) would carry instead a Berry-curvature-induced third-order anomalous Hall effect ($j_\mu \propto \sigma_{\mu\alpha\beta\gamma} E_\alpha E_\beta E_\gamma$) as the leading order response.

Because of the reported possibility that Ce$_2$NiGe$_3$~\cite{huo2001electric} and Ce$_2$RhSi$_3$~\cite{chevalier1984anew,das1994magnetic,kase2009antiferromagnetic,szlawska2009antiferromagnetic} may crystallize in other forms, we perform new experiments on Ce$_2$NiGe$_3$ and present our results in Supplementary Note 15 and Supplementary Figs.~24--27. We confirm that Ce$_2$NiGe$_3$ has the achiral SG no.~190 ($P \bar{6} 2c$), has a magnetic transition, and has a semimetallic-type resistivity. We further present new results for Ce$_2$NiGe$_3$ on the magnetoresistivity and Hall resistivity, which confirm the Kondo physics.

For the achiral SG no.~176 ($P 6_3 / m$), the primary candidate materials include Ce$_6$Rh$_{32}$P$_{17}$~\cite{pivan1988crystal} and Ce$_6$Co$_{2-\delta}$Si$_3$~\cite{gaudin2007on,chevalier2007the}.
If the time-reversal symmetry is broken, the Kondo state with SG no.~176 ($P 6_3 / m$) would carry a first-order anomalous Hall effect ($j_\mu \propto \sigma_{\mu\alpha} E_\alpha$) as the leading order response from the Berry curvature.

\noindent {\bf{Density functional theory calculations for $\rm CePt_2B$} and $\rm Ce_2NiGe_3$ ~~~}
Fig.~\ref{fig:DFT_band_structures_main_text}(b) provides the density functional theory (DFT) band structure of CePt$_2$B within an $f$-core calculation that includes the effect of spin-orbit couplings.
The degeneracies and nodal lines are consistent with our symmetry analysis. 
Fig.~\ref{fig:DFT_band_structures_main_text}(c) shows a zoomed-in plot of the Weyl points along $\Gamma-A$ and $K-H$ directions, where we determine Weyl charges for some representative Weyl points based on symmetry indicators~\cite{tsirkin2017composite}.
The detailed analysis can be found in Supplementary Note 7, where we verify that the Weyl charges between given two adjacent band indices sum to zero.
In Supplementary Note 10, we also show the DFT band structure of $\rm Ce_2NiGe_3$ in SG no.~190 ($P\bar{6}2c$) where symmetry enforced Weyl nodal lines are identified.
The consistency of the $f$-core band structure with our symmetry analysis provides confidence that the heavy fermion electronic states of these materials, which develop near the Fermi energy, will feature the type of Weyl-Kondo degeneracies advanced in our model calculations.

\noindent {\bf{Conclusion and outlook} ~~~}
In this work, we identify hexagonal topological Kondo semimetals and theoretically realize chiral Weyl-Kondo semimetals.
We formulate our theory in terms of periodic Anderson models, which contain a hybridization between the localized $f$ electron moments and itinerant $spd$ electrons together with the Coulomb repulsion for the $f$ electrons.
Solving for the heavy fermion state, we identify nodes of chiral Weyl semimetal types in the Kondo-driven low-energy excitations. 
Moreover, we develop the first materials search procedure of any kind that overcomes the limitations for strongly correlated settings.
This procedure adds to the usual materials database search the step of searching through experimental signatures of desired properties; while currently labor-intensive, the procedure is apt for further empowerment by artificial-intelligence methods.
Where the existing experimental literature is uncertain, we have carried out new materials synthesis and experimental measurements to resolve the ambiguity.
Already, guided by our theoretical results, we have applied the procedure to identify a candidate chiral Weyl-Kondo semimetal CePt$_2$B as well as its achiral counterparts Ce$_2$NiGe$_3$ and Ce$_6$Co$_{2-\delta}$Si$_3$. 
The chiral Weyl Kondo semimetals advanced here serve as a platform for novel electronic transport properties, including the spontaneous Hall response and CPGE.
Furthermore, the hexagonal crystals contain magnetic correlations that are geometrically frustrated, where in the ground state the magnetic moments of the localized electrons could have extremely enhanced quantum fluctuations.
Thus, our work raises the prospect to realize novel Kondo-driven topological metals in the setting of exotic quantum magnetism, thereby opening a new frontier in quantum materials research.

\medskip

\noindent{\bf\large Methods}
\\
\noindent {\bf{Periodic Anderson models} ~~~}
We base our investigation of correlated topological semimetals on the Anderson/Kondo-lattice models.
The Anderson lattice model contains $n_{sub}$ sub-lattices per unit cell, which could be related to each other by the nonsymmorphic operation with a fractional Bravais lattice translation.
On each sub-lattice, we place spin-1/2 itinerant $c$ electrons and localized $f$ electrons.
Hence, there are $4 n_{sub}$ basis states per unit cell.
We denote the fermionic annihilation and creation operators for $c$ [$f$] electrons as $c_{\mathbf{R},\alpha,\sigma}$ and $c^\dag_{\mathbf{R},\alpha,\sigma}$ [$f_{\mathbf{R},\alpha,\sigma}$ and $f^\dag_{\mathbf{R},\alpha,\sigma}$], where $\mathbf{R}$ indicates the unit cell (integer linear combination of the Bravais lattice vectors $\mathbf{a}_1$, $\mathbf{a}_2$, and $\mathbf{a}_3$ in Fig.~\ref{fig:hexagonal_Bravais_lattice_BZ_schematic_model}(a)), $\alpha$ indicates the sub-lattice, and $\sigma$ indicates the spin ($\uparrow$ or $\downarrow$).
The Hamiltonian is given by~\cite{Hewson1997}
\begin{align}
    H = & \sum_{\mathbf{R},\mathbf{R}',\alpha,\sigma,\alpha',\sigma'} c^\dagger_{\mathbf{R}'+\mathbf{R},\alpha,\sigma} t^c_{(\alpha,\sigma)\leftarrow(\alpha',\sigma')}(\mathbf{R}) c_{\mathbf{R}',\alpha',\sigma'} \nonumber \\
    & + E_f \sum_{\mathbf{R},\alpha,\sigma} n^f_{\mathbf{R},\alpha,\sigma} -\mu \sum_{\mathbf{R},\alpha,\sigma}\left( n^c_{\mathbf{R},\alpha,\sigma} + n^f_{\mathbf{R},\alpha,\sigma} \right) \nonumber \\
    & + V \sum_{\mathbf{R},\alpha,\sigma} \left( c^\dagger_{\mathbf{R},\alpha,\sigma} f_{\mathbf{R},\alpha,\sigma} + f^\dagger_{\mathbf{R},\alpha,\sigma} c_{\mathbf{R},\alpha,\sigma} \right) \nonumber \\
    & + U \sum_{\mathbf{R},\alpha} n^f_{\mathbf{R},\alpha,\uparrow} n^f_{\mathbf{R},\alpha,\downarrow}, \label{eq:periodic_Anderson_model}
\end{align}
where $n^c_{\mathbf{R},\alpha,\sigma} \equiv c^\dag_{\mathbf{R},\alpha,\sigma} c_{\mathbf{R},\alpha,\sigma}$ and $n^f_{\mathbf{R},\alpha,\sigma} \equiv f^\dag_{\mathbf{R},\alpha,\sigma} f_{\mathbf{R},\alpha,\sigma}$.
$t^c_{(\alpha,\sigma)\leftarrow(\alpha',\sigma')}(\mathbf{R}) \in \mathbb{C}$ is the hopping strength for a $c$ electron from sub-lattice $\alpha'$ with spin $\sigma'$ in an arbitrary unit cell $\mathbf{R}'$ to sub-lattice $\alpha$ with spin $\sigma$ in the unit cell $\mathbf{R}'+\mathbf{R}$.
Through the hopping strengths, the $c$ electrons form dispersive energy bands and Bloch states.
$E_f$ is the on-site energy of the localized $f$ electrons, which in our study will be chosen to be much lower than the energy bands of the itinerant $c$ electrons.
$\mu$ is the chemical potential which controls the total number of electrons, \textit{i.e.} filling, of our system.
Our approach highlights the role of space-group symmetry.
We consider the on-site Hubbard interaction with a strength $U > 0$ for the localized $f$ electrons.
For each nonsymmorphic hexagonal SG, the parameters of $H$ are chosen to respect the space-group symmetry as well as the spin-1/2 time-reversal symmetry.
A schematic representation of our model for $n_{sub} = 2$ with sub-lattices located at $(0,0,0)$ and $(0,0,1/2)$ within the unit cell is provided in Fig.~\ref{fig:hexagonal_Bravais_lattice_BZ_schematic_model}(d).
Through the Kondo effect, the localized $f$ moments are screened by the itinerant $c$ electrons, inducing Kondo singlets of the form $\frac{1}{\sqrt{2}}\left( | \uparrow \rangle_f | \downarrow \rangle_c - | \downarrow \rangle_f | \uparrow \rangle_c \right)$ [Fig.~\ref{fig:hexagonal_Bravais_lattice_BZ_schematic_model}(d)].
A demonstration of our hexagonal periodic Anderson model with the formation of Kondo singlets is shown in Fig.~\ref{fig:hexagonal_Bravais_lattice_BZ_schematic_model}(d), which leads to a paramagnetic phase with heavy-fermion excitations.
Without a loss of generality for the heavy fermion state, we obtain the excitation spectrum of this interacting Hamiltonian in the infinite-$U$ limit using the parton saddle-point calculation by introducing an auxiliary boson field $b$ that characterizes the valence fluctuations ($f^1 \rightleftharpoons f^0 + e^{-}$) of $f$ electrons.
We provide details of the numerical implementation of the parton saddle-point calculation in Supplementary Note 2.

\noindent {\bf{Density functional theory calculations} ~~~}
In order to validate the symmetry analysis encoded in our model calculation,  we performed DFT calculations.
First-principles calculations for CePt$_2$B and Ce$_2$NiGe$_3$ were carried out in the framework of the Perdew--Burke--Ernzerhof type generalized gradient approximation of the density functional theory~\cite{perdew1996generalized}. 
The projector augmented-wave (PAW) method was employed as implemented in the Vienna \textit{ab initio} simulation package (VASP)~\cite{kresse1993initiomolecular, kresse1996efficiency,kresse1996efficient, kresse1999ultrasoft}. 
A plane-wave cutoff energy of 500~eV was used throughout the calculations. 
The BZ was sampled using a $9 \times 9 \times 4$ $k$-point mesh for CePt$_2$B and a $5 \times 5 \times 4$ $k$-point mesh for Ce$_2$NiGe$_3$. 
The irreducible representations of the electronic bands were calculated using the Irvsp program~\cite{gao2021irvsp}.

\vskip 0.5 cm
\noindent{\bf\large Data availability}
\\
\R{The data that support the findings of this study are available from the corresponding author upon request.}

\vskip 0.5 cm
\noindent{\bf\large Code availability}
\\
The computer codes that were used to generate the data that support the findings of this study are available from the corresponding author upon request.

\vskip 0.5 cm
\noindent $^\dagger$ These authors contributed equally.

\bibliographystyle{naturemagallauthors}

\bibliography{reference}

\clearpage

\medskip

\noindent{\bf Acknowledgment:}~~
We thank Mounica Mahankali, Shouvik Sur, Eric Bauer, Lei Chen, Adolfo Grushin, and Joel Moore for useful discussions. Work at Rice has been supported by the National Science Foundation under Grant No. DMR-2220603 (K.-S.L., Y.F. and F.X.), and by the Robert A. Welch Foundation Grant No. C-1411 and the Vannevar Bush Faculty Fellowship ONR-VB N00014-23-1-2870 (Q.S.). K.-S.L. acknowledges the Carl and Lillian Illig Postdoctoral Fellowship from the Smalley-Curl Institute at Rice University. The majority of the computational calculations have been performed on the Shared University Grid at Rice funded by NSF under Grant EIA-0216467, a partnership between Rice University, Sun Microsystems, and Sigma Solutions, Inc., the Big-Data Private-Cloud Research Cyberinfrastructure MRI-award funded by NSF under Grant No. CNS-1338099, and the Extreme Science and Engineering Discovery Environment (XSEDE) by NSF under Grant No. DMR170109. H.F., A.P. and S.P. acknowledge funding by the European Union (ERC, CorMeTop, project 101055088). R. L. and M.G.V. acknowledge support to the  Spanish Ministerio de Ciencia e Innovacion (grant PID2022-142008NB-I00), partial support from European Research Council (ERC) grant agreement no. 101020833 and the European Union NextGenerationEU/PRTR-C17.I1, as well as by the IKUR Strategy under the collaboration agreement between Ikerbasque Foundation and DIPC on behalf of the Department of Education of the Basque Government and the Ministry for Digital Transformation and of Civil Service of the Spanish Government through the QUANTUM ENIA project call - Quantum Spain project, and by the European Union through the Recovery, Transformation and Resilience Plan - NextGenerationEU within the framework of the Digital Spain 2026 Agenda. M.G.V. and S.P. acknowledge funding from the Deutsche Forschungsgemeinschaft (DFG, German Research Foundation) and the Austrian Science Fund (FWF) through the project FOR 5249 (QUAST). M.G.V. received financial support from the Canada Excellence Research Chairs Program for Topological Quantum Materials. J.C. acknowledges the support of the National Science Foundation under Grant No. DMR-1942447, support from the Alfred P. Sloan Foundation through a Sloan Research Fellowship and the support of the Flatiron Institute, a division of the Simons Foundation. J.C. and Q.S. acknowledge the hospitality of the Aspen Center for Physics, which is supported by the National Science Foundation under Grant No. PHY-2210452.

\vspace{0.2cm}
\noindent{\bf Author contributions}\\
Q.S. and J.C. conceived the research. K.-S.L., Y.F., F.X., J.C. and Q.S. carried out theoretical model studies. R.L. and M.G.V. performed DFT calculations. K.-S.L., Y.F., H.F., A.P., S.P., Q.S. developed the new procedure for correlated topological materials search and applied it to
identify candidate materials for the proposed correlated topological semimetals. H.F., A.P., S.P. synthesized the materials and performed the physical properties measurements.  K.-S.L., Y.F., and Q.S. wrote the manuscript, with inputs from all authors.

\vspace{0.2cm}
\noindent{\bf Competing 
 interests}\\
The authors declare no competing interests.

\vspace{0.2cm}
\noindent{\bf Additional information}\\
Correspondence and requests for materials should be addressed to 
Q.S. (qmsi@rice.edu).

\clearpage
\begin{table}[th]
    \centering
    \begin{tabular}{c|cc|c|c|c|c|c}
        \multicolumn{8}{c}{Primary candidate heavy fermion material} \\
        \hline
        Material & SG \# & SG & $T_{\rm N}$ & $T_{\rm K}$ & Resistivity & Sommerfeld coefficient & Crossing \\
        \hline
        \hline
        \multicolumn{8}{c}{Chiral heavy fermion material} \\
        \hline
        \hline
        CePt$_2$B & $180$ & $P 6_2 22$ & $2.1$K & $3.5 - 5$K & Metallic & Large $C_p / T$ & WP \\
        \hline
        \hline
        \multicolumn{8}{c}{Achiral heavy fermion material} \\
        \hline
        \hline
        Ce$_2$NiGe$_3$ & $190$ & $P \bar{6} 2c$ & $3.2$K &  & Semimetallic & $25$ mJ/K$^2$ Ce-mol & WNL \\
        \hline
        Ce$_2$RhSi$_3$ & $190$ & $P \bar{6} 2c$ & $4.5-6.8$K & $9-12$K & Semimetallic & Large & WNL \\
        \hline
        Ce$_6$Co$_{2-\delta}$Si$_3$ & $176$ & $P 6_3 / m$ & \xmark &  &  & $162$ mJ/K$^2$ Ce-mol & DNL \\
        \hline
        Ce$_6$Rh$_{32}$P$_{17}$ & $176$ & $P 6_3 / m$ & \xmark &  & Semimetallic &  & DNL 
    \end{tabular}
    \caption{Summary of the physical properties of the primary candidate materials Ce$_6$Rh$_{32}$P$_{17}$~\cite{pivan1988crystal}, Ce$_6$Co$_{2-\delta}$Si$_3$~\cite{gaudin2007on,chevalier2007the}, CePt$_2$B~\cite{sologub2000newstructuretype,lackner2005lowtemperature}, Ce$_2$NiGe$_3$~\cite{huo2001electric,kalsi2014neutron}, and Ce$_2$RhSi$_3$~\cite{szytula1993neutron,leciejewicz1995antiferromagnetic,chevalier1984anew,das1994magnetic,kase2009antiferromagnetic,szlawska2009antiferromagnetic}. ``SG'' means ``space group''. $T_{\rm N}$ denotes the anti-ferromagnetic transition temperature. A ``\xmark'' mark in $T_{\rm N}$ means no magnetic transition is reported. 
    $T_{\rm K}$ denotes the Kondo temperature.
    $C_p$ denotes the specific heat. A blank entry means the physical property is not reported. ``DNL'' means ``Dirac nodal line'', ``WP'' means ``Weyl point'', and ``WNL'' means ``Weyl nodal line''. 
    We note that in addition to SG no.~190 ($P \bar{6} 2c$)~\cite{kalsi2014neutron,szytula1993neutron,leciejewicz1995antiferromagnetic}, Ce$_2$NiGe$_3$~\cite{huo2001electric} and Ce$_2$RhSi$_3$~\cite{chevalier1984anew,das1994magnetic,kase2009antiferromagnetic,szlawska2009antiferromagnetic} may crystallize in other forms.
    We also note that the chemical formula for Ce$_6$Co$_{2-\delta}$Si$_3$ is Ce$_6$Co$_{1.67}$Si$_3$, corresponding to $\delta = 0.33$~\cite{gaudin2007on,chevalier2007the}.
    We further note that the magnetic transition of Ce$_2$NiGe$_3$ at $T_{\rm N} = 3.2$K is suggested to be a transition into a spin-glass state~\cite{huo2001electric,kalsi2014neutron}.
    We refer readers to Supplementary Note 12 for a detailed description of the primary candidate materials, such as how different physical properties are determined.
    In Supplementary Note 15 we present the results of our new experiments on Ce$_2$NiGe$_3$, which not only confirm properties reported in the literature, such as the SG no.~190 ($P \bar{6} 2c$), but also provide new experimental data such as magnetoresistivity and Hall resistivity.
    We have also confirmed the presence of Weyl points and Weyl nodal lines in CePt$_2$B [Fig.~\ref{fig:DFT_band_structures_main_text}] and Ce$_2$NiGe$_3$ [Supplementary Fig.~S14], respectively, based on DFT calculations.}
    \label{tab:primary_candidate_materials_table_main_text}
\end{table}

\clearpage

\clearpage

\clearpage

\clearpage

\onecolumngrid 
\begin{center}
\textbf{\large Supplementary information for: Chiral Weyl-Kondo semimetals and hexagonal heavy fermion systems}
\end{center}

\setcounter{secnumdepth}{2}
\setcounter{equation}{0}
\setcounter{figure}{0}
\setcounter{table}{0}
\renewcommand{\theequation}{S\arabic{equation}}
\renewcommand{\figurename}{Supplementary Fig.}
\renewcommand{\thefigure}{S\arabic{figure}}
\renewcommand{\tablename}{Supplementary Table}
\renewcommand{\thetable}{S\arabic{table}}

\renewcommand{\theHfigure}{S\arabic{figure}}
\renewcommand{\theHequation}{S\arabic{equation}}
\renewcommand{\theHtable}{S\arabic{table}}

\renewcommand{\thesection}{Supplementary Note~\arabic{section}}
\setcounter{secnumdepth}{3}
\renewcommand{\thesubsection}{\arabic{subsection}}

\tableofcontents

\section{Details of the tight-binding models}

In this section, we provide the details of our tight-binding models. 
Our models are partly constructed using the open-source Python Tight Binding (PythTB) software package~\cite{pythtb}.

Hexagonal crystal systems describe crystal structures with a hexagonal Bravais lattice and a sixfold rotation (or roto-inversion) axis.
The Bravais lattice vectors of the hexagonal crystal system are $\mathbf{a}_1 = (a,0,0)$, $\mathbf{a}_2 = (-a/2,\sqrt{3}a/2,0)$, and $\mathbf{a}_3 = (0,0,c)$, where $a$ and $c$ are the hexagonal lattice constants.
Throughout our studies, all the position-space vectors are represented in the reduced coordinate, which means $(x_1,x_2,x_3)$ represents $x_1 \mathbf{a}_1 + x_2 \mathbf{a}_2 + x_3 \mathbf{a}_3$.

The hopping strength $t^c_{(\alpha,\sigma)\leftarrow(\alpha',\sigma')} (\mathbf{R}) \in \mathbb{C}$ for the $c$ electrons can be represented as a $2 \times 2$ matrix $[ t^c_{\alpha\leftarrow \alpha'} (\mathbf{R}) ]$ whose matrix element $[ t^c_{\alpha\leftarrow \alpha'} (\mathbf{R}) ]_{\sigma,\sigma'} \equiv t^c_{(\alpha,\sigma)\leftarrow(\alpha',\sigma')} (\mathbf{R})$.
Specifically, when $[ t^c_{\alpha\leftarrow \alpha'} (\mathbf{R}) ]$ does not represent an on-site potential, we require that $[ t^c_{\alpha\leftarrow \alpha'} (\mathbf{R}) ]$ takes the following form~\cite{po2017symmetry,bradlyn2019disconnected}
\begin{equation}
    t_0 \sigma_0 + i \vec{t} \cdot \vec{\sigma} = t_0 \sigma_0 + i\left( t_1 \sigma_1 + t_2 \sigma_2 + t_3 \sigma_3 \right)
\end{equation}
where $t_\mu \in \mathbb{R}$ ($\mu = 0,1,2,3$), $\sigma_0$ is the $2 \times 2$ identity matrix $\begin{pmatrix} 1 & 0 \\ 0 & 1 \end{pmatrix}$, $\sigma_j$ ($j = 1,2,3$) are the Pauli matrices given by
\begin{equation}
    \sigma_1 = \begin{pmatrix} 0 & 1 \\ 1 & 0 \end{pmatrix}, \  \sigma_2 = \begin{pmatrix} 0 & -i \\ i & 0 \end{pmatrix}, \  \sigma_3 = \begin{pmatrix} 1 & 0 \\ 0 & -1 \end{pmatrix}.
\end{equation}
$t_0$ represents the spin-independent part of the hopping strength, and $t_{j}$ ($j=1,2,3$) represents the spin-dependent part of the hopping strength.
This form of $[ t^c_{\alpha\leftarrow \alpha'} (\mathbf{R}) ]$ automatically preserves the spin-1/2 time-reversal symmetry.
When $\alpha = \alpha'$ and $\mathbf{R} = \mathbf{0}$, $[ t^c_{\alpha\leftarrow \alpha} (\mathbf{0}) ]$ represents an on-site potential which we require to take the form $m \sigma_0$ ($m \in \mathbb{R}$) to preserve the spin-1/2 time-reversal symmetry.
To ensure the Hermiticity of the Hamiltonian, $\left( t^c_{(\alpha,\sigma)\leftarrow(\alpha',\sigma')} (\mathbf{R}) \right)^* = t^c_{(\alpha',\sigma')\leftarrow(\alpha,\sigma)} (-\mathbf{R})$.

We here outline the procedure to construct a model respecting a certain space group symmetry.
Consider a space group $SG$ and a lattice with lattice translation $\mathbb{T}$, which is a normal subgroup of $SG$.
The most general $\mathbb{T}$-symmetric tight-binding Hamiltonian takes the form
\begin{equation}
    H = \sum_{\mathbf{R}'} \sum_{\mathbf{R},(\alpha,\sigma),(\alpha',\sigma')} t^c_{(\alpha,\sigma)\leftarrow(\alpha',\sigma')}(\mathbf{R}) c^\dagger_{\mathbf{R}'+\mathbf{R},\alpha,\sigma} c_{\mathbf{R}',\alpha',\sigma'} \,.
\end{equation}
Consider a coset-decomposition 
\begin{equation}
    SG = g_1 \mathbb{T} + g_2 \mathbb{T} + \dots g_n \mathbb{T} \,,
\end{equation}
where the representatives $g_1,\dots,g_n \in SG$ and they act on real space position vector $\mathbf{r}$,
\begin{equation}
    g_i = \{R_i|\mathbf{t}_i\} \,,\quad g_i \mathbf{r} =R_i \mathbf{r} + \mathbf{t}_i \,.
\end{equation}
Here $R_i$ and $\mathbf{t}_i$ stands for rotational and translational part of the symmetry element. 
Denote the set of coset-representatives $\mathcal{C}=\{g_1,\dots g_n\}$. Note $\mathcal{C}$ needs not be a group.

The symmetric Hamiltonians should satisfy conditions~\cite{po2017symmetry,bradlyn2019disconnected}
\begin{align}
    H &= \frac{1}{|\mathcal{C}|}\sum_{g\in \mathcal{C}} g H g^{-1} \,, \\ 
    H &= \frac12 \left( H+ \mathcal{T} H \mathcal{T}^{-1} \right)  \,, \\ 
    H &= \frac12 \left( H + H^\dagger \right) \,. 
\end{align}
By examining these constraints, we can derive the symmetric hopping terms including SOC terms for real space tight-binding models.

We list the parameters of our model with space group no.~173 ($P6_3$) in Supplementary Table~\ref{tab:SG_173_parameters},
space group no.~176 ($P6_3/m$) in Supplementary Table~\ref{tab:SG_176_parameters},
space group no.~180 ($P6_2 22$) in Supplementary Table~\ref{tab:SG_180_parameters},
and space group no.~190 ($P\bar{6}2c$) in Supplementary Table~\ref{tab:SG_190_parameters}.

\begin{table}[ht]
    \centering
    \begin{tabular}{|c|c|c|c|c|c|c|}
         \hline
         $(\alpha,\sigma)$ & $(\alpha',\sigma')$ & $n_1$ & $n_2$ & $n_3$ & $t^c_{(\alpha,\sigma)\leftarrow (\alpha',\sigma')} (\mathbf{R})$ & \# of related hoppings \\
         \hline
         $(1,\uparrow)$ & $(0,\uparrow)$ & $0$ & $0$ & $-1$ & $1+i$ &8 \\ 
         \hline 
         $(0,\uparrow)$ & $(0,\uparrow)$ & $-1$ & $0$ & $0$ & $0.5+i$ &24 \\ 
         \hline
         $(0,\uparrow)$ & $(0,\uparrow)$ & $-2$ & $-1$ & $0$ & $0.08i$ &24 \\ 
         \hline 
         $(0,\downarrow)$ & $(0,\uparrow)$ & $-2$ & $-1$ & $0$ & $0.08$ &24 \\ 
         \hline
         $(0,\uparrow)$ & $(0,\uparrow)$ & $-1$ & $-1$ & $-1$ & $0.16$ &24 \\ 
         \hline
         $(0,\downarrow)$ & $(0,\uparrow)$ & $-1$ & $-1$ & $-1$ & $0.16+0.16i$ &24 \\ 
         \hline
         $(1,\uparrow)$ & $(0,\uparrow)$ & $-1$ & $-1$ & $0$ & $0.16$ &24 \\ 
         \hline
         $(1,\downarrow)$ & $(0,\uparrow)$ & $-1$ & $-1$ & $0$ & $0.16+0.16i$ &24 \\ 
         \hline
    \end{tabular}
    \caption{Independent tight-binding parameters for the model with space group no.~173 ($P 6_3$).}
    \label{tab:SG_173_parameters}
\end{table}

\begin{table}[ht]
    \centering
    \begin{tabular}{|c|c|c|c|c|c|c|}
         \hline
         $(\alpha,\sigma)$ & $(\alpha',\sigma')$ & $n_1$ & $n_2$ & $n_3$ & $t^c_{(\alpha,\sigma)\leftarrow (\alpha',\sigma')} (\mathbf{R})$ & \# of related hoppings \\
         \hline
         $(1,\uparrow)$ & $(0,\uparrow)$ & $0$ & $0$ & $-1$ & $1.0$ &8 \\
         \hline
         $(1,\uparrow)$ & $(0,\uparrow)$ & $-1$ & $-1$ & $-1$ & $0.6i$ &48\\
         \hline
         $(1,\uparrow)$ & $(0,\downarrow)$ & $-1$ & $-1$ & $-1$ & $0.6$ &48\\
         \hline
         $(1,\uparrow)$ & $(0,\uparrow)$ & $-2$ & $-1$ & $-1$ & $0.6i$ &48\\
         \hline
         $(1,\uparrow)$ & $(0,\downarrow)$ & $-2$ & $-1$ & $-1$ & $0.6$ &48\\
         \hline
    \end{tabular}
    \caption{Independent tight-binding parameters for the model with space group no.~176 ($P 6_3/m$).}
    \label{tab:SG_176_parameters}
\end{table}

\begin{table}[ht]
    \centering
    \begin{tabular}{|c|c|c|c|c|c|c|}
         \hline
         $(\alpha,\sigma)$ & $(\alpha',\sigma')$ & $n_1$ & $n_2$ & $n_3$ & $t^c_{(\alpha,\sigma)\leftarrow (\alpha',\sigma')} (\mathbf{R})$ & \# of related hoppings \\
         \hline
         $(2,\uparrow)$ & $(0,\uparrow)$ & $0$ & $0$ & $-1$ & $1.0$ &12 \\
         \hline
         $(0,\uparrow)$ & $(0,\uparrow)$ & $-1$ & $-1$ & $0$ & $0.1$ &24\\
         \hline
         $(0,\uparrow)$ & $(0,\downarrow)$ & $-1$ & $-1$ & $0$ & $0.2$ &24\\
         \hline
         $(1,\uparrow)$ & $(1,\downarrow)$ & $-1$ & $-1$ & $0$ & $0.2$ &12\\
         \hline
    \end{tabular}
    \caption{Independent tight-binding parameters for the model with space group no.~180 ($P 6_2 22$).}
    \label{tab:SG_180_parameters}
\end{table}

\begin{table}[ht]
    \centering
    \begin{tabular}{|c|c|c|c|c|c|c|}
         \hline
         $(\alpha,\sigma)$ & $(\alpha',\sigma')$ & $n_1$ & $n_2$ & $n_3$ & $t^c_{(\alpha,\sigma)\leftarrow (\alpha',\sigma')} (\mathbf{R})$ & \# of related hoppings \\
         \hline
         $(1,\uparrow)$ & $(0,\uparrow)$ & $0$ & $0$ & $-1$ & $1.0$ &8 \\
         \hline
         $(0,\uparrow)$ & $(0,\downarrow)$ & $-1$ & $-1$ & $0$ & $0.2$ &24\\
         \hline
         $(1,\uparrow)$ & $(0,\downarrow)$ & $-1$ & $-1$ & $-1$ & $0.5$ &48\\
         \hline
         $(1,\uparrow)$ & $(0,\uparrow)$ & $-1$ & $-1$ & $0$ & $0.2$ &48\\
         \hline
    \end{tabular}
    \caption{Independent tight-binding parameters for the model with space group no.~190 ($P \bar6 2c$).}
    \label{tab:SG_190_parameters}
\end{table}

\section{Parton formalism}
In this section, we review the parton formalism, which could be employed to study mixed-valence phenomena in Kondo-lattice systems, typically formulated microscopically as periodic Anderson models.
We will also provide our numerical procedure, as well as the assumptions we made, in obtaining the saddle-point solutions of the periodic Anderson models~\cite{Hewson1997}.

We begin by writing down a periodic Anderson model with some generality.
The itinerant $c$ electrons and the localized $f$ electrons form such a system.
We will denote the unit cell by $\mathbf{R} = \sum_{j=1}^3 n_j \mathbf{a}_j$ ($n_j \in \mathbb{Z}$), the sub-lattice within the unit cell by $\alpha = 1 \ldots n_{sub}$, and the spin by $\sigma = \uparrow,$ $\downarrow$.
We assume that both the $c$ and $f$ electrons are specified by $\mathbf{R}$, $\alpha$, and $\sigma$.
In other words, each sub-lattice hosts both spin-1/2 $c$ and $f$ orbitals.
The periodic Anderson model we consider here can be written as Eq.~\eqref{eq:periodic_Anderson_model}.

In the infinite-$U$ limit, the number of $f$ electrons on each site (with fixed $\mathbf{R}$ and $\alpha$) can only be $0$ or $1$, leading to the non-holomorphic constraint $\sum_{\sigma} \left\langle f^{\dagger}_{\mathbf{R},\alpha,\sigma} f_{\mathbf{R},\alpha,\sigma} \right\rangle = \sum_{\sigma} \langle n^f_{\mathbf{R},\alpha,\sigma} \rangle \leq 1$.
Hereafter, $\langle \cdots \rangle$ is understood as the ground state expectation value at zero temperature or the thermal average at finite temperature.
To treat the constraint $\sum_{\sigma} \langle n^f_{\mathbf{R},\alpha,\sigma} \rangle \leq 1$, an auxiliary boson operator $b$ and a Lagrange multiplier $\lambda$ are introduced such that the periodic Anderson model in the infinite-$U$ limit in the auxiliary boson representation is
\begin{align}
    & H_{\mathrm{auxiliary-boson}} = \sum_{\mathbf{R},\mathbf{R}',\alpha,\sigma,\alpha',\sigma'} c^\dagger_{\mathbf{R}'+\mathbf{R},\alpha,\sigma} t^c_{(\alpha,\sigma)\leftarrow(\alpha',\sigma')}(\mathbf{R}) c_{\mathbf{R}',\alpha',\sigma'} \nonumber \\
    & + E_f \sum_{\mathbf{R},\alpha,\sigma} n^f_{\mathbf{R},\alpha,\sigma} -\mu \sum_{\mathbf{R},\alpha,\sigma}\left( n^c_{\mathbf{R},\alpha,\sigma} + n^f_{\mathbf{R},\alpha,\sigma} \right) \nonumber \\
    & + V \sum_{\mathbf{R},\alpha,\sigma} \left( c^\dagger_{\mathbf{R},\alpha,\sigma} \left( b^\dagger f_{\mathbf{R},\alpha,\sigma} \right) + \left( f^\dagger_{\mathbf{R},\alpha,\sigma} b \right) c_{\mathbf{R},\alpha,\sigma} \right) \nonumber \\
    & + \lambda \sum_{\mathbf{R},\alpha} \left( \left[ \sum_{\sigma} f^\dagger_{\mathbf{R},\alpha,\sigma} f_{\mathbf{R},\alpha,\sigma}  \right] + b^\dagger b -1 \right),
\end{align}
where in the $c$-$f$ hybridization term with $V$ the localized $f$ electron operators $f^\dagger_{\mathbf{R},\alpha,\sigma}$ and $f_{\mathbf{R},\alpha,\sigma}$ are replaced by $f^\dagger_{\mathbf{R},\alpha,\sigma} b$ and $b^\dagger f_{\mathbf{R},\alpha,\sigma}$, respectively.
At the saddle-point level, a solution can be obtained by considering the condensation of the auxiliary boson operator, which means effectively replacing $b$ by $\langle b \rangle$.
We may absorb the complex phase of $\langle b \rangle$ by the $f$ electron operators such that we further replace $\langle b \rangle$ by $r$, where $r$ is a real number.
This leads to
\begin{align}
    & H_{\mathrm{auxiliary-boson}}^{\mathrm{saddle-point}} = \sum_{\mathbf{R},\mathbf{R}',\alpha,\sigma,\alpha',\sigma'} c^\dagger_{\mathbf{R}'+\mathbf{R},\alpha,\sigma} t^c_{(\alpha,\sigma)\leftarrow(\alpha',\sigma')}(\mathbf{R}) c_{\mathbf{R}',\alpha',\sigma'} \nonumber \\
    & + E_f \sum_{\mathbf{R},\alpha,\sigma} n^f_{\mathbf{R},\alpha,\sigma} -\mu \sum_{\mathbf{R},\alpha,\sigma}\left( n^c_{\mathbf{R},\alpha,\sigma} + n^f_{\mathbf{R},\alpha,\sigma} \right) \nonumber \\
    & + r V \sum_{\mathbf{R},\alpha,\sigma} \left( c^\dagger_{\mathbf{R},\alpha,\sigma}   f_{\mathbf{R},\alpha,\sigma} +  f^\dagger_{\mathbf{R},\alpha,\sigma}  c_{\mathbf{R},\alpha,\sigma} \right) \nonumber \\
    & + \lambda \sum_{\mathbf{R},\alpha} \left( \left[ \sum_{\sigma} f^\dagger_{\mathbf{R},\alpha,\sigma} f_{\mathbf{R},\alpha,\sigma}  \right] + r^2 -1 \right).
\end{align}
In $H_{\mathrm{auxiliary-boson}}^{\mathrm{saddle-point}}$, the hybridization strength $V$ is re-normalized by a factor of $r \in \mathbb{R}$.
The term with the Lagrange multiplier $\lambda$ is to implement the constraint $\sum_{\sigma} \left\langle f^{\dagger}_{\mathbf{R},\alpha,\sigma} f_{\mathbf{R},\alpha,\sigma} \right\rangle = \sum_{\sigma} \langle n^f_{\mathbf{R},\alpha,\sigma} \rangle \leq 1$ via $1 = \left[ \sum_{\sigma} \left\langle f^\dagger_{\mathbf{R},\alpha,\sigma} f_{\mathbf{R},\alpha,\sigma} \right\rangle  \right] + r^2$, which means that $r^2$ is equal to how much charge is lost from the localized $f$ electron moment per site due to the interaction with the itinerant $c$ electrons.
The number of $f$ electrons on each site is then equal to $1- r^2$.

In order to solve for the excitation spectrum, a set of self-consistent equations is required in order to determine $r$, $\lambda$, and $\mu$.
The saddle-point solution for $r$ and $\lambda$ can be obtained by requiring
\begin{equation}
    \left\langle \frac{\delta H_{\mathrm{auxiliary-boson}}^{\mathrm{saddle-point}} }{\delta \lambda} \right\rangle =  0,\ \left\langle \frac{\delta H_{\mathrm{auxiliary-boson}}^{\mathrm{saddle-point}} }{\delta r} \right\rangle =  0,
\end{equation}
which leads to the following self-consistent equations
\begin{align}
    & r = \sqrt{1 - \frac{1}{n_{sub}}\sum_{\alpha,\sigma} \left\langle f^\dagger_{\mathbf{R},\alpha,\sigma} f_{\mathbf{R},\alpha,\sigma} \right\rangle}, \\
    & \lambda = \frac{-V}{r \cdot n_{sub}} \sum_{\alpha,\sigma} \mathrm{Re} \left\langle c^\dagger_{\mathbf{R},\alpha,\sigma} f_{\mathbf{R},\alpha,\sigma} \right\rangle,
\end{align}
respectively.
In obtaining the expressions of $r$ and $\lambda$, we have assumed that the saddle-point solution preserves the original translation symmetry of $H$, such that $\left\langle f^\dagger_{\mathbf{R},\alpha,\sigma} f_{\mathbf{R},\alpha,\sigma} \right\rangle$ and $\left\langle c^\dagger_{\mathbf{R},\alpha,\sigma} f_{\mathbf{R},\alpha,\sigma} \right\rangle$ are independent of $\mathbf{R}$. 
We have also assumed that the saddle-point solution preserves the original space-group symmetry in $H$, including the spin-1/2 time-reversal symmetry.
We have further assumed that all the sub-lattices are related to each other by the symmetry in the space group, otherwise one would need to consider different parameters $r_\alpha$ and $\lambda_\alpha$ ($\alpha = 1\ldots n_{sub}$), each of which has a self-consistent equation, for different sub-lattices.

$H_{\mathrm{auxiliary-boson}}^{\mathrm{saddle-point}}$ can also be written as
\begin{equation}
    H_{\mathrm{auxiliary-boson}}^{\mathrm{saddle-point}} = \left( \sum_{\mathbf{k}} \Psi^\dagger_{\mathbf{k}} [H_{eff} (\mathbf{k}) ] \Psi_{\mathbf{k}} \right) + \lambda N_{cell} n_{sub} (r^2 -1),
\end{equation}
where 
\begin{widetext}
\begin{equation}
    \Psi^\dagger_{\mathbf{k}} = \begin{pmatrix} c^\dagger_{\mathbf{k},1,\uparrow} & c^\dagger_{\mathbf{k},1,\downarrow} & \cdots & c^\dagger_{\mathbf{k},n_{sub},\uparrow} & c^\dagger_{\mathbf{k},n_{sub},\downarrow} & f^\dagger_{\mathbf{k},1,\uparrow} & f^\dagger_{\mathbf{k},1,\downarrow} & \cdots & f^\dagger_{\mathbf{k},n_{sub},\uparrow} & f^\dagger_{\mathbf{k},n_{sub},\downarrow} \end{pmatrix}.
\end{equation}
We have adopted the following conventions for the Bloch basis operators
\begin{align}
    & c^\dagger_{\mathbf{k},\alpha,\sigma} = \frac{1}{\sqrt{N_{cell}}} \sum_{\mathbf{R}} e^{i\mathbf{k} \cdot \left( \mathbf{R} + \mathbf{r}^c_{\alpha} \right)} c^\dagger_{\mathbf{R},\alpha,\sigma},\  c_{\mathbf{k},\alpha,\sigma} = \frac{1}{\sqrt{N_{cell}}} \sum_{\mathbf{R}} e^{-i\mathbf{k} \cdot \left( \mathbf{R} + \mathbf{r}^c_{\alpha} \right)} c_{\mathbf{R},\alpha,\sigma}, \\
    & f^\dagger_{\mathbf{k},\alpha,\sigma} = \frac{1}{\sqrt{N_{cell}}} \sum_{\mathbf{R}} e^{i\mathbf{k} \cdot \left( \mathbf{R} + \mathbf{r}^f_{\alpha} \right)} f^\dagger_{\mathbf{R},\alpha,\sigma},\ f_{\mathbf{k},\alpha,\sigma} = \frac{1}{\sqrt{N_{cell}}} \sum_{\mathbf{R}} e^{-i\mathbf{k} \cdot \left( \mathbf{R} + \mathbf{r}^f_{\alpha} \right)} f_{\mathbf{R},\alpha,\sigma},
\end{align}
and their relation to the position-space operators
\begin{align}
    & c^\dagger_{\mathbf{R},\alpha,\sigma} = \frac{1}{\sqrt{N_{cell}}} \sum_{\mathbf{k}} e^{-i\mathbf{k} \cdot \left( \mathbf{R} + \mathbf{r}^c_{\alpha} \right)} c^\dagger_{\mathbf{k},\alpha,\sigma},\  c_{\mathbf{R},\alpha,\sigma} = \frac{1}{\sqrt{N_{cell}}} \sum_{\mathbf{k}} e^{i\mathbf{k} \cdot \left( \mathbf{R} + \mathbf{r}^c_{\alpha} \right)} c_{\mathbf{k},\alpha,\sigma}, \\
    & f^\dagger_{\mathbf{R},\alpha,\sigma} = \frac{1}{\sqrt{N_{cell}}} \sum_{\mathbf{k}} e^{-i\mathbf{k} \cdot \left( \mathbf{R} + \mathbf{r}^f_{\alpha} \right)} f^\dagger_{\mathbf{k},\alpha,\sigma},\ f_{\mathbf{R},\alpha,\sigma} = \frac{1}{\sqrt{N_{cell}}} \sum_{\mathbf{k}} e^{i\mathbf{k} \cdot \left( \mathbf{R} + \mathbf{r}^f_{\alpha} \right)} f_{\mathbf{k},\alpha,\sigma},
\end{align}
\end{widetext}
where $\mathbf{r}^c_\alpha$ and $\mathbf{r}^f_\alpha$ are the positions of the $c$ orbitals and $f$ orbitals, respectively, on the $\alpha$ sub-lattice within the unit cell.
Note that the positions of the basis states do not depend on the spin index $\sigma$.
In our model, we have also assumed $\mathbf{r}^c_\alpha = \mathbf{r}^f_\alpha$.
The effective Bloch Hamiltonian matrix is
\begin{equation}
    [H_{eff} (\mathbf{k}) ] = \begin{pmatrix}
        [H^c (\mathbf{k})] - \mu \mathbb{I} & rV \mathbb{I} \\
        rV \mathbb{I} & \left( E_f - \mu + \lambda \right) \mathbb{I}
    \end{pmatrix},
\end{equation}
where $[H^c (\mathbf{k})]$ is the Bloch Hamiltonian matrix constructed from the hopping strengths $t^c_{(\alpha,\sigma)\leftarrow(\alpha',\sigma')}(\mathbf{R})$ for the $c$ electron through
\begin{equation}
    [H^c (\mathbf{k})]_{(\alpha,\sigma),(\alpha',\sigma')} = \sum_{\mathbf{R}} t^c_{(\alpha,\sigma)\leftarrow(\alpha',\sigma')}(\mathbf{R}) e^{-i\mathbf{k} \cdot \left( \mathbf{R} + \mathbf{r}^c_\alpha - \mathbf{r}^c_{\alpha'} \right)},
\end{equation}
and $\mathbb{I}$ is an identity matrix with an appropriate shape.

We now describe the numerical procedure for determining $(\mu,r,\lambda)$.
To begin, let us first cast the self-consistent equations using the eigenvalues and eigenvectors of $[H_{eff} (\mathbf{k}) ]$:
\begin{equation}
    [H_{eff} (\mathbf{k}) ] | u_{n,\mathbf{k}} \rangle = \epsilon_{n,\mathbf{k}} | u_{n,\mathbf{k}} \rangle,
\end{equation}
where
\begin{widetext}
\begin{equation}
    | u_{n,\mathbf{k}} \rangle = \begin{pmatrix}
    u_{n,\mathbf{k}}^{c,1,\uparrow} & u_{n,\mathbf{k}}^{c,1,\downarrow} & \cdots & u_{n,\mathbf{k}}^{c,n_{sub},\uparrow} & u_{n,\mathbf{k}}^{c,n_{sub},\downarrow} & u_{n,\mathbf{k}}^{f,1,\uparrow} & u_{n,\mathbf{k}}^{f,1,\downarrow} & \cdots & u_{n,\mathbf{k}}^{f,n_{sub},\uparrow} & u_{n,\mathbf{k}}^{f,n_{sub},\downarrow}
    \end{pmatrix}^\mathrm{T}.
\end{equation}
\end{widetext}
The self-consistent equations can be written as
\begin{align}
    & r = \sqrt{1 - \frac{1}{n_{sub}}\frac{1}{N_{cell}}\sum_{\alpha,\sigma} \sum_{\mathbf{k}} \left\langle f^\dagger_{\mathbf{k},\alpha,\sigma} f_{\mathbf{k},\alpha,\sigma} \right\rangle} \\
    & \lambda = \frac{-V}{r \cdot n_{sub}} \frac{1}{N_{cell}} \sum_{\alpha,\sigma} \sum_{\mathbf{k}} \mathrm{Re} \left\langle c^\dagger_{\mathbf{k},\alpha,\sigma} f_{\mathbf{k},\alpha,\sigma} \right\rangle
\end{align}
where we have used the assumption $\mathbf{r}^c_{\alpha} = \mathbf{r}^{f}_{\alpha}$ and that the translation symmetry is preserved.
In terms of the eigenvalues and eigenvectors of $[H_{eff} (\mathbf{k}) ]$, the self-consistent equations can then be written as
\begin{align}
    & r = \sqrt{1 - \frac{1}{n_{sub}} \frac{1}{N_{cell}} {\sum_{n,\mathbf{k}}}'\sum_{\alpha,\sigma}\left|u_{n,\mathbf{k}}^{f,\alpha,\sigma}\right|^2}, \\
    & \lambda = \frac{-V}{r \cdot n_{sub}} \frac{1}{N_{cell}} \mathrm{Re}\left( {\sum_{n,\mathbf{k}}}' \sum_{\alpha,\sigma} \left( u_{n,\mathbf{k}}^{c,\alpha,\sigma} \right)^* u_{n,\mathbf{k}}^{f,\alpha,\sigma} \right),
\end{align}
where thereafter the summation ${\sum_{n,\mathbf{k}}}'$ is over all the occupied eigenvectors $| u_{n,\mathbf{k}} \rangle$.
We will impose the filling constraint such that each sub-lattice in the unit cell has the fermion number expectation value (or thermal average) equal to $1$:
\begin{equation}
    \sum_{\sigma} \left( \left\langle c^\dagger_{\mathbf{R},\alpha,\sigma} c_{\mathbf{R},\alpha,\sigma} \right\rangle + \left\langle f^\dagger_{\mathbf{R},\alpha,\sigma} f_{\mathbf{R},\alpha,\sigma} \right\rangle \right) = 1.
\end{equation}
Note the right-hand side of the above equation is independent of $\mathbf{R}$, since we have assumed that the translation symmetry is preserved.
Since we also assume that different sub-lattices are related to each other by the symmetry in the space group, we have that $\left\langle c^\dagger_{\mathbf{R},\alpha,\sigma} c_{\mathbf{R},\alpha,\sigma} \right\rangle = \left\langle c^\dagger_{\mathbf{R},\alpha',\sigma} c_{\mathbf{R},\alpha',\sigma} \right\rangle$ and $\left\langle f^\dagger_{\mathbf{R},\alpha,\sigma} f_{\mathbf{R},\alpha,\sigma} \right\rangle = \left\langle f^\dagger_{\mathbf{R},\alpha',\sigma} f_{\mathbf{R},\alpha',\sigma} \right\rangle$ with $\alpha \neq \alpha'$, which allows us to cast the filling constraint into
\begin{equation}
    \sum_{\alpha,\sigma} \left( \left\langle c^\dagger_{\mathbf{R},\alpha,\sigma} c_{\mathbf{R},\alpha,\sigma} \right\rangle + \left\langle f^\dagger_{\mathbf{R},\alpha,\sigma} f_{\mathbf{R},\alpha,\sigma} \right\rangle \right) = n_{sub}.
\end{equation}
This means that the total filling of the system is $n_{sub}$ per unit cell.
Written using the eigenvalues and eigenvalues of $[H_{eff} (\mathbf{k}) ]$, the filling constraint becomes
\begin{align}
    \frac{1}{N_{cell}}{\sum_{n,\mathbf{k}}}' \sum_{\alpha,\sigma} \left( \left| u_{n,\mathbf{k}}^{c,\alpha,\sigma} \right|^2 + \left| u_{n,\mathbf{k}}^{f,\alpha,\sigma} \right|^2 \right) & = \frac{1}{N_{cell}}{\sum_{n,\mathbf{k}}}' 1 \nonumber \\
    & = n_{sub},
\end{align}
which means
\begin{equation}
    {\sum_{n,\mathbf{k}}}' 1 = n_{sub}N_{cell}.
\end{equation}
This allows us to consistently determine the chemical potential $\mu$, below which are occupied eigenvectors, and which is used to define the summation ${\sum_{n,\mathbf{k}}}'$.
Since the number of $\mathbf{k}$ points used in the numerical calculation is $N_{cell}$, we have that the number of occupied eigenvectors below the chemical potential is equal to $n_{sub} N_{cell}$.

We will determine $(\mu,r,l)$ iteratively in the following manner.
In the $i^\mathrm{th}$ step, we use a set of $(\mu_i,r_i,\lambda_i)$ to construct $[H_{eff} (\mathbf{k}) ]$ and obtain its eigenvalues $\epsilon_{n,\mathbf{k}}$ and eigenvectors $| u_{n,\mathbf{k}} \rangle$.
We then obtain a new chemical potential $\mu_{i+1}$ that satisfies the filling constraint, and from which we also can obtain the eigenvectors that are occupied.
Next, we use the occupied eigenvectors to obtain a new set of $r$ and $\lambda$ via
\begin{align}
    & \tilde{r}_i = \sqrt{1 - \frac{1}{n_{sub}} \frac{1}{N_{cell}} {\sum_{n,\mathbf{k}}}'\sum_{\alpha,\sigma}\left|u_{n,\mathbf{k}}^{f,\alpha,\sigma}\right|^2}, \\
    & \tilde{\lambda}_i = \frac{-V}{\tilde{r}_i \cdot n_{sub}} \frac{1}{N_{cell}} \mathrm{Re}\left( {\sum_{n,\mathbf{k}}}' \sum_{\alpha,\sigma} \left( u_{n,\mathbf{k}}^{c,\alpha,\sigma} \right)^* u_{n,\mathbf{k}}^{f,\alpha,\sigma} \right).
\end{align}
Note that since the value of $\lambda$ depends on $r$, one may also obtain $\tilde{\lambda}_i$ via
\begin{equation}
    \tilde{\lambda}_i = \frac{-V}{r_i \cdot n_{sub}} \frac{1}{N_{cell}} \mathrm{Re}\left( {\sum_{n,\mathbf{k}}}' \sum_{\alpha,\sigma} \left( u_{n,\mathbf{k}}^{c,\alpha,\sigma} \right)^* u_{n,\mathbf{k}}^{f,\alpha,\sigma} \right),
\end{equation}
using $r_i$ instead of $\tilde{r}_i$.
We then check if $\mu_{i+1}$, $\tilde{r}_i$, and $\tilde{\lambda}_i$ are close enough to $\mu_{i}$, $r_i$, and $\lambda_i$ with certain convergent criteria, which may differ from systems to systems.
If yes, then the iteration stops and $(\mu_{i+1},\tilde{r}_i,\tilde{\lambda}_i)$ is the final result of $(\mu,r,l)$.
If not, we perform a mixing and obtain
\begin{align}
    & r_{i+1} = (1 - \gamma) r_{i} + \gamma \tilde{r}_i , \\
    & \lambda_{i+1} = (1 - \gamma) \lambda_i + \gamma \tilde{\lambda}_i ,
\end{align}
and proceed to the next $(i+1)^\mathrm{th}$ iteration step using the updated parameters $(\mu_{i+1},r_{i+1},\lambda_{i+1})$.
The parameter $\gamma$ controls how smoothly we would like $r$ and $l$ to be updated.
For example, one may choose $\gamma = 0.2$ to ensure that $r_{i+1}$ and $\lambda_{i+1}$ do not deviate significantly from $r_i$ and $\lambda_i$.
The iteration will continue until the convergence is reached.

Once the convergence is reached, we use the final results of $\mu$, $r$, $\lambda$ to construct $[H_{eff} (\mathbf{k}) ]$.
The eigen-spectrum of this $[H_{eff} (\mathbf{k}) ]$ is the excitation spectrum of the periodic Anderson model solved using the parton saddle-point calculation.
In Supplementary Table~\ref{tab:parton_saddle_point_paramaters} we provide the parton saddle-point parameters that we obtain self-consistently for our periodic Anderson models.

\begin{table}[ht]
    \centering
    \begin{tabular}{|c|c|c|c|c|c|c|c|}
        \hline
        Model's Space Group &$E_f$  &$V$ & $\mu$ & $r$ & $\lambda$ \\
        \hline
        no.~173 ($P 6_3$) &$-30.0$ &$20.0$ & $-19.09$ & $0.5378$ & $16.802$ \\
        \hline
        no.~176 ($P 6_3/m$) &$-30.0$ &$20.0$ & $-20.06$ & $0.5692$ & $16.400$ \\
        \hline
        no.~180 ($P 6_2 22$) &$-4.0$ &$2.0$ & $-3.00$ & $0.5730$ & $1.5048$ \\
        \hline
        no.~190 ($P \bar{6} 2c$)  &$-6.0$ &$3.0$ & $-5.00$ & $0.6281$ & $1.8523$ \\
        \hline
    \end{tabular}
    \caption{Parton saddle-point parameters $\mu$, $r$, $\lambda$ obtained self-consistently for our hexagonal periodic Anderson models with space groups no.~173 ($P 6_3$), no.~176 ($P 6_3/m$), no.~180 ($P 6_2 22$), and no.~190 ($P \bar{6} 2c$). All parameters except $r$ are in unit of nearest hopping amplitude $t$.}
    \label{tab:parton_saddle_point_paramaters}
\end{table}

As a recap, we have made use of the following assumptions in the derivation and practical implementation of the parton saddle-point calculation.
First, we assume that both the $c$ and $f$ electrons carry the same labels $\mathbf{R}$, $\alpha$, and $\sigma$.
Second, we assume that both the $c$ and $f$ electrons are $s$ orbitals (or other orbitals) such that an isotropic hybridization between them is symmetry-allowed.
Third, we assume that all the sub-lattices are related to each other by some symmetries in the space group.
Fourth, we assume that the mean-field solution preserves the original space group symmetry of the interacting Hamiltonian.
Fifth, we assume that the model is solved with the periodic boundary condition and with finite number of unit cells.

\section{Extended result for the Anderson lattice model with the chiral space group no.~173}
We now describe the space-group symmetries of the hexagonal space groups and provide extended results for the corresponding models that are shown in the main text.

The chiral hexagonal space group $P 6_3 $ is generated by the hexagonal lattice translations and the screw rotation symmetry $\{ C_{6z} | 0,0,1/2 \}$.
In Supplementary Fig.~\ref{fig:173view2} we show how the chiral crystal in this group breaks inversion and mirror symmetries.
Supplementary Fig.~\ref{fig:173view2}(a,c) and (b,d) correspond to crystals of opposite chiralities in this space group, and they are related by a mirror reflection symmetry.

In Supplementary Fig.~\ref{fig:SG_173_excitation_spectrum_and_Weyl_points}(a) we show the low-energy heavy-fermion sector with a predominant $f$-electron character of the excitation spectrum of our hexagonal periodic Anderson model with the chiral space group no.~173 ($P 6_3$).
For this chiral WKSM phase, in Supplementary Fig.~\ref{fig:SG_173_excitation_spectrum_and_Weyl_points}(b) we show the distribution of the hourglass-type heavy-fermion Weyl points in the 3D Brillouin zone (BZ) and their corresponding topological chiral charges, which take values of $+3$ and $-1$ verified using the numerical method in \ref{sec:numerical_methods_Weyl_point_charge}.

\begin{figure}[ht]
    \centering
    \includegraphics[width=0.95\linewidth]{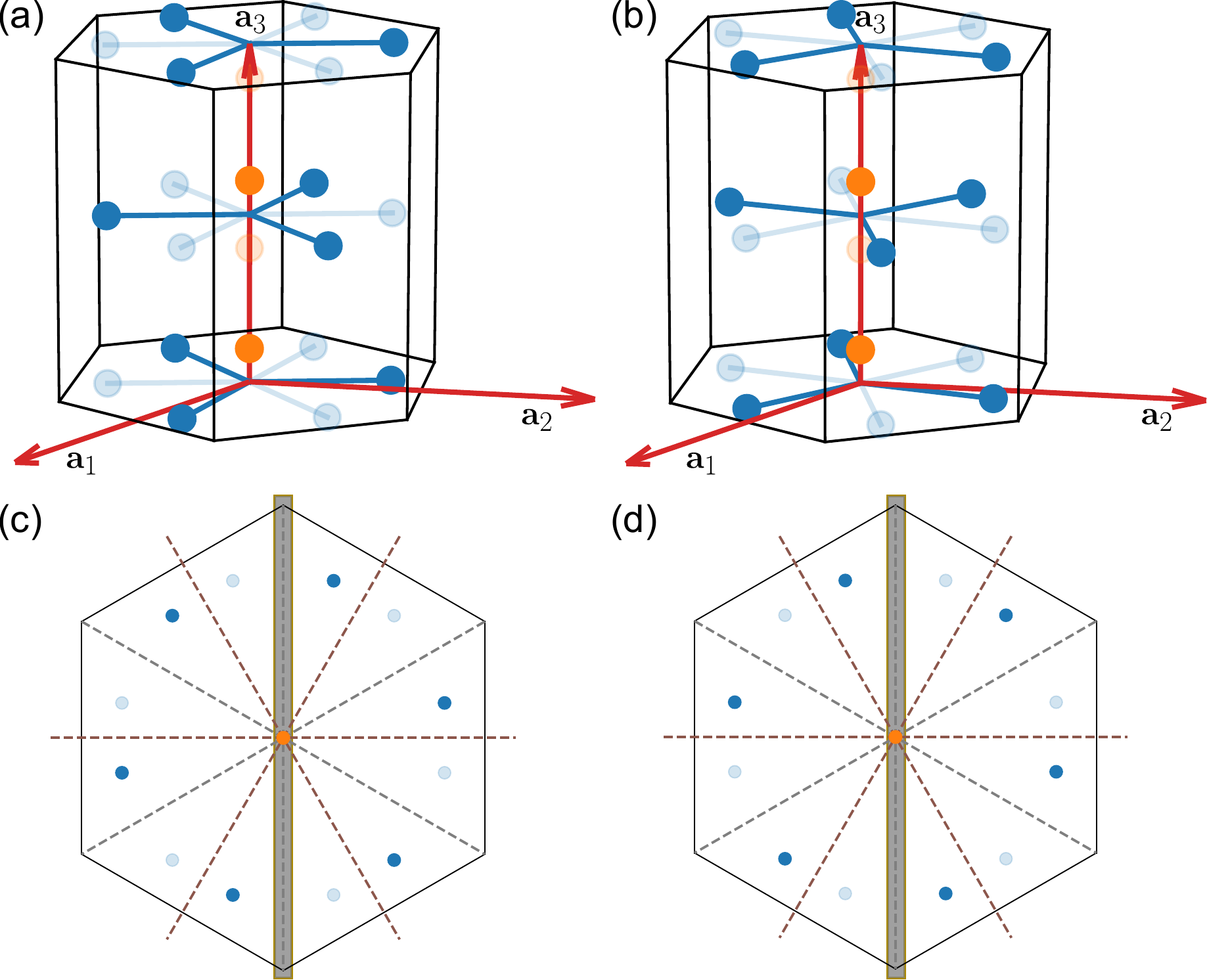}
    \caption{(a)~Unit cell in chiral space group no.~173. Solid dots are the atoms and transparent dots are their inversion symmetric positions. 
    If inversion is restored by placing atoms at these inversion-related positions, the space group symmetry rise to no.~176. 
    (b)~Unit cell in chiral space group no.~173 with mirrored atomic configuration of (a). 
    (c)~Top view of the unit cell in (a) showing broken mirror symmetries. Dashed lines indicate the potential mirror symmetries in hexagonal lattice. Solid dots are the atoms and transparent dots are their mirror symmetric positions. 
    (d)~Top view of (b).
    The mirror symmetry that maps between left panels with right panels is illustrated as the vertical gray bars in (c) and (d).
    }
    \label{fig:173view2}
\end{figure}

\begin{figure}[ht]
    \centering
    \includegraphics[width=0.64\linewidth]{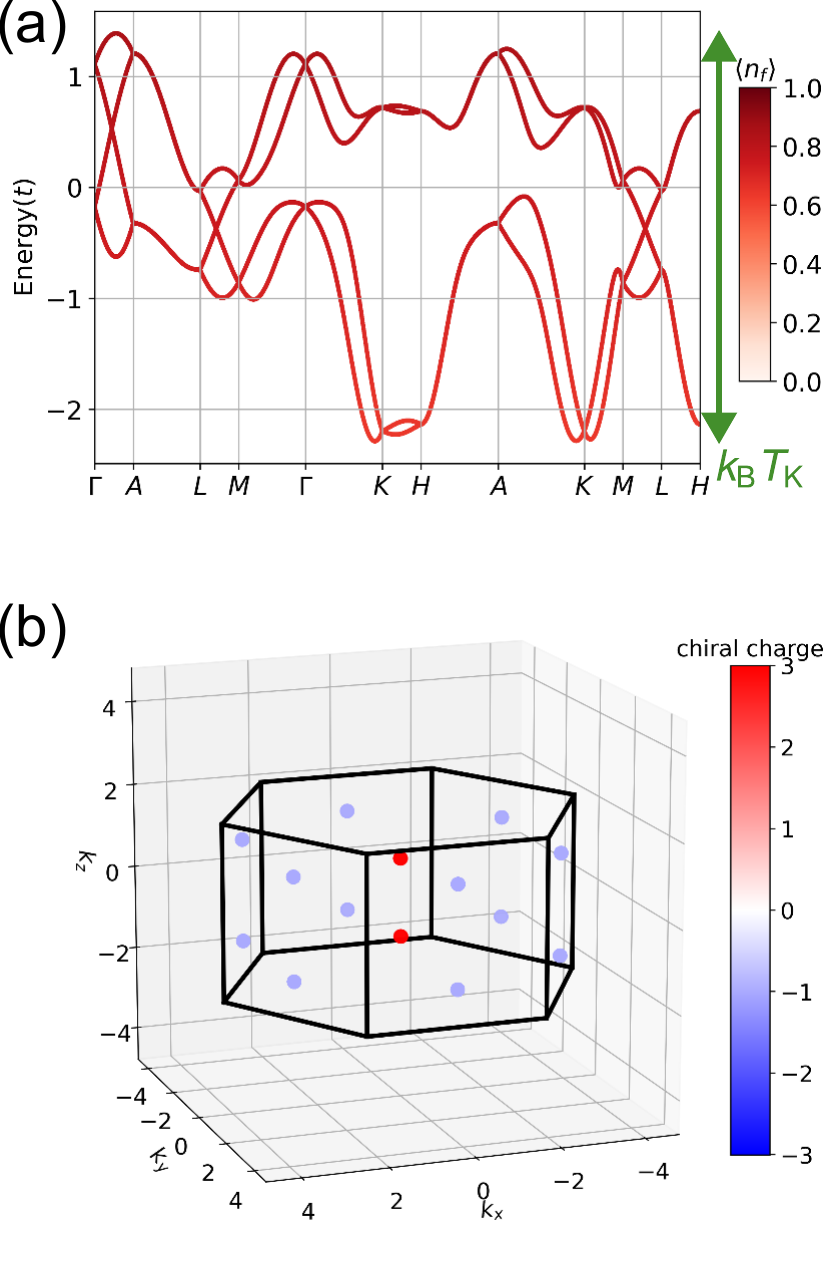}
    \caption{(a)~The excitation spectrum of the hexagonal periodic Anderson model with space group no.~173 ($P 6_3$).
    The Kondo scale $k_{\rm B} T_{\rm K}$ is also denoted.
    The color map indicates the portion of the $f$ electron for a given excitation.
    Due to the presence of $\{ C_{2z} \mathcal{T} | 0,0,1/2 \}$ symmetry, where $\mathcal{T}$ is the spin-1/2 time-reversal operation, all the energy bands on the top or bottom BZ boundary [Fig.~1(b) in main text], \textit{e.g.} along $A-L$ and $H-A$ high-symmetry lines, are twofold-degenerate. 
    We also provide a derivation of such twofold degeneracy in \ref{sec:implication_C2zT_0_0_one_half_SG_173}.
    (b)~The distribution of hourglass-type Weyl points together with their topological charges (the color map) in the three-dimensional BZ.
    Due to the presence of spin-1/2 time-reversal symmetry, given a Weyl point at momentum $\mathbf{k}_*$ with a topological charge $C$, there will be another Weyl point at momentum $-\mathbf{k}_*$ with the same topological charge $C$.}
    \label{fig:SG_173_excitation_spectrum_and_Weyl_points}
\end{figure}

\section{Extended result for the Anderson lattice model with the chiral space group no.~180}
The chiral hexagonal space group $P 6_2 22$ contains screw rotation $\{ C_{6z} | 0,0,1/3 \}$ and two $180^\circ$ rotations with rotation axes on the $xy$ plane. 
In Supplementary Fig.~\ref{fig:180view2}(a,c) we show how the chiral crystal in this group breaks inversion and mirror symmetries.
Supplementary Fig.~\ref{fig:173view2}(b,d) corresponds to the crystal of opposite chirality in space group no.~181, and it is related to the crystal in (a,c) by a mirror reflection symmetry.

In Supplementary Fig.~\ref{fig:SG180WP}, we show the Weyl points on the high-symmetry lines $\Gamma-A$ and $K-H$ in our periodic Anderson model with the chiral space group no.~180 ($P 6_2 22$).

\begin{figure}[ht]
    \centering
    \includegraphics[width=0.95\linewidth]{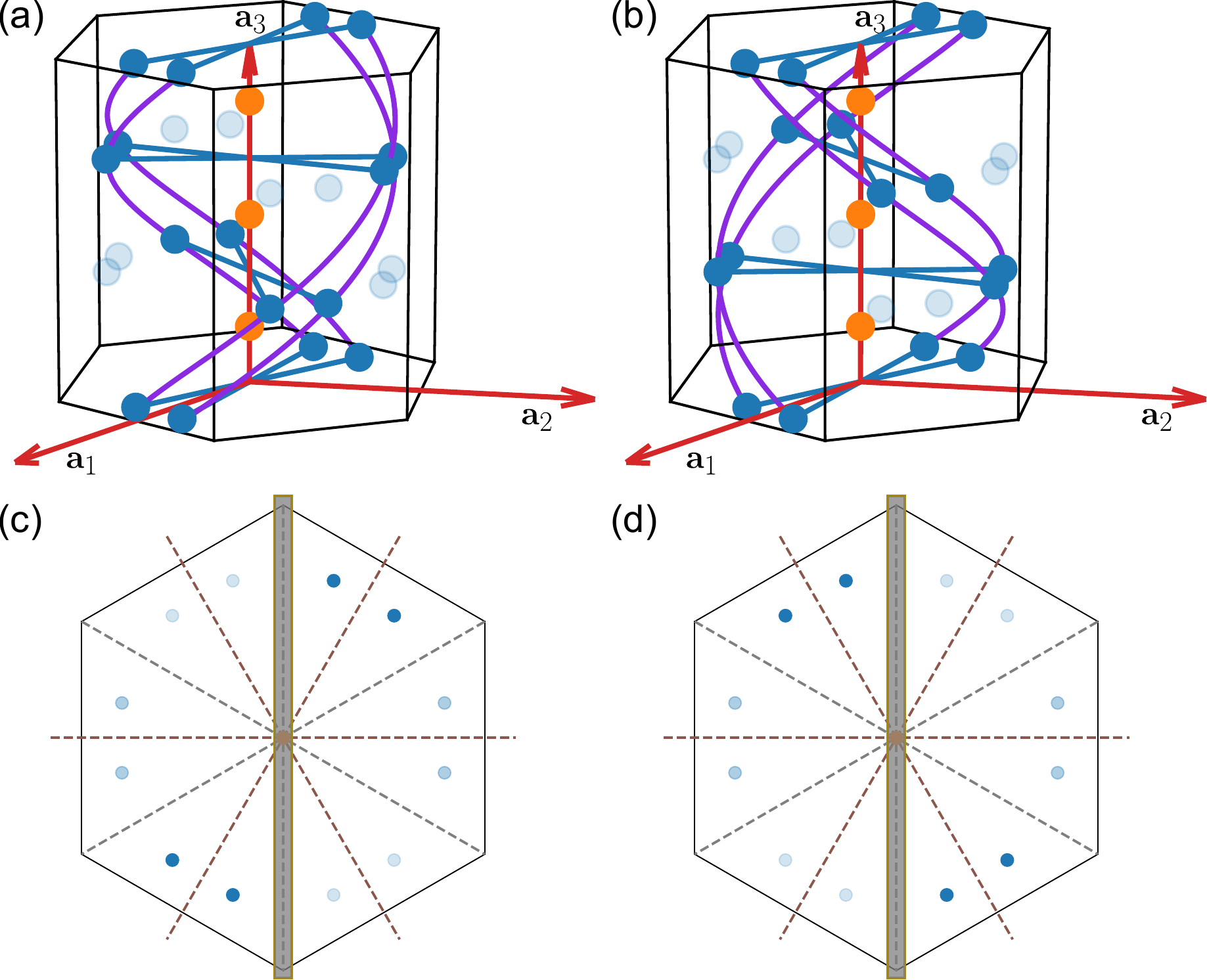}
    \caption{(a)~Unit cell in chiral space group no.~180. Solid dots are the atoms and transparent dots are their inversion-related positions.
    Applying screw operation $\{C_{6z}|0,0,\frac13\}$ to atoms generates a right-handed trajectory.
    The violet curves represent this helical structure.
    (b)~Unit cell in chiral space group no.~181. 
    Applying screw operation $\{C_{6z}^{-1}|0,0,\frac13\}$ to atoms generates a left-handed trajectory, as illustrated by the violet curves.
    (c)~Top view of the middle layer in (a) showing broken mirror symmetries. Dashed lines indicate the potential mirror symmetries in hexagonal lattice. Solid dots are the atoms and transparent dots are their mirror symmetric positions. 
    (d)~Top view of the middle layer in (b).
    The mirror symmetry that maps between left panels with right panels is illustrated as the vertical gray bars in (c) and (d).
    }
    \label{fig:180view2}
\end{figure}

\begin{figure}
    \centering
    \includegraphics[width=0.7\linewidth]{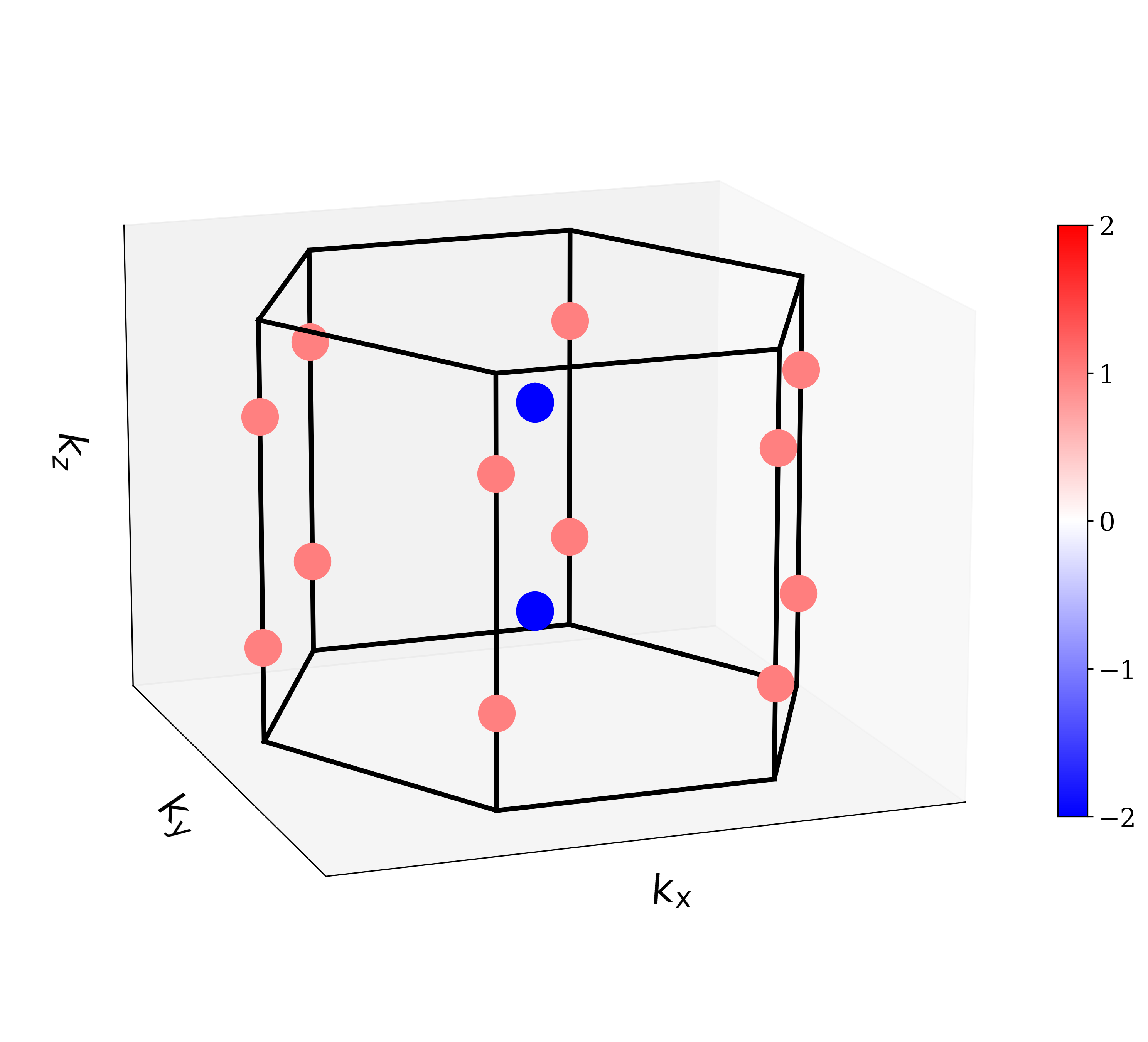}
    \caption{Weyl points and their chiral charges in the hexagonal periodic Anderson model with space group no.~180 ($P 6_2 22$).}
    \label{fig:SG180WP}
\end{figure}

\section{Anderson lattice model with the achiral topological Kondo semimetals for the achiral space group no.~176}
In the following section, we present model calculations for the two achiral space groups that are discussed in the main text.

The achiral hexagonal space group $P 6_3/m $ is generated by the hexagonal lattice translations, the screw rotation symmetry $\{ C_{6z} | 0,0,1/2 \}$ and inversion symmetry.
The excitation spectrum of the hexagonal periodic Anderson model with space group no.~176 ($P 6_3 / m$) is shown in Supplementary Fig.~\ref{fig:Fig_achiral}(a).
In the low-energy heavy-fermion sector, the excitations have a predominant $f$-electron character.
Due to the presence of inversion $\mathcal{I}$ and spin-1/2 time-reversal $\mathcal{T}$ symmetry, every band is twofold-degenerate.
On the top (or bottom) BZ boundary, we identify fourfold-degenerate, \textit{i.e.} Dirac, nodal lines forming a sixfold-symmetric pattern, as shown in Supplementary Fig.~\ref{fig:Fig_achiral}(b).
In the Supplementary Note 8 and 9, we further derive the quantized non-Abelian Berry phases in space group $P 6_3 / m$ and show the existence of drumhead surface states~\cite{chan20163,muechler2020modular} bound by the Dirac nodal lines.
We therefore demonstrate the Dirac-Kondo semimetal phase in the paramagnetic state of a system with the achiral hexagonal space group no.~176 ($P 6_3 / m$).
We further emphasize that the Dirac-Kondo semimetal in the strongly-correlated setting serves as a vintage point to investigate the broad landscape of correlated topological phases by symmetry breaking~\cite{chen2022topologicalsemimetal}.

\begin{figure}[t]
    \centering
    \includegraphics[width=\linewidth]{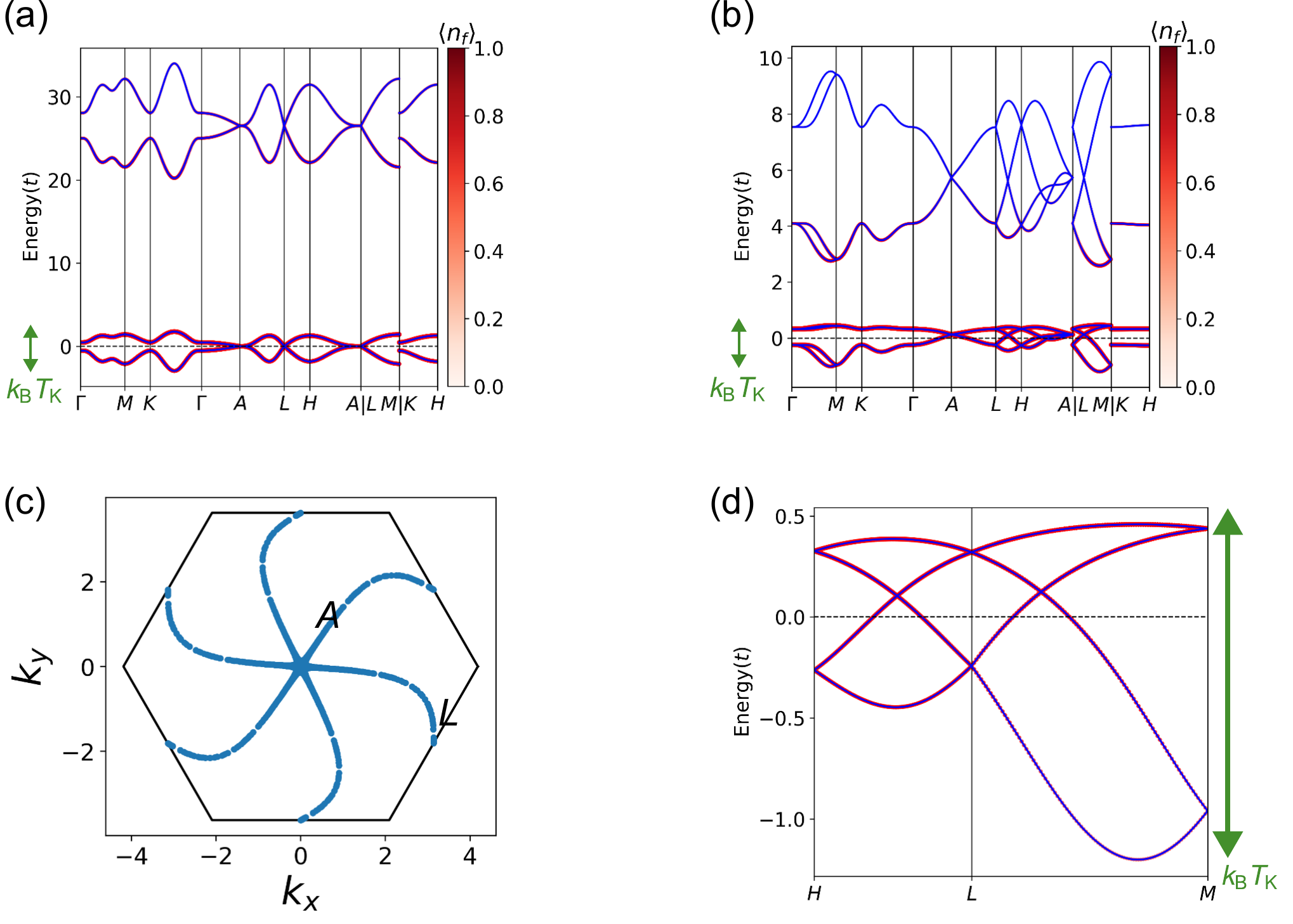}
    \caption{(a)~The excitation spectrum of the hexagonal periodic Anderson model with achiral space group no.~176 ($P 6_3 / m$) and
    (b)~the excitation spectrum of the hexagonal periodic Anderson model with achiral space group no.~190 ($P \bar{6} 2c$). 
    The color map indicates the portion of the $f$ electron for a given excitation.
    In (c) we show the Dirac nodal lines on the top (or bottom) BZ boundary corresponding to (a).
    In (d) we show the enlarged view of the band crossings consistent with the existence of low-energy heavy Weyl nodal lines in (b).
    The Kondo scale is denoted as $k_{\rm B} T_{\rm K}$.}
    \label{fig:Fig_achiral}
\end{figure}

In Supplementary Fig.~\ref{fig:SG_176_excitation_spectrum_and_fourfold_nodal_lines_and_slab_excitation_spectrum}(a) we show the low-energy heavy-fermion sector of the excitation spectrum of our hexagonal periodic Anderson model with the achiral space group no.~176 ($P 6_3 / m$).
The low-energy heavy-fermion sector has a predominant $f$-electron character.
We also show the fourfold-degenerate, \textit{i.e.} Dirac, nodal lines with a sixfold-symmetric pattern on the top (or bottom) BZ boundary in Supplementary Fig.~\ref{fig:SG_176_excitation_spectrum_and_fourfold_nodal_lines_and_slab_excitation_spectrum}(b).
We further find that when the system is cut into a slab with a finite size along $\mathbf{a}_3$, there are heavy-fermion surface states bound by the projected in-plane momenta of the Dirac nodal lines, as shown in Supplementary Fig.~\ref{fig:SG_176_excitation_spectrum_and_fourfold_nodal_lines_and_slab_excitation_spectrum}(c).
These surface states could be established based on the quantized non-Abelian Berry phases in \ref{sec:quantized_Berry_phases_SG_176}.

\begin{figure}[ht]
    \centering
    \includegraphics[width=\linewidth]{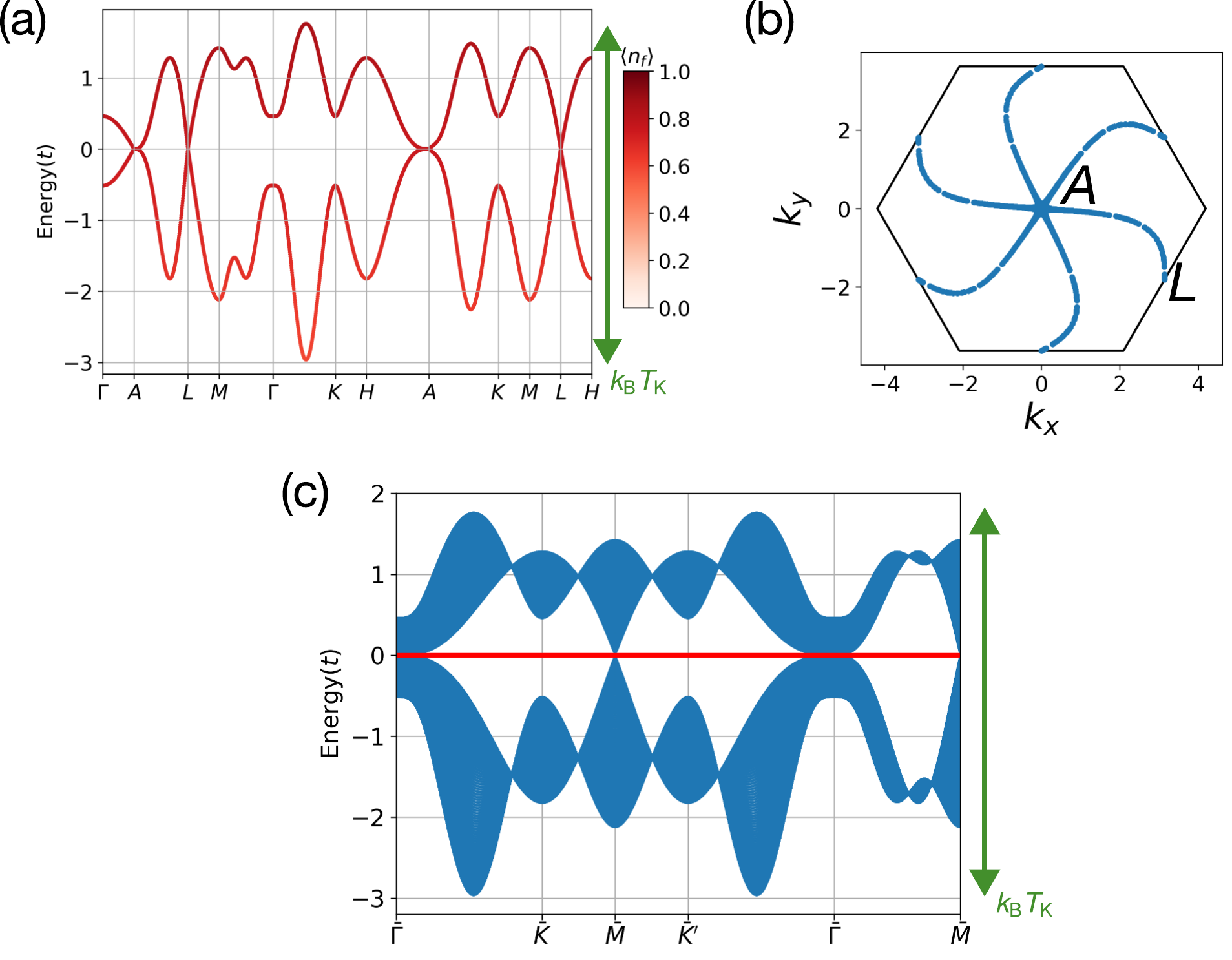}
    \caption{(a)~The excitation spectrum of the hexagonal periodic Anderson model with space group no.~176 ($P 6_3 / m$).
    The Kondo scale $k_{\rm B} T_{\rm K}$ is also denoted.
    The color map indicates the portion of the $f$ electron for a given excitation.
    (b)~The fourfold-degenerate, \textit{i.e.} Dirac, nodal lines on the top or bottom boundary of the BZ [Fig.~1(b) in main text].
    (c)~The slab band structure demonstrating the existence of surface states bound by the projected momenta of the Dirac nodal lines.
    The bulk continuum is colored by blue and the surface states are colored by red.}
    \label{fig:SG_176_excitation_spectrum_and_fourfold_nodal_lines_and_slab_excitation_spectrum}
\end{figure}

\section{Achiral topological Kondo semimetals for the achiral space group no.~190 ($P \bar{6} 2c$)}
The achiral hexagonal space group $P \bar{6} 2c$ contains roto-inversion $\{ C_{6z}\mathcal{I} | 0,0,1/2 \}$ and glide mirror symmetries such as $\{ m_{1,-1,0} | 0,0,1/2 \}$, where $m_{1,-1,0}$ indicates the mirror reflection that flips the position-space vector $\mathbf{a}_1 - \mathbf{a}_2$.

The excitation spectrum of the periodic Anderson model with space group no.~190 ($P \bar{6} 2c$) is shown in Supplementary Fig.~\ref{fig:Fig_achiral}(c). 
The signature of the Weyl nodal line can be seen from the hourglass-type band crossing along $M-L$ in Supplementary Fig.~\ref{fig:Fig_achiral}(d)~\cite{zhang2018topological}. 
Through the space-group symmetry, such symmetry-enforced Weyl nodal lines also appear on the other side-surfaces of the BZ.

In Supplementary Fig.~\ref{fig:SG190WNL}(a), we show the Weyl nodal lines in our periodic Anderson model with the achiral space group no.~190 ($P \bar{6} 2c$).
There are three different Weyl nodal lines:
(i)~Weyl nodal lines on the BZ boundary containing $M$, $L$, $K$, $H$ high-symmetry points, which are shown in Supplementary Fig.~\ref{fig:SG190WNL}(b).
They are enforced by the nonsymmorphic symmetries in this group.
(ii)~Weyl nodal lines on $k_3=\pi$ plane, these are accidental crossings protected by $\{m_z|00\frac12\}$.
(iii)~Weyl nodal lines on $k_1=k_2$ plane, these are accidental crossings protected by the glide symmetry $\{m_{1\bar{1}0}|00\frac12\}$.
Interestingly, these three types of Weyl nodal lines coexist and are linked to each other in our model.

\begin{figure}[ht]
    \centering
    \includegraphics[width=\linewidth]{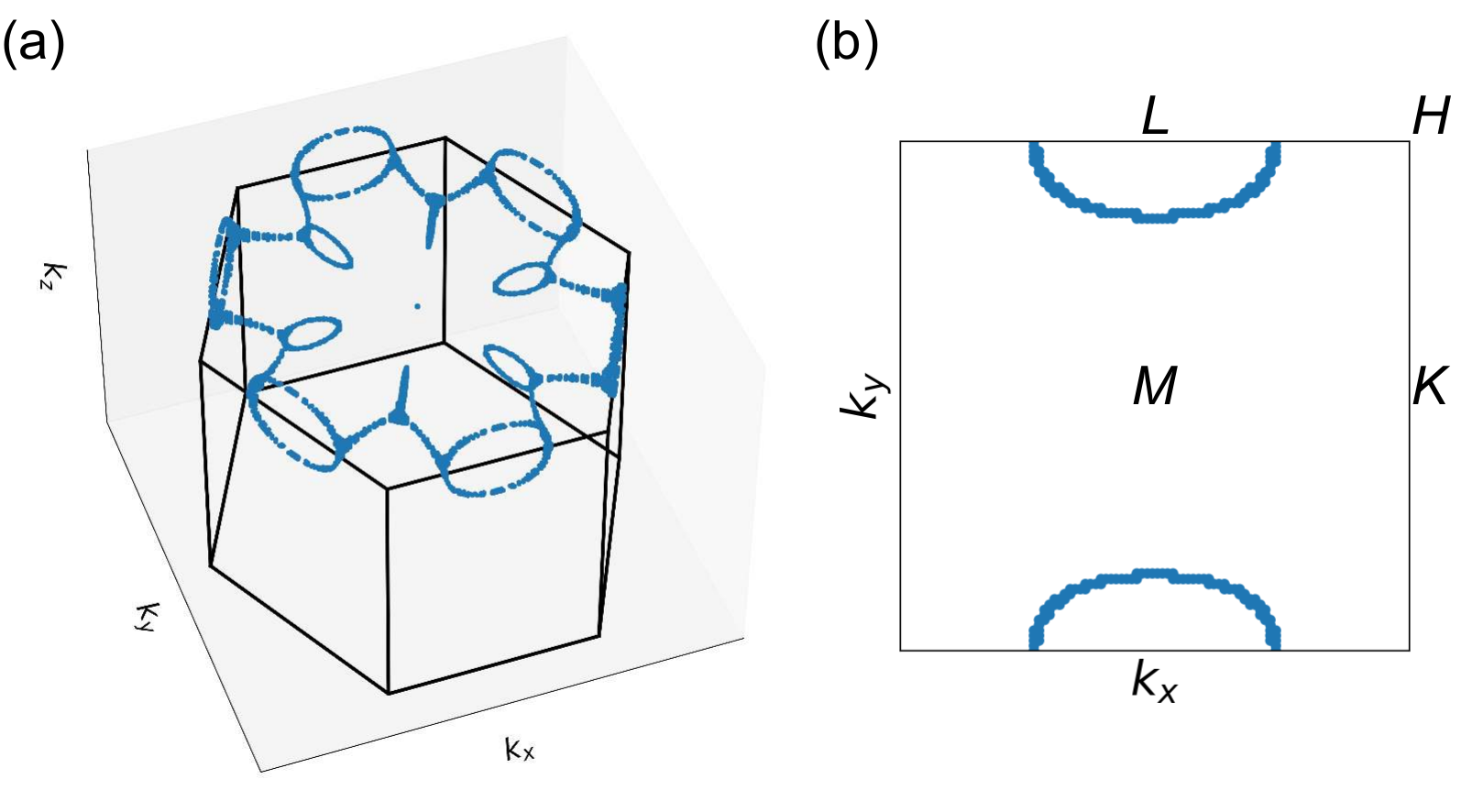}
    \caption{(a)~Weyl nodal line in the excitation spectrum of our periodic Anderson model with space group no.~190 ($P \bar{6} 2c$). (b)~Symmetry enforced Weyl nodal line on the $M-K-H-L$ plane.}
    \label{fig:SG190WNL}
\end{figure}

\section{Numerical methods for computing topological charges of Weyl points}
\label{sec:numerical_methods_Weyl_point_charge}

In this section, we provide the details of the two numerical methods and one symmetry indicator method that we use to compute the topological charges of Weyl points, which yield consistent results.

\setcounter{subsection}{0}
\subsection{Discrete jump of Chern number}

The Brillouin zone of a 3D system can be thought of as a collection of 2D planes in the $(k_1,k_2)$-momentum space with a constant $k_3$ value, where $k_3$ varies from $-0.5$ to $0.5$ (in the reduced coordinate), see Supplementary Fig.~\ref{fig:SG_173_Wilson_loop}.
Provided that the energy gaps of the $\mathbf{k}$ points on a 2D plane are opened, one could compute the Chern number of that 2D plane.
Hence, one could evaluate the Chern number as a function of $k_3$, as $k_3$ moves from $-0.5$ to $0.5$.
Before and after the 2D plane passes through some $\mathbf{k}$ points on which the energy gaps are closed, the Chern number of the 2D plane may experience a discrete change, see Supplementary Fig.~\ref{fig:SG_173_Wilson_loop}.
Specifically, the Chern number of a 2D plane in the momentum space can be computed as the winding number of the Wilson-loop spectrum.

In our hexagonal periodic Anderson model with space group no.~173 ($P 6_3$), we have computed the Chern number for the lowest two bands as a function of $k_3$ in Supplementary Fig.~\ref{fig:SG_173_Wilson_loop}, in which we identify various discrete jumps of magnitude $3$.
This is consistent with the distribution of the Weyl points in the 3D Brillouin zone in Supplementary Fig.~\ref{fig:SG_173_excitation_spectrum_and_Weyl_points}, where there are three Weyl points with $-1$ chiral charge all on the same $k_3$-constant plane, and there is one Weyl point with $+3$ chiral charge on another $k_3$-constant plane.

\begin{figure}[ht]
    \centering
    \includegraphics[width=0.8\linewidth]{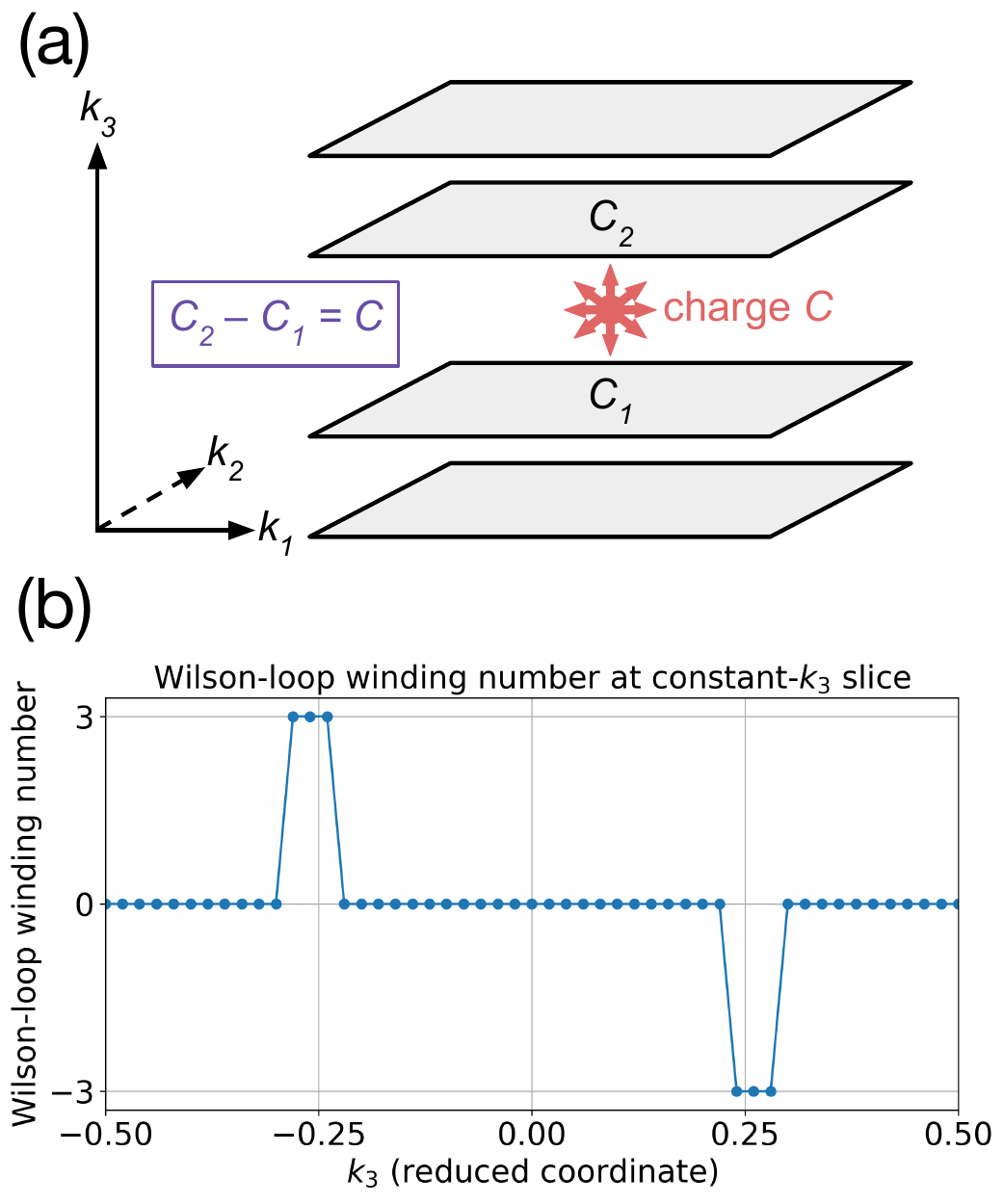}
    \caption{(a)~3D Brillouin zone viewed as slices of 2D planes. If there is a Weyl point with topological charge $C$ between two constant-$k_3$ planes, then the Chern numbers of the two planes will differ by $C$. (b)~Chern number (Wilson-loop winding number) as a function of $k_3$ for our hexagonal periodic Anderson model with space group no.~173 ($P 6_3$). There are discrete jumps of magnitude $3$ which are consistent with the presence of (i)~a single Weyl point carrying a topological charge of $+3$ and (ii)~three Weyl points with the same $k_3$ coordinate and each carrying a topological charge of $-1$.}
    \label{fig:SG_173_Wilson_loop}
\end{figure}

\subsection{Direct computation of Berry flux}

In addition to using the discrete jump of the Chern number of 2D planes in the 3D Brillouin zone as the 2D plane slides through the entire 3D Brillouin zone to infer the topological charges of Weyl points, one could also compute the topological charge directly.
This method usually requires that one already has an estimate about where the target Weyl point is located in the Brillouin zone, which can be done by (1)~a numerical minimization procedure, or (2)~an exhaustive search over grids in 3D Brillouin zone, to identify $\mathbf{k}$ points with small gaps.
The definition of the topological charge for a Weyl point is through a surface integral
\begin{equation}
    C = \frac{1}{2\pi} \int_{S} d^2\mathbf{k} \cdot \bm{\Omega} (\mathbf{k}),
\end{equation}
where $S$ is the surface enclosing the Weyl point and $\bm{\Omega} (\mathbf{k})$ is the Berry curvature.
Such a surface integral over a closed surface $S$ with the Berry curvature being the integrand yields the total Berry flux emitted by the Weyl point enclosed by the surface $S$.
In order to compute this integral, one could enclose the target Weyl point by a cube, divide the surfaces of the cube into small plaquettes, compute and sum over the Berry phases for all the small plaquettes. 

Suppose we sort the band index $n$ for a given $\mathbf{k}$ via $E_{n,\mathbf{k}} \leq E_{n+1,\mathbf{k}}$ with $n=1$ being the lowest-energy band index.
In our hexagonal periodic Anderson model with space group no.~173 ($P 6_3$), using this method of direct computation we verify that the Weyl points (between $n=2$ and $n=3$ bands) located along the $\Gamma-A$ line each carries a topological chiral charge of $+3$, and that the Weyl points (between $n=2$ and $n=3$ bands) located along the $M-L$ line each carries a topological chiral charge of $-1$, leading to Fig.~2 in the main text. 
We further verify that exactly at the time-reversal-invariant momenta $\Gamma$ and $M$ the Kramers degeneracies (between $n=1$ and $n=2$ bands) are in fact Kramers-Weyl points carrying a topological chiral charge of $-1$ and $+1$, respectively.

\subsection{Symmetry indicator for chiral charges}

\begin{figure}[ht]
    \centering
    \includegraphics[width=0.7\linewidth]{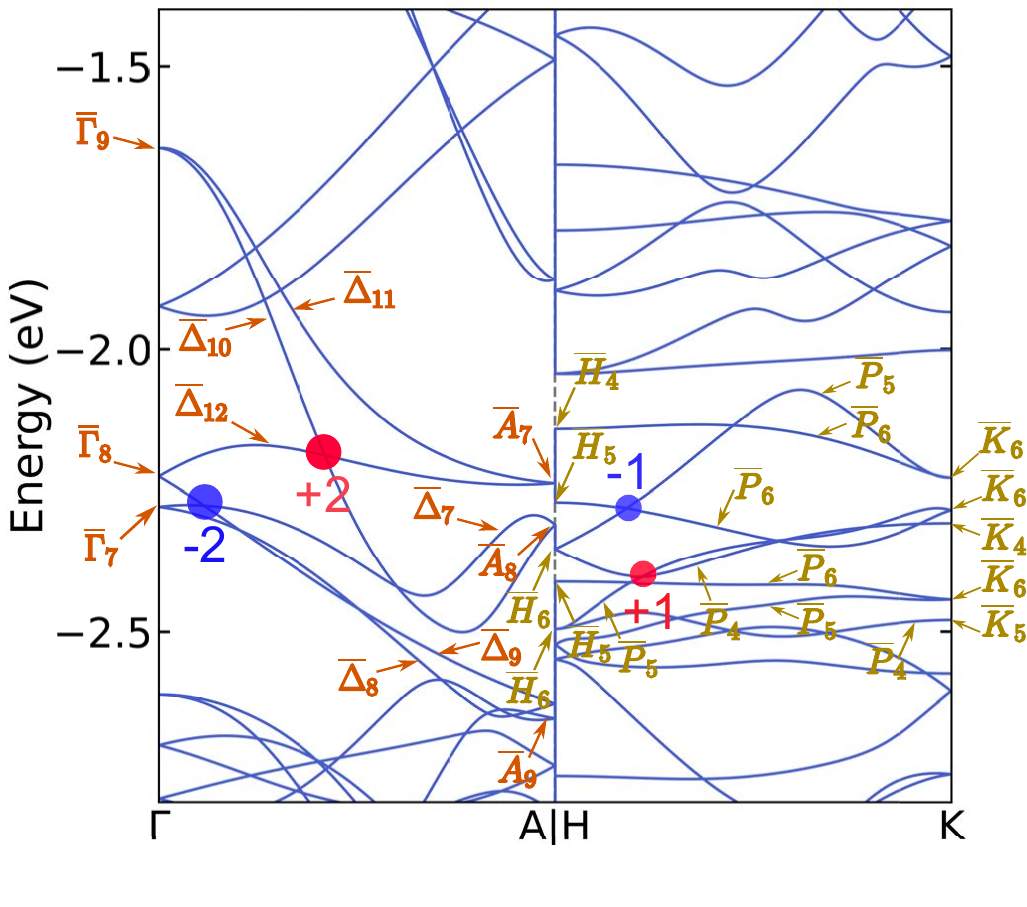}
    \caption{Irreps of CePt$_2$B DFT bands along high symmetry line $\Delta$ ($\Gamma-A$) and $P$ ($K-H$).}
    \label{fig:DFT180irrep}
\end{figure}

Symmetry eigenvalues of the Bloch states at the band crossing point can be used to determine the chiral charge of a Weyl point~\cite{tsirkin2017composite}.
We adopt this method to determine the chiral charges of the Weyl points in the DFT band structure of CePt$_2$B.
CePt$_2$B has the chiral space group $P6_2 22$ that contains the screw symmetry $\{C_{6z}|0,0,1/3\}$.
Along the $\Gamma-A$ high-symmetry line, we could label the Bloch state by its $\{C_{6z}|0,0,1/3\}$ eigenvalue, which is $\exp(i\frac{\pi}{3}j_z + i k_z/3)$, with $j_z=\ \frac12,\frac32,\frac52,-\frac52,-\frac32,-\frac12$. 
This corresponds to the irreducible representation (irrep) labels $\overline{\Delta}_{11},\overline{\Delta}_{8},\overline{\Delta}_{9},\overline{\Delta}_{12},\overline{\Delta}_{7},\overline{\Delta}_{10}$, respectively. 
Along the $K-H$ high-symmetry line, we could label the Bloch state by its $\{C_{3z}|0,0,2/3\}$ eigenvalue, which is $\exp(i\frac{\pi}{3}j_z + 2i k_z/3)$, with $j_z=\ \frac12,\frac32,-\frac12$. 
This corresponds to the irreducible representation labels $\overline{P}_6$, $\overline{P}_4$, $\overline{P}_5$, respectively.

The crossing of two irreps at $k_0=(0,0,k_z)$ ($\Gamma-A$) or $k_0=(\frac13,\frac13,k_z)$ ($K-H$) with $j_z(k_0-\delta k_z)$ and $j_z(k_0+\delta k_z)$ has Weyl charge~\cite{tsirkin2017composite}
\begin{equation}
    C = j_z(k_0+\delta k_z) - j_z(k_0-\delta k_z) 
\end{equation}
In Supplementary Fig.~\ref{fig:DFT180irrep}, we label the DFT band structure of CePt$_2$B along $\Gamma-A$ by such irreducible representation labels.
For example, there is a crossing point between the bands labeled as $\overline{\Delta}_{12}$ and $\overline{\Delta}_{10}$, meaning that these two bands carry $\{C_{6z}|0,0,1/3\}$ eigenvalues $j_z=-\frac52$ and $j_z=-\frac12$, respectively.
Hence, the ratio between the $\{C_{6z}|0,0,1/3\}$ eigenvalues at this band crossing point is $2\mod 6$.
The chiral charges of all other band crossings highlighted in Supplementary Fig.~\ref{fig:DFT180irrep} are also determined in this way.
We verify that the Weyl charges between given two adjacent band indices sum to zero.

\section{Quantized non-Abelian Berry phases in space group no.~176 ($P 6_3 / m$) with spin-1/2 time-reversal symmetry}
\label{sec:quantized_Berry_phases_SG_176}

In this section, we show that a crystalline system with space group no.~176 ($P 6_3 / m$) and spin-1/2 time-reversal symmetry exhibits quantized non-Abelian Berry phases, which could be used to establish the presence of surface states when the corresponding hexagonal crystal is cut finite along the $\mathbf{a}_3$ direction.

\setcounter{subsection}{0}
\subsection{Derivation of quantized non-Abelian Berry phases}

The modern development of electronic structure theory has established the connection between Berry phase and electronic polarization, which is closely related to the presence of surface states and surface charges.
For our multi-band hexagonal crystal systems, we define the $k_3$-directed Wilson-loop matrix $[\mathcal{W}^{\mathbf{G}_3}_{k_1,k_2}]$ whose matrix elements are given by (see Supplementary Fig.~\ref{fig:hexagonal_BZ_with_Wilson_loop_path} for a schematic diagram of $[\mathcal{W}^{\mathbf{G}_3}_{k_1,k_2}]$)
\begin{equation}
    [\mathcal{W}^{\mathbf{G}_3}_{k_1,k_2}]_{mn} = \langle u_{m,\mathbf{k}+\mathbf{G}_3} | \left( \prod_{\mathbf{q}}^{\mathbf{k}+\mathbf{G}_3 \leftarrow \mathbf{k}} [P(\mathbf{q})] \right) | u_{n,\mathbf{k}} \rangle,
\end{equation}
where $[P(\mathbf{k})]$ is the matrix projection operator onto the vector space spanned by the $N_{occ}$ occupied eigenvectors of the Bloch Hamiltonian matrix, namely
\begin{equation}
    [P(\mathbf{k})] = \sum_{n \in \mathrm{occ}} | u_{n,\mathbf{k}} \rangle \langle u_{n,\mathbf{k}} |,
\end{equation}
and $| u_{n,\mathbf{k}} \rangle$ is the eigenvector of the Bloch Hamiltonian matrix. 
Specifically, the eigenvector $| u_{n,\mathbf{k}} \rangle$ satisfies the periodic gauge such that $| u_{n,\mathbf{k}+\mathbf{G}} \rangle = [V(\mathbf{G})]^{-1} | u_{n,\mathbf{k}} \rangle = [V(\mathbf{G})]^{\dagger} | u_{n,\mathbf{k}} \rangle$. 
For the case of our effective Bloch Hamiltonian matrix $[H_{eff}(\mathbf{k})]$ in the auxiliary-boson representation at the saddle-point level we have 
\begin{equation}
    [V(\mathbf{G})] = \begin{pmatrix}
        [\tilde{V}(\mathbf{G})] & 0 \\
        0 & [\tilde{V}(\mathbf{G})]
    \end{pmatrix},
\end{equation}
with $[\tilde{V}(\mathbf{G})]_{(\alpha,\sigma),(\alpha',\sigma')} = e^{i \mathbf{G} \cdot \mathbf{r}_{\alpha}} \delta_{\alpha,\alpha'} \delta_{\sigma,\sigma'}$.
The eigenspectrum of $[\mathcal{W}^{\mathbf{G}_3}_{k_1,k_2}]$ takes the unimodular form $e^{i \gamma_{j} (k_1 ,k_2)}$ for $j = 1 \ldots N_{occ}$.
The eigenphases $\gamma_{j} (k_1 ,k_2) \in \mathbb{R}$, also known as the non-Abelian Berry phases, are well-defined module $2\pi$.
Importantly, $\frac{\gamma_{j} (k_1 ,k_2)}{2\pi} \mathbf{a}_3$ is the localized position along $\mathbf{a}_3$ of the hybrid Wannier function localized along $\mathbf{a}_3$ while extended along $\mathbf{a}_1$ and $\mathbf{a}_2$.
Intuitively, one could think of the eigenphases $\gamma_{j} (k_1 ,k_2)$ of the Wilson-loop matrix $[\mathcal{W}^{\mathbf{G}_3}_{k_1,k_2}]$ as equivalent to the position along $\mathbf{a}_3$ as a function of the remaining momentum components $(k_1 , k_2)$.
Hence, the symmetry transformation properties of the set of eigenphases $\{ \gamma_{j} (k_1 ,k_2) \ | \ j = 1\ldots N_{occ} \}$ can be derived as follows, by using the fact that $\gamma_{j}$ transforms in the same manner as the position along $\mathbf{a}_3$.

We recall here that space group no.~176 ($P 6_3 / m$) contains inversion symmetry $\mathcal{I}$, screw rotation symmetry $\{ C_{6z} | 0,0,1/2 \}$, and we also assume spin-1/2 time-reversal symmetry $\mathcal{T}$.
For spin-1/2 time-reversal symmetry $\mathcal{T}$, which flips the momentum while leaving the position invariant, we have the symmetry constraint
\begin{align}
    & \{ \gamma_{j} (k_1 ,k_2) \ | \ j = 1\ldots N_{occ} \} \nonumber \\
    & = \{ \gamma_{j} (- k_1 , - k_2) \ | \ j = 1\ldots N_{occ} \}.
\end{align}
Hereafter, the equality between the set of eigenphases is well-defined module $2\pi$.
For inversion symmetry $\mathcal{I}$, which flips both the momentum and position, we have the symmetry constraint
\begin{align}
    & \{ \gamma_{j} (k_1 ,k_2) \ | \ j = 1\ldots N_{occ} \} \nonumber \\
    & = \{ - \gamma_{j} (- k_1 , - k_2) \ | \ j = 1\ldots N_{occ} \}.
\end{align}
For $\{ C_{6z} | 0,0,1/2 \}$ symmetry, which rotates the momentum by $C_{6z}$, and rotates the position by $C_{6z}$ followed by a fractional translation $\frac{1}{2}\mathbf{a}_3$, we have the symmetry constraint
\begin{align}
    & \{ \gamma_{j} (k_1 ,k_2) + \pi \ | \ j = 1\ldots N_{occ} \} \nonumber \\
    & = \{ \gamma_{j} ( C_{6z}(k_1,k_2) ) \ | \ j = 1\ldots N_{occ} \},
\end{align}
where the $+\pi$ on the left-hand side is due to the fractional translation $\frac{1}{2}\mathbf{a}_3$.

Our goal in below would be to combine the above symmetry constraints such that we obtain a set of constraints for $\{ \gamma_{j} (k_1 ,k_2) \ | \ j = 1\ldots N_{occ} \}$ right on the same momentum $(k_1 , k_2)$.
From $\mathcal{T}$ and $\{ C_{6z} | 0,0,1/2 \}$ symmetries, we can derive
\begin{align}
    & \{ \gamma_{j} (k_1 ,k_2) \ | \ j = 1\ldots N_{occ} \} \nonumber \\
    & = \{ \gamma_{j} ( -k_1,-k_2 ) \ | \ j = 1\ldots N_{occ} \} \nonumber \\
    & = \{ \gamma_{j} ( C_{6z}^3(k_1,k_2) ) \ | \ j = 1\ldots N_{occ} \} \nonumber \\
    & = \{ \gamma_{j} (C_{6z}^2(k_1,k_2)) + \pi \ | \ j = 1\ldots N_{occ} \} \nonumber \\
    & = \{ \gamma_{j} (C_{6z}(k_1,k_2)) + 2\pi \ | \ j = 1\ldots N_{occ} \} \nonumber \\
    & = \{ \gamma_{j} (k_1,k_2) + 3 \pi \ | \ j = 1\ldots N_{occ} \},
\end{align}
which means
\begin{align}
    & \{ \gamma_{j} (k_1 ,k_2) \ | \ j = 1\ldots N_{occ} \} \nonumber \\
    & = \{ \gamma_{j} (k_1,k_2) + \pi \ | \ j = 1\ldots N_{occ} \}.
\end{align}
This means that the set of eigenphases at any momentum $(k_1 , k_2)$ must be equal to (up to a module of $2\pi$) itself plus $\pi$.
Going further, we define the summed Berry phase as
\begin{equation}
    \gamma (k_1 ,k_2) \equiv \sum_{j=1}^{N_{occ}} \gamma_{j} (k_1 ,k_2) \mod \ 2\pi .
\end{equation}
From the symmetry constriants $\{ \gamma_{j} (k_1 ,k_2) \ | \ j = 1\ldots N_{occ} \} = \{ \gamma_{j} (k_1,k_2) + \pi \ | \ j = 1\ldots N_{occ} \}$ we first obtain the following result
\begin{equation}
    \gamma (k_1 ,k_2) = \gamma (k_1 ,k_2) + \pi N_{occ} \mod \ 2\pi.
\end{equation}
We can then deduce that $N_{occ}$ can not be equal to an odd integer.
This means that the Wilson loop matrix $[\mathcal{W}^{\mathbf{G}_3}_{k_1,k_2}]$ is only well-defined when $N_{occ}$ is equal to an even integer, otherwise there is a contradiction.
Hence, in the following derivation of quantized Berry phases in space group no.~176 ($P 6_3 / m$) we only consider an even integer $N_{occ}$.

One could also derive the constraint of $\gamma_{j} (k_1 ,k_2)$ at the same momentum due to $\mathcal{I} \mathcal{T}$ symmetry. 
The presence of $\mathcal{I} \mathcal{T}$ symmetry is guaranteed since both $\mathcal{I}$ and $\mathcal{T}$ are symmetries of the system.
Note that $(\mathcal{I} \mathcal{T})^2 = -1$ since $(\mathcal{I} \mathcal{T})^2 = \mathcal{I} \mathcal{T} \mathcal{I} \mathcal{T} = \mathcal{I} \mathcal{I} \mathcal{T} \mathcal{T} = \mathcal{I}^2 \mathcal{T}^2 = -1$ due to $\mathcal{I}^2 = 1$ and $\mathcal{T}^2=-1$.
By combining the constraints from $\mathcal{I}$ and $\mathcal{T}$ symmetries we have
\begin{align}
    & \{ \gamma_{j} (k_1 ,k_2) \ | \ j = 1\ldots N_{occ} \} \\
    & = \{ -\gamma_{j} (-k_1 ,-k_2) \ | \ j = 1\ldots N_{occ} \} \\
    & = \{ -\gamma_{j} (k_1 ,k_2) \ | \ j = 1\ldots N_{occ} \}.
\end{align}
Hence, $\mathcal{I} \mathcal{T}$ symmetry leads to the constraint
\begin{align}
    & \{ \gamma_{j} (k_1 ,k_2) \ | \ j = 1\ldots N_{occ} \} \nonumber \\
    & = \{ - \gamma_{j} (k_1 ,k_2) \ | \ j = 1\ldots N_{occ} \},
\end{align}
which means that the summed Berry phase satisfies
\begin{equation}
    \gamma (k_1 ,k_2) = - \gamma (k_1 ,k_2) \mod \ 2\pi
\end{equation}
such that 
\begin{equation}
    \gamma (k_1 ,k_2) = n \pi
\end{equation}
where $n \in \mathbb{Z}$.
From $(\mathcal{I} \mathcal{T})^2 = -1$, the Kramers theorem further implies that for a fixed $(k_1 , k_2)$ if there is an eigenphase $\gamma$, there will be another independent eigenphase $-\gamma$. 
In other words, $(\mathcal{I} \mathcal{T})^2 = -1$ means that the eigenphases must form pairs of the form $\{-\gamma,\gamma \}$.

\begin{figure}[ht]
    \centering
    \includegraphics[width=\linewidth]{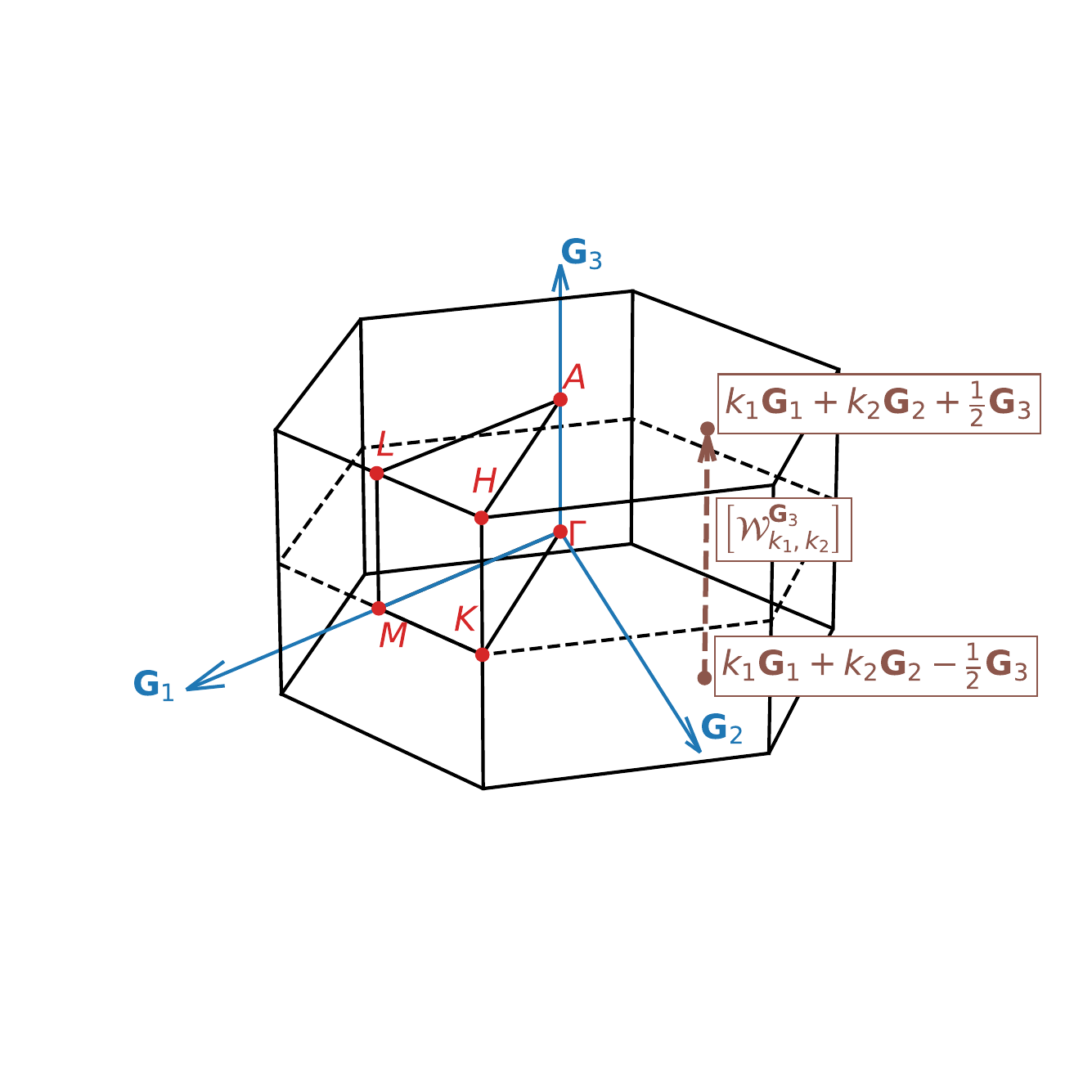}
    \caption{Schematic diagram of the Wilson loop $[\mathcal{W}^{\mathbf{G}_3}_{k_1,k_2}]$ for a non-contractible loop (the dashed arrow) in the Brillouin zone from $\mathbf{k}$ to $\mathbf{k}+\mathbf{G}_3$. This Wilson loop $[\mathcal{W}^{\mathbf{G}_3}_{k_1,k_2}]$ is used to derive the quantized non-Abelian Berry phases in the space group no.~176 ($P 6_3 / m$) with the spin-1/2 time-reversal symmetry $\mathcal{T}$.}
    \label{fig:hexagonal_BZ_with_Wilson_loop_path}
\end{figure}

\begin{figure}[ht]
    \centering
    \includegraphics[width=\linewidth]{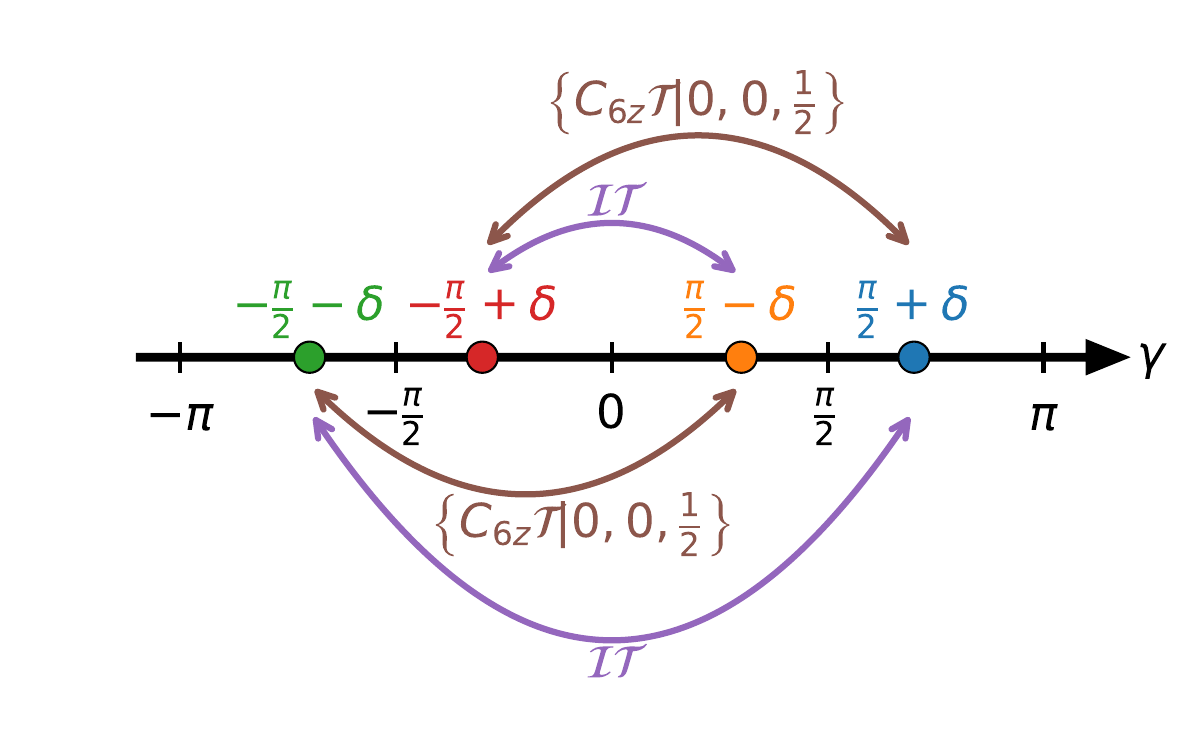}
    \caption{The most generic distribution of the Wilson-loop eigenphases $\gamma_j (k_1,k_2)$ at a specific momentum $(k_1,k_2)$ from the symmetry constraints of the space group no.~176 ($P 6_3 / m$) with the spin-1/2 time-reversal symmetry $\mathcal{T}$. Starting from a generic value of the $k_3$-directed Wilson-loop eigenphase $\frac{\pi}{2}+\delta$ at a specific momentum $(k_1,k_2)$, the $\mathcal{I}\mathcal{T}$ symmetry guarantees that $-\frac{\pi}{2}-\delta$ is also a $k_3$-directed Wilson-loop eigenphase at the same momentum $(k_1,k_2)$. Building on this, the $\left\{ C_{6z}\mathcal{T} | 0,0, 1/2 \right\}$ symmetry further guarantees that $-\frac{\pi}{2}+\delta$ and $\frac{\pi}{2}-\delta$ are also $k_3$-directed Wilson-loop eigenphases at the same momentum $(k_1,k_2)$.}
    \label{fig:SG_176_with_TR_Berry_phase_symmetry_constraint_schematic_diagram}
\end{figure}

We could now combine all the constraints derived above and obtain
\begin{align}
    & \{ \gamma_{j} (k_1 ,k_2) \ | \ j = 1\ldots N_{occ} \} \nonumber \\
    & = \{ \gamma_{j} (k_1,k_2) + \pi \ | \ j = 1\ldots N_{occ} \} \nonumber \\
    & = \{ - \gamma_{j} (k_1 ,k_2) \ | \ j = 1\ldots N_{occ} \},
\end{align}
together with the constraint that eigenphases must form pairs of the form $\{ -\gamma,\gamma \}$.
Supplementary Fig.~\ref{fig:SG_176_with_TR_Berry_phase_symmetry_constraint_schematic_diagram} demonstrates the most generic distribution of the eigenphases $\gamma_{j} (k_1 ,k_2)$ at a specific momentum $(k_1 ,k_2)$ due to the above symmetry constraints and how each eigenphase is related to the other by a certain symmetry, including $\mathcal{I}\mathcal{T}$ and $\left\{ C_{6z}\mathcal{T} | 0,0,1/2 \right\}$. Although Supplementary Fig.~\ref{fig:SG_176_with_TR_Berry_phase_symmetry_constraint_schematic_diagram} tentatively shows eigenphases $-\frac{\pi}{2}-\delta$, $-\frac{\pi}{2}+\delta$, $\frac{\pi}{2}-\delta$, and $\frac{\pi}{2}+\delta$ on the general ground at a specific momentum $(k_1,k_2)$, we emphasize that it is possible that the set of eigenphases only contains two elements as $\{ -\frac{\pi}{2},\frac{\pi}{2}\}$, because $\{ -\frac{\pi}{2},\frac{\pi}{2}\}$ still satisfies all the above symmetry constraints. Specifically, this is consistent with the case where $\delta = 0$, and is the only case satisfying all the symmetry constraints when the number of eigenphases is equal to $2$.
Hence, we note here that if $N_{occ}=2$, then $\{ \gamma_{j} (k_1 ,k_2) \ | \ j = 1,2 \}$ must equal $\{ -\frac{\pi}{2},\frac{\pi}{2}\}$ and is $(k_1,k_2)$-independent. On the other hand, when $N_{occ} = 4$, we will have $\{ \gamma_{j} (k_1 ,k_2) \ | \ j = 1\ldots 4 \} = \{ -\frac{\pi}{2}-\delta(k_1,k_2),-\frac{\pi}{2}+\delta(k_1,k_2),\frac{\pi}{2}-\delta(k_1,k_2),\frac{\pi}{2}+\delta(k_1,k_2) \}$ where $\delta(k_1,k_2)$ is $(k_1,k_2)$-dependent.
Based on Supplementary Fig.~\ref{fig:SG_176_with_TR_Berry_phase_symmetry_constraint_schematic_diagram}, we conclude that when $N_{occ} = 2$ $\times$ an odd integer there must be two non-Abelian Berry phases (out of the in total $N_{occ}$ eigenphases) taking values of $(k_1 , k_2)$-independent $\{ -\frac{\pi}{2} , \frac{\pi}{2}\}$, while when $N_{occ} = 2$ $\times$ an even integer then all $N_{occ}$ non-Abelian Berry phases are in general not quantized and $(k_1 , k_2)$-dependent.

As a final remark, we note that a system with space group no.~176 ($P 6_3 / m$) and spin-1/2 time-reversal symmetry must have fourfold degenerate nodal lines at the top (or equivalently the bottom) boundary of the BZ~\cite{zhang2018topological}. 
Hence, when the occupied space and $(k_1,k_2)$ are chosen such that the path going from $\mathbf{k}$ to $\mathbf{k}+\mathbf{G}_3$ in the definition of the Wilson loop matrix passes through gap-closing points from the fourfold degenerate nodal lines, then the Wilson loop matrix is no longer well-defined, as it requires that the matrix projection operator is well-defined along the path.
In such a situation, the non-Abelian Berry phases can not be obtained.
Our above results hence apply to the situation when the Wilson loop matrix is well-defined.
In Supplementary Fig.~\ref{fig:SG_176_Wilson_loop}, we numerically demonstrate the quantized non-Abelian Berry phases $\{ -\frac{\pi}{2},\frac{\pi}{2}\}$ we derived above.

\begin{figure}
    \centering
    \includegraphics[width=0.75\linewidth]{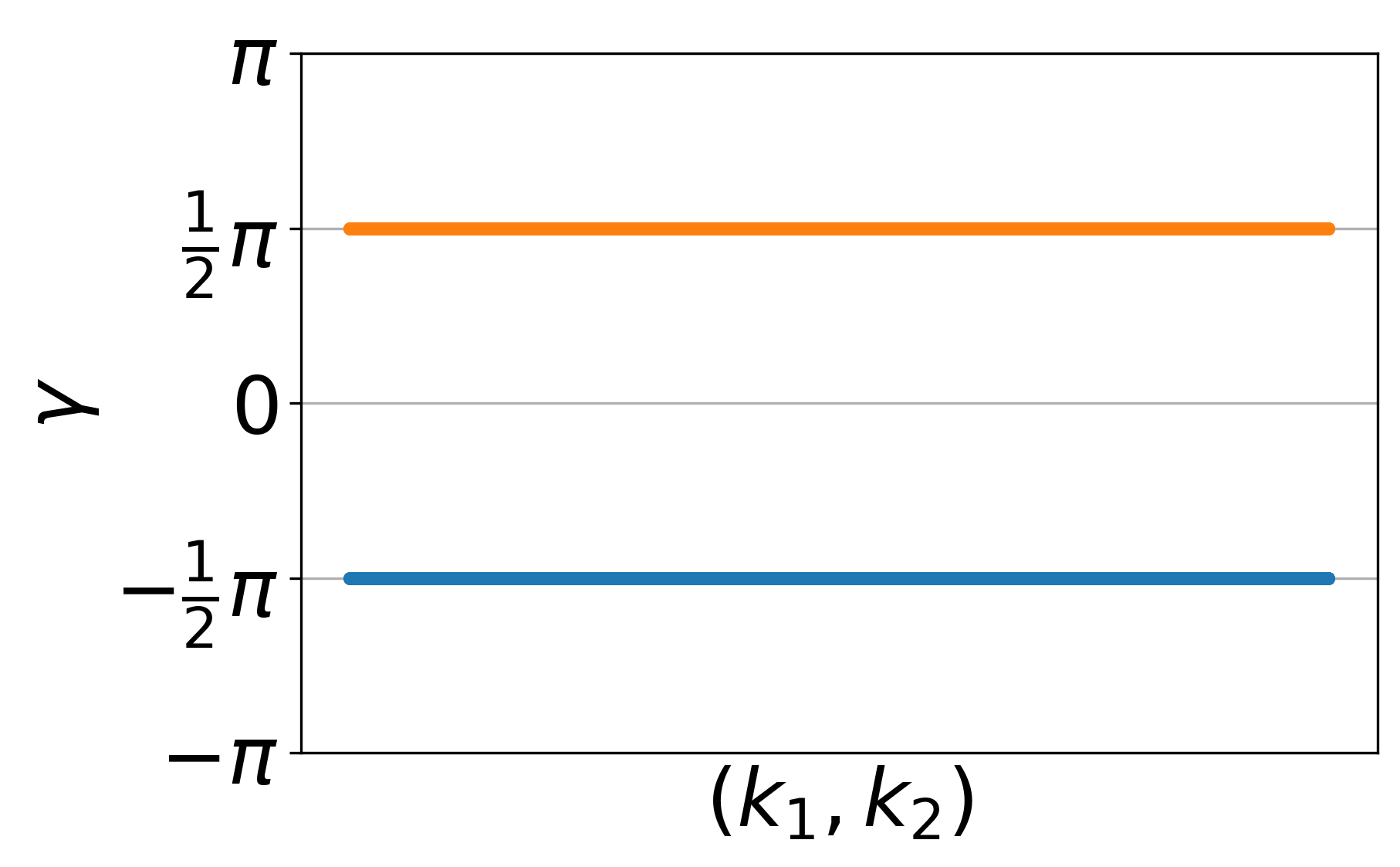}
    \caption{Non-Abelian Berry phase ($k_3$-directed Wilson-loop) spectrum as a function of $(k_1,k_2)$ for the lowest two bands in the low-energy sector of the excitation spectrum for our hexagonal periodic Anderson model with space group no.~176 ($P 6_3 / m$). $(k_1,k_2)$ points are sampled uniformly in the 2D BZ. Importantly, we remove the results of non-Abelian Berry phase when $(k_1,k_2) = (0,0)$, $(0.5,0)$, $(0,0.5)$ and $(0.5,0.5)$ in the reduced coordinate, since their $k_3$-directed Wilson loops pass through the Dirac nodal line, making the corresponding Wilson loop matrices ill-defined. After removing those $(k_1,k_2)$ points that make the $k_3$-directed Wilson loop matrices ill-defined, we see that the non-Abelian Berry phase spectrum contains only values of $-\frac{\pi}{2}$ and $\frac{\pi}{2}$.}
    \label{fig:SG_176_Wilson_loop}
\end{figure}

\subsection{Surface states from Wannier centers}

In this section, we illustrate how the quantized non-Abelian Berry phases with values $-\frac{\pi}{2}$ and $\frac{\pi}{2}$ in space group no.~176 ($P 6_3 / m$) with spin-1/2 time-reversal symmetry could be used to establish the presence of surface states.

When a hexagonal crystal structure is cut into finite size along $\mathbf{a}_3$, the resulting finite-size slab is still extended along $\mathbf{a}_1$ and $\mathbf{a}_2$. 
Hence, the BZ of the slab structure is specified by $(k_1 , k_2)$.
Let us specifically focus on a given value of $(k_1 , k_2)$, at which the Bloch Hamiltonian of the slab describes an effective 1D system finite along $\mathbf{a}_3$.
The eigenphases of the Wilson-loop matrix $[\mathcal{W}^{\mathbf{G}_3}_{k_1,k_2}]$ are then the non-Abelian Berry phases of that effective 1D system, and hence specifies where its Wannier centers are located.
With this in mind, we can now explore the consequence of the $(k_1,k_2)$-independent Wilson-loop eigenphases $\{ -\frac{\pi}{2} , \frac{\pi}{2} \}$.
For simplicity, we will specialize to the case where the number of occupied bands is $2$ and the system contains two sub-lattices, one at $\mathbf{0}$ and the other at $\frac{1}{2}\mathbf{a}_3$.
Our analysis will carry over to any $(k_1 , k_2)$, provided that the occupied energy bands along $(k_1,k_2,k_3)$ to $(k_1 , k_2 , k_3) + \mathbf{G}_3$ are gapped from the other bands such that the Wilson-loop matrix $[\mathcal{W}^{\mathbf{G}_3}_{k_1,k_2}]$ is well-defined.
The set of non-Abelian Berry phases $\{ -\frac{\pi}{2} , \frac{\pi}{2} \}$ of the effective 1D system means that the 1D Wannier centers are localized at $\{ -\frac{1}{4} \mathbf{a}_3 , \frac{1}{4} \mathbf{a}_3 \}$.
This means that the Wannier centers are always located between the two sub-lattices, as illustrated in Supplementary Fig.~\ref{fig:Wannier_centers_between_two_sublattices}.
Hence, when we cut the system into finite size along $\mathbf{a}_3$, there is always a Wannier center that is exposed outside the sample, leading to the presence of boundary states.
Such boundary states are then the surface states at momentum $(k_1,k_2)$ where the $[\mathcal{W}^{\mathbf{G}_3}_{k_1,k_2}]$ is well-defined -- namely those $(k_1,k_2)$ points that do not correspond to the projected momenta of the fourfold-degenerate nodal line.
From this, we can further state that such surface states are bound by the projected momenta of the fourfold-degenerate nodal line.
We emphasize that this is a new type of drumhead surface state, as it is a consequence of the fourfold-degenerate (Dirac) nodal line.

\begin{figure}[ht]
    \centering
    \includegraphics[width=1\linewidth]{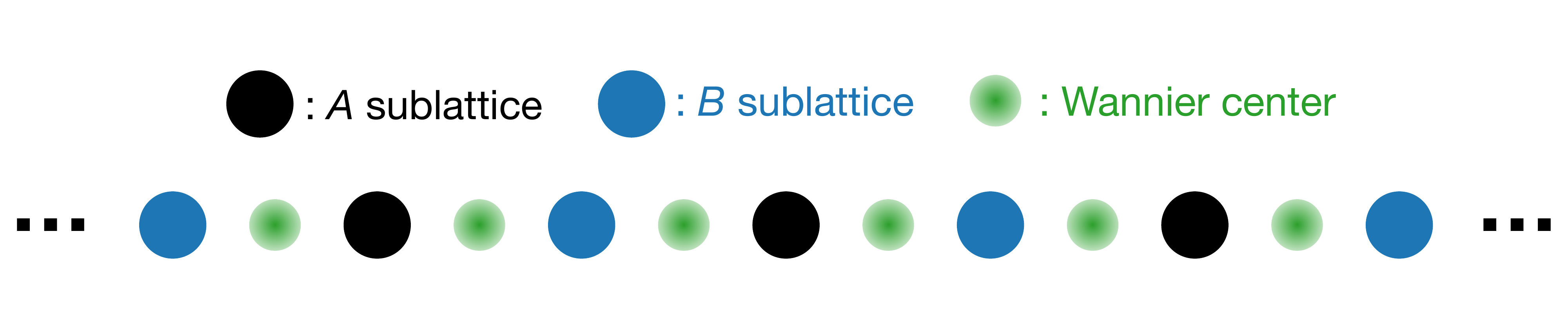}
    \caption{Schematic representation of Wannier centers located in the middle between the two sublattices. $A$ sublattices are at $n \mathbf{a}_3$, $B$ sublattices are at $\left( n + \frac{1}{2} \right) \mathbf{a}_3$, and the Wannier centers are at $\left( n - \frac{1}{4} \right) \mathbf{a}_3$ and $\left( n + \frac{1}{4} \right) \mathbf{a}_3$, where $n \in \mathbb{Z}$. These Wannier centers correspond to the quantized non-Abelian Berry phases in the space group no.~176 ($P 6_3 / m$) with the spin-1/2 time-reversal symmetry $\mathcal{T}$.}
    \label{fig:Wannier_centers_between_two_sublattices}
\end{figure}

\section{Implication of $\{ C_{2z} \mathcal{T}|0,0,1/2 \}$ symmetry in space group no.~173 ($P 6_3$) with spin-1/2 $\mathcal{T}$ symmetry}\label{sec:implication_C2zT_0_0_one_half_SG_173}

In the presence of spin-1/2 time-reversal symmetry $\mathcal{T}$, space group no.~173 ($P 6_3$) contains a group element $\{ C_{2z} \mathcal{T}|0,0,1/2 \}$. In this section, we derive the consequence of such a symmetry --- enforced twofold degeneracy on a Brillouin zone boundary.

We start by writing down a Bloch state, \textit{i.e.} energy eigenstate of the Hamiltonian, with crystal momentum $(k_1,k_2,0.5)$ in the reduced coordinate as $| \psi_{k_1,k_2,0.5} \rangle$.
This crystal momentum $(k_1,k_2,0.5)$ is located on the top (or equivalently bottom) Brillouin zone boundary.
Since $\{ C_{2z} \mathcal{T}|0,0,1/2 \} $ is a symmetry of the Hamiltonian, $\{ C_{2z}\mathcal{T}|0,0,1/2 \}  | \psi_{k_1,k_2,0.5} \rangle$ is also an eigenstate of the Hamiltonian with the same energy eigenvalue as $| \psi_{k_1,k_2,0.5} \rangle$.
In addition, $\{ C_{2z}\mathcal{T}|0,0,1/2 \}  | \psi_{k_1,k_2,0.5} \rangle$ and $| \psi_{k_1,k_2,0.5} \rangle$ carry the same crystal momentum $(k_1,k_2,0.5)$.
This is because $C_{2z}$ maps $(k_1,k_2,k_3)$ to $(-k_1,-k_2,k_3)$, $\mathcal{T}$ maps $(k_1,k_2,k_3)$ to $(-k_1,-k_2,-k_3)$, the translation part of the space-group operation does not change the momentum, and $k_3 = -0.5$ and $k_3 = 0.5$ are equivalent $k_3$-components.
We can evaluate the inner product of $\{ C_{2z}\mathcal{T}|0,0,1/2 \}  | \psi_{k_1,k_2,0.5} \rangle$ and $| \psi_{k_1,k_2,0.5} \rangle$ as
\begin{align}
    & \langle \psi_{k_1,k_2,0.5} | \{ C_{2z}\mathcal{T}|0,0,1/2 \}  \psi_{k_1,k_2,0.5} \rangle \nonumber \\
    & = \langle \{ C_{2z}\mathcal{T}|0,0,1/2 \}  \psi_{k_1,k_2,0.5} | \left( \{ C_{2z}\mathcal{T}|0,0,1/2 \}  \right)^2 \psi_{k_1,k_2,0.5} \rangle^* \nonumber \\
    & = \langle \{ C_{2z}\mathcal{T}|0,0,1/2 \}  \psi_{k_1,k_2,0.5} | \{ E | 0,0,1 \} \psi_{k_1,k_2,0.5} \rangle^* \nonumber \\
    & = -\langle \{ C_{2z}\mathcal{T}|0,0,1/2 \}  \psi_{k_1,k_2,0.5} | \psi_{k_1,k_2,0.5} \rangle^* \nonumber \\
    & = -\langle \psi_{k_1,k_2,0.5} | \{ C_{2z}\mathcal{T}|0,0,1/2 \}  \psi_{k_1,k_2,0.5} \rangle
\end{align}
where we have used the fact that $\{ C_{2z}\mathcal{T}|0,0,1/2 \} $ is an anti-unitary symmetry,
\begin{align}
    & \left( \{ C_{2z} \mathcal{T}|0,0,1/2 \} \right)^2 \nonumber \\
    & = \{ C_{2z} \mathcal{T}|0,0,1/2 \} \{ C_{2z} \mathcal{T}|0,0,1/2 \} \nonumber \\
    & = \{ C_{2z} \mathcal{T} C_{2z} \mathcal{T}|0,0,1 \} \nonumber \\
    & = \{ \left( C_{2z} \mathcal{T} \right)^2 |0,0,1 \} \nonumber \\
    & = \{ E | 0,0,1 \},
\end{align}
with $E$ being the identity, and $\{ E | 0,0,1 \} | \psi_{k_1,k_2,0.5} \rangle = -| \psi_{k_1,k_2,0.5} \rangle$.
This implies
\begin{equation}
    \langle \psi_{k_1,k_2,0.5} | \{ C_{2z} \mathcal{T}|0,0,1/2 \}  \psi_{k_1,k_2,0.5} \rangle = 0,
\end{equation}
meaning that $\{ C_{2z}\mathcal{T}|0,0,1/2 \}  | \psi_{k_1,k_2,0.5} \rangle$ and $| \psi_{k_1,k_2,0.5} \rangle$ are orthogonal states with the same crystal momentum $(k_1,k_2,0.5)$ and energy.
Hence, due to the $\{ C_{2z}\mathcal{T}|0,0,1/2 \} $ symmetry, the energy bands with crystal momentum $(k_1,k_2,0.5)$ is twofold degenerate for a system with space group no.~173 ($P 6_3$) and spin-1/2 time-reversal symmetry.

\section{DFT results for $\rm Ce_2NiGe_3$}

\begin{figure}[th]
    \centering
    \includegraphics[width=\linewidth]{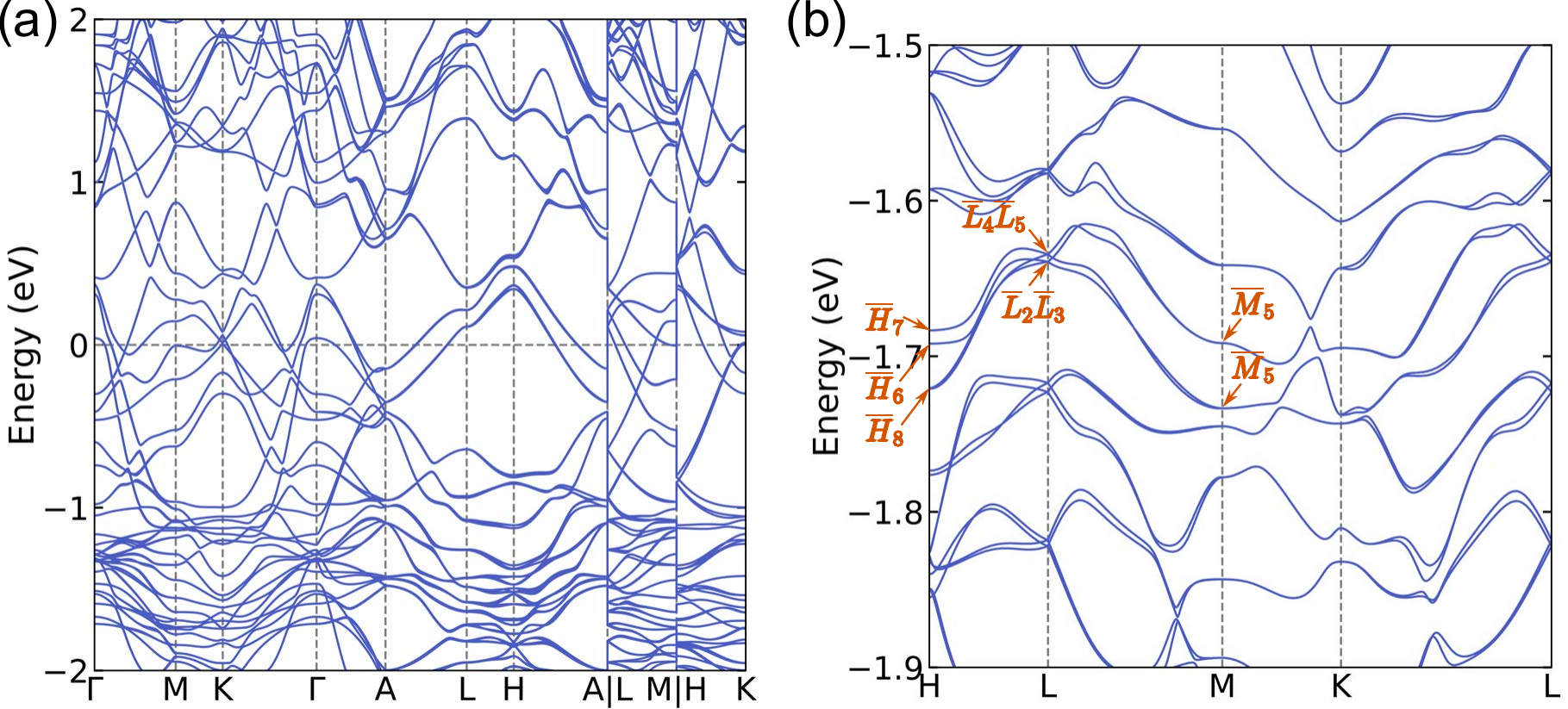}
    \caption{(a)~DFT band structures (with spin-orbit couplings) of $\rm Ce_2NiGe_3$ with the achiral space group no.~190 ($P \bar{6} 2c$), and (b)~the corresponding zoomed-in band structure along high symmetry lines $H-L-M-K-L$. 
    The Weyl-nodal lines near $L$ point is consistent with symmetry requirement.
    }
    \label{fig:DFT_band_structures_SI}
\end{figure}
Supplementary Fig.~\ref{fig:DFT_band_structures_SI}(a) provides the DFT band structure of $\rm Ce_2NiGe_3$, where the effect of spin-orbit couplings is included.
The degeneracies and nodal lines that happen away from Fermi energy are consistent with our symmetry analysis. 
Supplementary Fig.~\ref{fig:DFT_band_structures_SI}(b) shows a zoomed-in plot of the hourglass shaped crossings (part of the symmetry enforced Weyl-nodal line) near $-2$eV along high symmetry lines $H-L-M-K-L$. 
The irreducible representations are also analyzed,
which agrees with the symmetry analysis~\cite{zhang2018topological}.

\section{Procedure for searching candidate materials}

In this section, we provide an expanded description of the procedure we used to search for candidate materials of topological Kondo semimetals in hexagonal crystal systems.

In our procedure, we start by obtaining cerium-, uranium-, and ytterbium-based materials with hexagonal space groups that support band-crossing features~\cite{zhang2018topological} in the Inorganic Crystal Structure Database (ICSD).
We next perform a literature search on those materials about their measured physical properties, which can be categorized as follows:
\begin{enumerate}
    \item Electrical resistivity ($\rho$): Measurements of $\rho$ can establish if a material has a semimetallic-type resistivity and, in some cases, if the material exhibits Kondo physics. 
    If $\rho(T)$ is nearly independent of $T$, or does not show a significant difference between high and low $T$, where $T$ is the temperature, then the system exhibits a semimetallic-type resistivity.
    This can be quantified by evaluating the residual resistivity ratio (RRR) $\equiv \frac{\rho (\mathrm{high}\ T)}{\rho (\mathrm{low}\ T)}$. 
    A material with the semimetallic-type resistivity would exhibit an order $O(1)$ RRR.
    On the other hand, a system with band-crossing features could also be a metal, where the Fermi-liquid type resistivity exhibits a $T^2$-behavior. Alternatively, if $\rho(T)$ decreases as $T$ decreases, we may also associate the system with metallic-type resistivity. On the other hand, Kondo physics can lead to various features in $\rho(T)$. The Kondo-lattice coherence can lead to a local maximum in $\rho(T)$. The magnetic contribution to the resistivity, denoted as $\rho_\mathrm{mag} (T)$, also exhibits a $-\ln (T)$-behavior at relatively high temperatures due to the Kondo effect.
    \item Magnetic properties: Magnetic measurements such as the temperature dependence of the magnetic susceptibility ($\chi$) can establish whether the magnetic coupling in the material is of the antiferromagnetic type or of the ferromagnetic type, and if the material has magnetic orders. Magnetic transitions typically appear as anomalies in $\chi(T)$. Since we focus on the paramagnetic states with spin-1/2 time-reversal symmetry, materials with low magnetic transition temperatures, or no magnetic orders observed in the experiments, are favorable.
    From $\chi(T)$, the paramagnetic Curie temperature ($\theta_p$) can be determined through the Curie-Weiss law $\chi(T) = \chi_0 + \frac{C}{T - \theta_p}$ where both $\chi_0$ and $C$ are constants~\cite{mugiraneza2022tutorial}. The constant $C$ can be further used to determine the effective magnetic moment $\mu_\mathrm{eff}$.
    A negative [positive] value of $\theta_p$ indicates antiferromagnetic [ferromagnetic] interactions. 
    \item Specific heat ($C_p$): Specific-heat measurements can establish if the material is a heavy-fermion compound that has a moderate or strong correlations, which typically lead to an enhanced value of the Sommerfeld coefficient $\gamma$. The Sommerfeld coefficient $\gamma$ is defined as the linear coefficient of $C_p$ on $T$.
    It is the magnetic contribution to the specific heat, denoted as $C_\mathrm{mag}$, that is used to determine the Sommerfeld coefficient via the linear-in-$T$ coefficient.
    \item Kondo temperature ($T_{\rm K}$): $T_{\rm K}$ denotes the temperature scale for the initial onset of Kondo screening as the temperature is lowered.
    The information of $T_{\rm K}$ can be extracted from measurements such as electrical resistivity, specific heat or thermopower.
\end{enumerate}

In Supplementary Fig.~\ref{fig:ICSD_Ce_U_Yb_bar_fig}, we show the amount of cerium-, uranium-, and ytterbium-based compounds in ICSD that have hexagonal space groups with topological band crossings identified in Ref.~\citenum{zhang2018topological}.
We see that space groups no.~173 ($P 6_3 $), no.~176 ($P 6_3 / m$), no.~180 ($P 6_2 22$), and no.~190 ($P \bar{6} 2c$) are relatively abundant.
In the following sections, we provide the details of the primary candidate materials and other materials that are candidate materials we identified through the search procedure described above.
In Supplementary Table~\ref{tab:notation_physical_properties} we provide the notations of relevant physical properties in our studies.

\begin{figure}[ht]
    \centering
    \includegraphics[width=\linewidth]{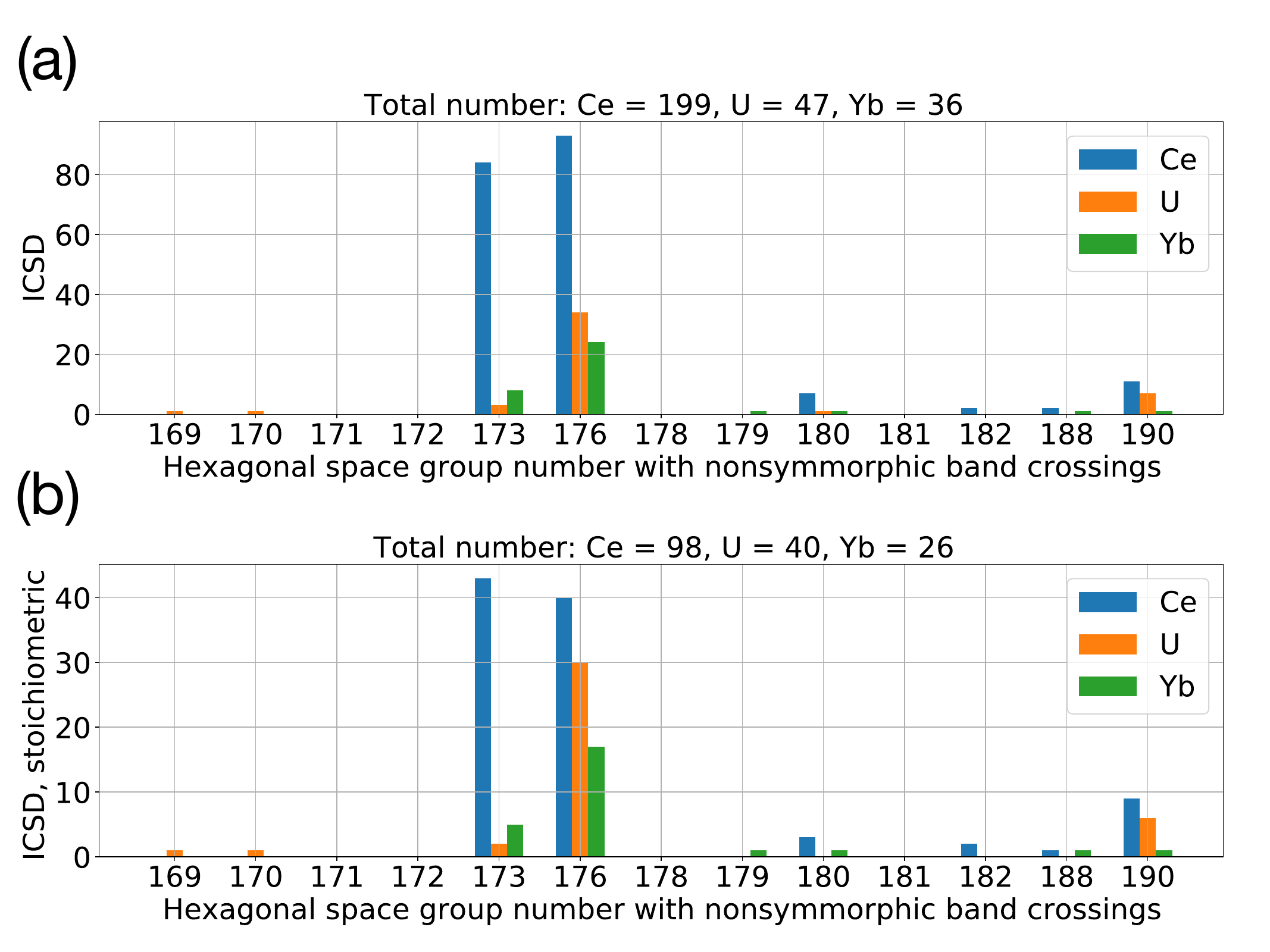}
    \caption{
    The amount of material entries in ICSD that (1)~contain either cerium (Ce), uranium (U), or ytterbium (Yb) element, and (2)~has a hexagonal space group that exhibits topological band crossings enforced by nonsymmorphic symmetries as classified in Ref.~\citenum{zhang2018topological}.
    (a)~includes both stoichiometric and non-stoichiometric compounds while (b) includes only the stoichiometric ones.
    }
    \label{fig:ICSD_Ce_U_Yb_bar_fig}
\end{figure}

\begin{table}[ht]
    \centering
    \begin{tabular}{|c|c|c|c|c|c|c|c|c|c|c|}
        \hline
        Temperature & $T$ \\
        \hline
        Magnetic susceptibility & $\chi$ \\
        \hline
        Paramagnetic Curie temperature or Curie-Weiss temperature & $\theta_p$ \\
        \hline
        Ne\'{e}l temperature (anti-ferromagnetic transition temperature) & $T_{\rm N}$ \\
        \hline
        Curie temperature (ferromagnetic transition temperature) & $T_\mathrm{C}$ \\
        \hline
        Bohr magneton & $\mu_\mathrm{B}$ \\
        \hline
        (Experimentally-determined) effective magnetic moment & $\mu_\mathrm{eff}$ \\
        \hline
        Specific heat & $C_p$ \\
        \hline
        Magnetic contribution to the specific heat & $C_\mathrm{mag}$ \\
        \hline
        Sommerfeld coefficient & $\gamma$ \\
        \hline
        Electrical resistivity & $\rho$ \\
        \hline
        Magnetic contribution to the electrical resistivity & $\rho_\mathrm{mag}$ \\
        \hline
        Residual resistivity ratio & RRR $\equiv \frac{\rho (\mathrm{high}\ T)}{\rho (\mathrm{low}\ T)}$ \\
        \hline
        Kondo temperature & $T_{\rm K}$ \\
        \hline
    \end{tabular}
    \caption{Notations related to the physical properties.}
    \label{tab:notation_physical_properties}
\end{table}

\section{Primary candidate materials}\label{sec:SI_prime_candidate}

\setcounter{subsection}{0}
\subsection{CePt$_2$B}

\begin{figure}
    \centering
    \includegraphics[width=1\linewidth]{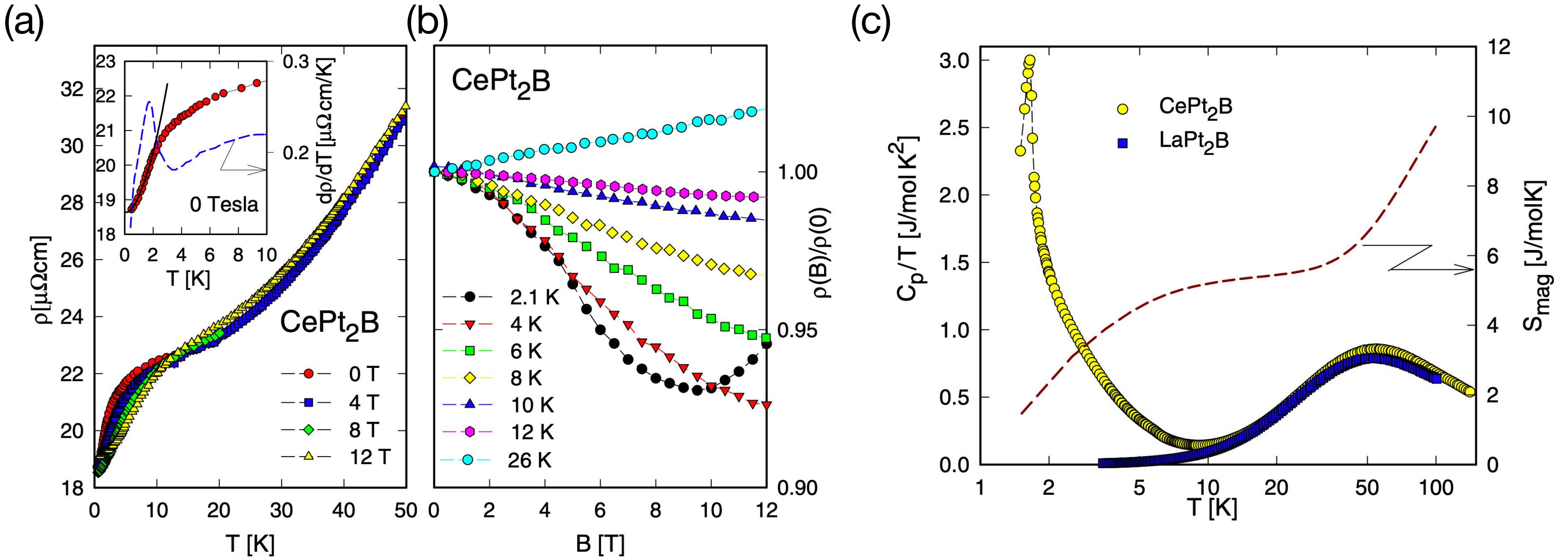}
    \caption{(a)~Electrical resistivity as a function of temperature at different applied field strengths, (b)~magnetoresistance as a function of applied field at different temperatures, and (c)~specific heat divided by temperature as a function of temperature together with the magnetic entropy as a function of temperature for CePt$_2$B copied from Ref.~\citenum{lackner2005lowtemperature}.}
    \label{fig:Ce_Pt2_B_experimental_figures_from_literature}
\end{figure}

CePt$_2$B~\cite{sologub2000newstructuretype,lackner2005lowtemperature} has the chiral space group no.~180 ($P 6_2 22$) and could be a potential candidate for the paramagnetic chiral WKSM with Weyl points~\cite{zhang2018topological}, provided that the temperature is above the magnetic transition temperature and below (or around) the characteristic temperature $T_{\rm K}$.
We now summarize the supporting physical properties of CePt$_2$B reported in Ref.~\citenum{lackner2005lowtemperature}, see also Supplementary Fig.~\ref{fig:Ce_Pt2_B_experimental_figures_from_literature} for the experimental figures we copied from Ref.~\citenum{lackner2005lowtemperature}.
The magnetic transition temperature is reported as $2.1$K from the observation of a $\lambda$-like anomaly in the $C_p / T$ result.
The characteristic temperature $T_{\rm K} = 3.5 - 5$K for CePt$_2$B is obtained from the results of specific heat and magnetic entropy.
In addition, the measured electrical resistivity $\rho$ of CePt$_2$B shows metallic behavior, where $\rho(T)$ decreases as $T$ decreases, and $\rho$ takes values around $20 - 30$ $\mu\Omega$cm from the lowest experimental temperature to $50$K.
Specifically, $\rho (T=5\mathrm{K}) \approx 22$ $\mu\Omega$cm for CePt$_2$B.
Regarding the electron correlation in CePt$_2$B, the experimental value of $C_p / T$ is large ($> 100$ mJ/K$^2$ mol), which we here interpret as a potential indication of heavy-fermion behavior.
Finally, as stated in Ref.~\citenum{lackner2005lowtemperature}, the magnetoresistance $\rho(B)/\rho(0)$, where $B$ is the applied magnetic field, of CePt$_2$B exhibits the qualitative form of the Kondo model~\cite{schlottmann1983betheansatz,dzsaber2021giant}.

\subsection{Ce$_6$Rh$_{32}$P$_{17}$}

Ce$_6$Rh$_{32}$P$_{17}$~\cite{pivan1988crystal} has the achiral space group no.~176 ($P 6_3 / m$) and it could be a potential candidate for the paramagnetic Dirac-Kondo semimetal with fourfold degenerate nodal lines on the BZ boundary~\cite{zhang2018topological}.
In Supplementary Fig.~\ref{fig:Ce6_Rh32_P17_experimental_figures_from_literature} we provide the experimental figures copied from Ref.~\citenum{pivan1988crystal}.
The measurements of the magnetic susceptibility yield the effective magnetic moment $\mu_\mathrm{eff} = 1.43 \mu_\mathrm{B}$ and the paramagnetic Curie temperature $\theta_p = -2$K. The negative value of $\theta_p$ suggests that Ce$_6$Rh$_{32}$P$_{17}$ has anti-ferromagnetic coupling. There is no signature of magnetic transition in the magnetic measurements reported in Ref.~\citenum{pivan1988crystal}, which means that Ce$_6$Rh$_{32}$P$_{17}$ remains paramagnetic down to the lowest temperature of the experiments.
The measured electrical resistivity $\rho(T)$ is $580000$ $\mu\Omega$cm at $T=293$K and is $495000$ $\mu\Omega$cm at $T = 2$K. Although these values of the electrical resistivity are relatively high, the ratio $\frac{\rho(T=293K)}{\rho(T=2K)} \approx 1.17$ indicates that the temperature dependence of $\rho(T)$ is of the semimetallic type, meaning that there is no significant difference between the high-temperature and low-temperature resistivities.
Furthermore, $\rho(T)$ shows a small upturn around $T=30$K as the temperature is lowered, which indicates potential Kondo physics such as the Kondo scattering.
We note that in ICSD, the P elements at the $4f$ and $2b$ positions in Ce$_6$Rh$_{32}$P$_{17}$ (ICSD Collection Code 52890) are reported to have half occupancy.
Since the element with half occupancy is not Ce, we may expect that this will not affect the lattice-coherent behavior formed from the Ce elements.

\begin{figure}
    \centering
    \includegraphics[width=0.65\linewidth]{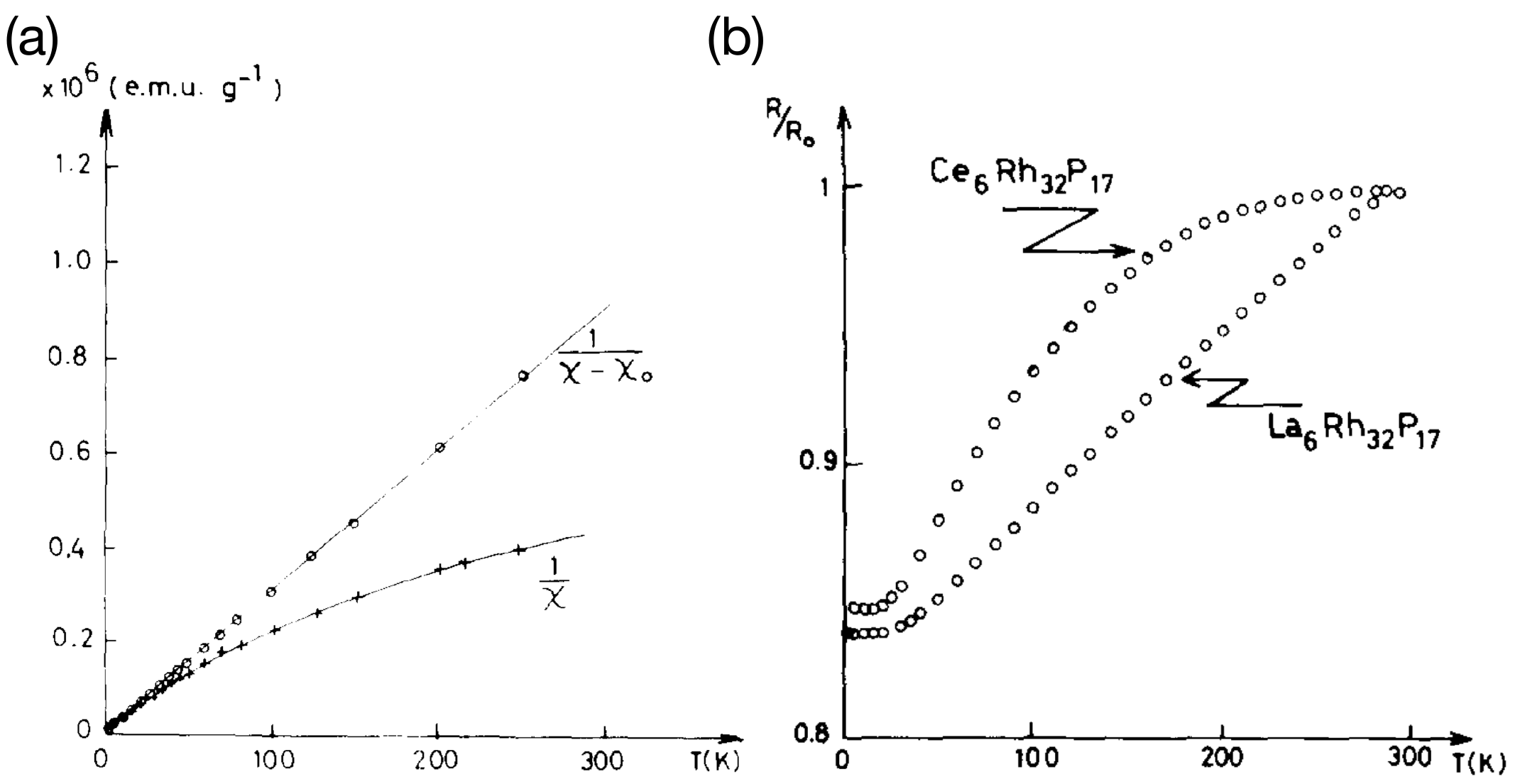}
    \caption{The temperature dependence of (a)~inverse magnetic susceptibility and (b)~normalized electrical resistivity for Ce$_6$Rh$_{32}$P$_{17}$ copied from Ref.~\citenum{pivan1988crystal}.}
    \label{fig:Ce6_Rh32_P17_experimental_figures_from_literature}
\end{figure}

\subsection{Ce$_6$Co$_{2-\delta}$Si$_3$ ($\delta = 0.33$)}

The chemical formula of Ce$_6$Co$_{2-\delta}$Si$_3$ we consider in this work is Ce$_6$Co$_{1.67}$Si$_3$~\cite{gaudin2007on,chevalier2007the}, which has the achiral space group no.~176 ($P 6_3 / m$) and it could be a potential candidate for the paramagnetic Dirac-Kondo semimetal with fourfold degenerate nodal lines on the BZ boundary~\cite{zhang2018topological}. 
In Supplementary Fig.~\ref{fig:Ce6_Co1p67_Si3_experimental_figures_from_literature} we provide the experimental figures of Ce$_6$Co$_{1.67}$Si$_3$ copied from Ref.~\citenum{gaudin2007on}.
The measurements of the magnetic susceptibility yield the effective magnetic moment $\mu_\mathrm{eff} = 2.59 \mu_\mathrm{B} / $Ce and the paramagnetic Curie temperature $\theta_p = -77$K. The relatively large negative value of $\theta_p$ indicates relatively strong anti-ferromagnetic couplings, suggesting that Ce$_6$Co$_{1.67}$Si$_3$ may be a Kondo compound.
The magnetic measurements also show that there is no magnetic ordering at least above $1.8$K, which is also supported by the specific-heat measurements showing no anomalies.
This indicates that Ce$_6$Co$_{1.67}$Si$_3$ remains paramagnetic at least above $1.8$K.
The Sommerfeld coefficient $\gamma$ is determined as $\gamma = 162$ mJ/K$^2$ Ce-mol using the specific-heat data in the temperature range between $T = 10$K and $22$K, suggesting that Ce$_6$Co$_{1.67}$Si$_3$ is a moderate heavy-fermion compound.
In addition, as reported in Ref.~\citenum{gaudin2007on}, the colour of Ce$_6$Co$_{1.67}$Si$_3$ is metallic light grey, which suggests that Ce$_6$Co$_{1.67}$Si$_3$ might be non-insulating.
We note that in Ce$_6$Co$_{1.67}$Si$_3$ some Co elements have partial occupancy, as reported in Ref.~\citenum{gaudin2007on}.
Since the element with partial occupancy is not Ce, we may expect that this will not affect the lattice-coherent behavior formed from the Ce elements.

\begin{figure}
    \centering
    \includegraphics[width=1\linewidth]{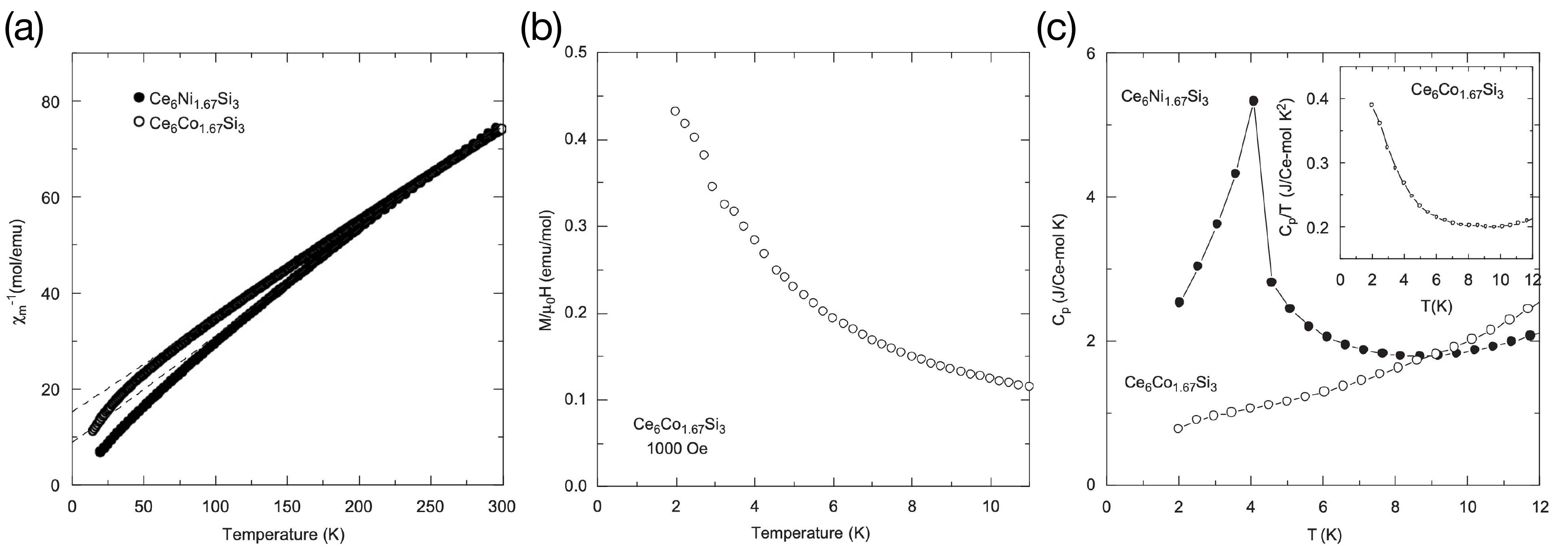}
    \caption{The temperature dependence of (a)~inverse magnetic susceptibility, (b)~magnetization divided by the applied field, and (c)~specific heat (inset: $C_p / T$ as a function of $T$) for Ce$_6$Co$_{1.67}$Si$_3$ (Ce$_6$Co$_{2-\delta}$Si$_3$ with $\delta = 0.33$) copied from Ref.~\citenum{gaudin2007on}.}
    \label{fig:Ce6_Co1p67_Si3_experimental_figures_from_literature}
\end{figure}

\subsection{Ce$_2$NiGe$_3$}
\label{sec:Ce2_Ni_Ge3_experimental_information_from_literature}

\begin{figure}
    \centering
    \includegraphics[width=1\linewidth]{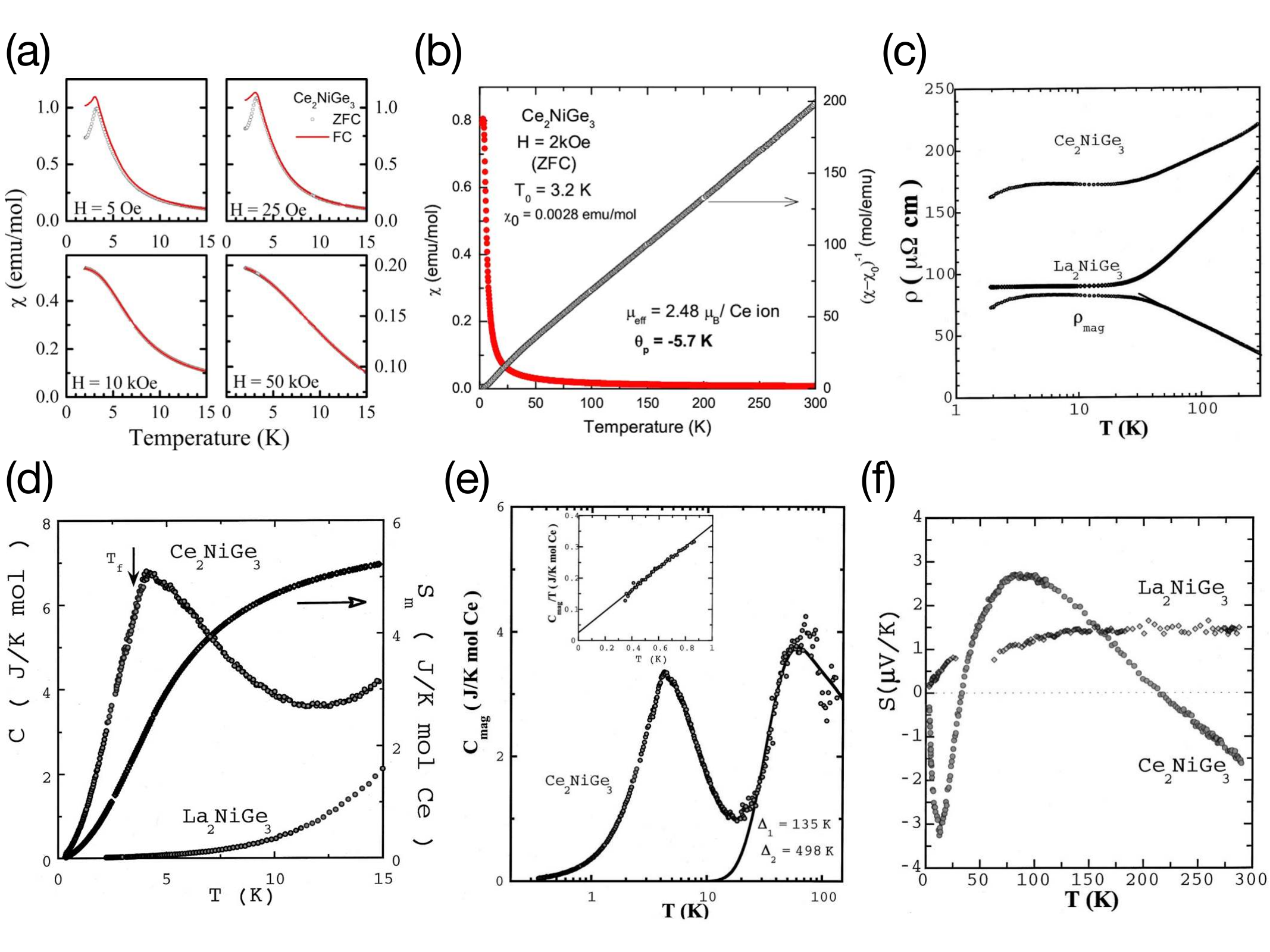}
    \caption{(a)\&(b)~Magnetic susceptibility of Ce$_2$NiGe$_3$ copied from Ref.~\citenum{kalsi2014neutron}. The temperature dependence of (c)~electrical resistivity and magnetic resistivity, (d)~specific heat and magnetic entropy, (e)~magnetic specific heat (inset: $C_\mathrm{mag}/T$ as a function of $T$), and (f) thermoelectric power for Ce$_2$NiGe$_3$ copied from Ref.~\citenum{huo2001electric}.}
    \label{fig:Ce2_Ni_Ge3_experimental_figures_from_literature}
\end{figure}

Ce$_2$NiGe$_3$~\cite{kalsi2014neutron} has the achiral space group no.~190 ($P \bar{6} 2c$) and it could be a potential candidate for the paramagnetic WKSM with Weyl nodal lines on the BZ boundary~\cite{zhang2018topological}, provided that the temperature is above the magnetic transition temperature. The measurements of the magnetic susceptibility yield the effective magnetic moment $\mu_\mathrm{eff} = 2.48 \mu_\mathrm{B}$ and the paramagnetic Curie temperature $\theta_p = -5.7$K. The negative value of $\theta_p = -5.7$K suggests anti-ferromagnetic couplings in Ce$_2$NiGe$_3$. The magnetic measurements further suggest a spin-glass state at $T_{\rm N} = 3.2$K. This means that above $T_{\rm N} = 3.2$K Ce$_2$NiGe$_3$ could be in the paramagnetic state, see also Supplementary Fig.~\ref{fig:Ce2_Ni_Ge3_experimental_figures_from_literature} for the experimental figures we copied from Ref.~\citenum{kalsi2014neutron}.

Although Ref.~\citenum{huo2001electric} reports that Ce$_2$NiGe$_3$ crystallizes in the hexagonal AlB$_2$-type structure, we here include the physical properties of Ce$_2$NiGe$_3$ reported in Ref.~\citenum{huo2001electric}, as they further supports that Ce$_2$NiGe$_3$ could be a paramagnetic WKSM, see also Supplementary Fig.~\ref{fig:Ce2_Ni_Ge3_experimental_figures_from_literature} for the experimental figures we copied from Ref.~\citenum{huo2001electric}. Specifically, the $\rho(T)$ data reported in Ref.~\citenum{huo2001electric} indicates that $\rho$($T=2$K) is about $160$ $\mu\Omega$cm, $\rho$($T=10$K) is about $170$ $\mu\Omega$cm, $\rho$($T=100$K) is about $200$ $\mu\Omega$cm, and $\rho$($T=200$K) is about $210$ $\mu\Omega$cm, which indicate a semimetallic-type resistivity as $\frac{\rho(T=200\mathrm{K})}{\rho(T=10\mathrm{K})} \approx 1.24$.
In addition, the magnetic contribution to the resistivity $\rho_\mathrm{mag}(T)$, which is obtained in Ref.~\citenum{huo2001electric} by subtracting the $\rho(T)$ data of La$_2$NiGe$_3$ from the $\rho(T)$ data of Ce$_2$NiGe$_3$, shows $-\ln T$-behaviors from the room temperature to $T = 50$K and from $T = 20$K to $T = 5$K, suggesting the Kondo effect in the paramagnetic phase of Ce$_2$NiGe$_3$.
The Sommerfeld coefficient is obtained as $\gamma = 25$ mJ/K$^2$ Ce-mol in Ref.~\citenum{huo2001electric} using the data below $T = 1$K of the magnetic contribution to the specific heat $C_\mathrm{mag}(T)$, which is obtained in Ref.~\citenum{huo2001electric} by subtracting the $C_p (T)$ data of La$_2$NiGe$_3$ from the $C_p (T)$ data of Ce$_2$NiGe$_3$.
The thermoelectric power of Ce$_2$NiGe$_3$ is also reported in Ref.~\citenum{huo2001electric}, showing a positive peak and a negative peak together with a sign change, which are significantly different behaviors compared with La$_2$NiGe$_3$.

\subsection{Ce$_2$RhSi$_3$}

Ce$_2$RhSi$_3$~\cite{szytula1993neutron,leciejewicz1995antiferromagnetic} has the achiral space group no.~190 ($P \bar{6} 2c$) and it could be a potential candidate for the paramagnetic WKSM with Weyl nodal lines on the BZ boundary~\cite{zhang2018topological}, provided that the temperature is above the magnetic transition temperature. 
The measurements of the magnetic susceptibility in the temperature range $100-300$K yield the effective magnetic moment $\mu_\mathrm{eff} = 2.4 \mu_\mathrm{B}/$Ce and the paramagnetic Curie temperature $\theta_p = -80$K.
In addition, in the temperature range $8-20$K, $\mu_\mathrm{eff} =  1.54 \mu_\mathrm{B}/$Ce and $\theta_p = -6.5$K.
The magnetic measurements further suggest an anti-ferromagnetic ordering at $T_{\rm N} = 6.8$K.
This means that above $T_{\rm N}$ Ce$_2$RhSi$_3$ could be in the paramagnetic state.

Although Ce$_2$RhSi$_3$ could crystallize in other hexagonal structures~\cite{chevalier1984anew,das1994magnetic,kase2009antiferromagnetic,szlawska2009antiferromagnetic}, in the following we include physical properties reported in some of these references, as they further support our prediction that Ce$_2$RhSi$_3$ could be a paramagnetic WKSM.
Importantly, some of these references also report the Sommerfeld coefficients $\gamma$ and the estimates of the Kondo temperature $T_{\rm K}$.
Hence, above $T_{\rm N}$ and below (or around) $T_{\rm K}$, Ce$_2$RhSi$_3$ could be in the paramagnetic Kondo-driven heavy-fermion phase.
In Supplementary Fig.~\ref{fig:Ce2_Rh_Si3_experimental_figures_from_literature_magnetic}, Supplementary Fig.~\ref{fig:Ce2_Rh_Si3_experimental_figures_from_literature_electrical_resistivity}, Supplementary Fig.~\ref{fig:Ce2_Rh_Si3_experimental_figures_from_literature_specific_heat}, and Supplementary Fig.~\ref{fig:Ce2_Rh_Si3_experimental_figures_from_literature_magnetoresistivity} we provide experimental figures for Ce$_2$RhSi$_3$ copied from the literature.

We begin with Ref.~\citenum{das1994magnetic}. 
Magnetic susceptibility measurements yield $\mu_\mathrm{eff} = 2.48 \mu_\mathrm{B}$ and $\theta_p = -65$K in the temperature range $100-300$K, and $\theta_p = -10$K in the temperature range $10-40$K.
The antiferromagnetic transition temperature is reported as $T_{\rm N} = 6$K.
The electrical resistivity of Ce$_2$RhSi$_3$ is also reported to be about $800$ $\mu\Omega$cm in the temperature range $8 - 20$K and $\frac{\rho(300\mathrm{K})}{\rho(20\mathrm{K})} \approx 1.48$, suggesting a semimetallic-type resistivity. 
Furthermore, the Sommerfeld coefficient is extracted as $\gamma = 100$ mJ/K$^2$ mol from the specific-heat data, suggesting heavy-fermion behavior.

Next, we consider Ref.~\citenum{kase2009antiferromagnetic}.
It reports $\mu_\mathrm{eff}=2.57\mu_\mathrm{B}/$Ce and $\theta_p = -79$K for the applied field $H \parallel [100]$ and $\mu_\mathrm{eff}=2.68\mu_\mathrm{B}/$Ce and $\theta_p =-127$K for $H \parallel [001]$ from the magnetic susceptibility above $100$K. 
The antiferromagnetic transition temperature of Ce$_2$RhSi$_3$ is also extracted as $T_{\rm N} = 4.5$ or $5$K.
The electrical resistivity of Ce$_2$RhSi$_3$ is consistent with a semimetal, with no significant difference between high-temperature and low-temperature values.
In addition, the magnetic contribution $\rho_\mathrm{mag}$ to the electrical resistivity, which is obtained by subtracting $\rho$ of La$_2$RhSi$_3$ from $\rho$ of Ce$_2$RhSi$_3$, exhibits $-\ln T$-behaviors from room temperature to $100$K and from $20$K to $5$K, suggesting Kondo physics.
Below $T_{\rm N}$, the Sommerfeld coefficient is determined as $\gamma = 290$ mJ/K$^2$ Ce-mol, which suggests that Ce$_2$RhSi$_3$ is a heavy-fermion system.
At $T_{\rm N}$, the magnetic entropy is extracted as $0.56 R \ln 2$, which suggests that a moderate Kondo effect could be present in Ce$_2$RhSi$_3$.
Importantly, the Kondo temperature is estimated as $T_{\rm K} = 12$K, which is the mean value of different estimates.

Finally, we have Ref.~\citenum{szlawska2009antiferromagnetic} for Ce$_2$RhSi$_3$. 
Magnetic susceptibility measurements yield $\mu_\mathrm{eff}^c = 2.55 \mu_\mathrm{B}$, $\theta_p^c = -102$K, $\mu_\mathrm{eff}^a = 2.6\mu_\mathrm{B}$, and $\theta_p^a = -78$K above $200$K, where the superscripts $a$ and $c$ mean that the applied magnetic field is along the $a$ and $c$ axes, respectively.
The magnetic transition temperature is reported as $T_{\rm N} = 4.5$K.
The X-ray photoemission spectroscopy measurements suggest that the Ce ions are trivalent.
The electrical resistivity at room temperature is reported as $\rho^a = 254$ $\mu\Omega$cm and $\rho^c = 272$ $\mu\Omega$cm, and at $2$K as $\rho^a = 222$ $\mu\Omega$cm and $\rho^c = 217$ $\mu\Omega$cm, which could be classified as the semimetallic-type resistivity.
$\rho^a$ and $\rho^c$ mean the electrical resistivity measured along the $a$ and $c$ axes.
$-\ln T$-behaviors are also observed in the magnetic contribution to the electrical resistivity.
In addition, as reported in Ref.~\citenum{szlawska2009antiferromagnetic}, the magnetoresistivity of Ce$_2$RhSi$_3$ demonstrates a data collapse corresponding to the scaling behavior of Kondo systems~\cite{schlottmann1983betheansatz,dzsaber2021giant}.
Below $3$K, the Sommerfeld coefficient is extracted as $\gamma = 580$ mJ/K$^2$ mol using the data of $C_p$(Ce$_2$RhSi$_3$)$-$$C_p$(La$_2$RhSi$_3$).
The Kondo temperature $T_{\rm K}$ is estimated to be $9$K or $9.5$K.

\begin{figure}
    \centering
    \includegraphics[width=1\linewidth]{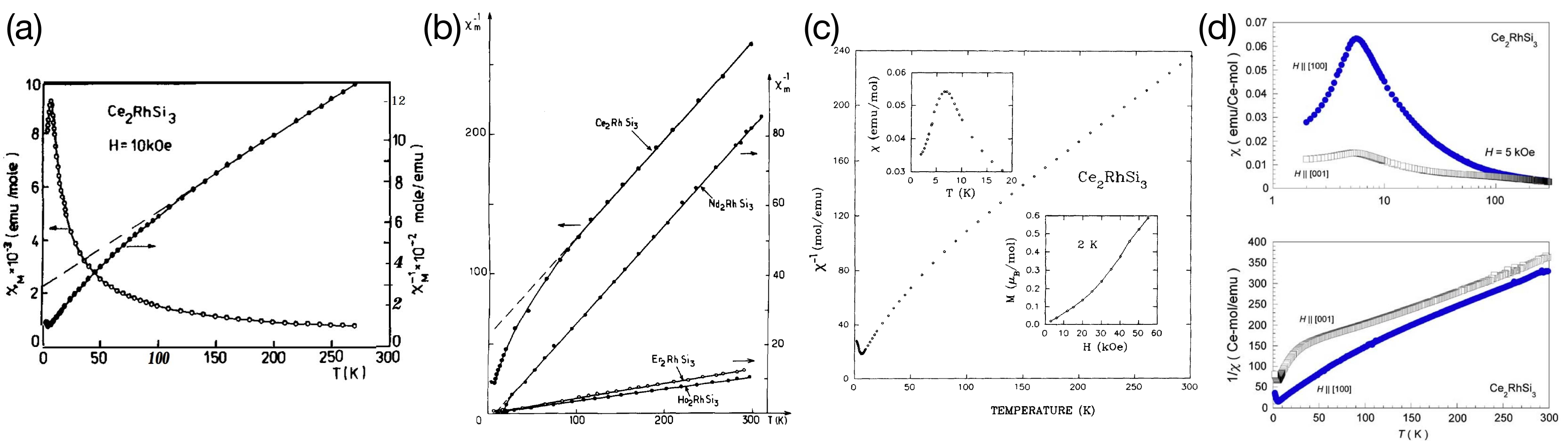}
    \caption{Temperature dependence of the magnetic susceptibility and/or the inverse magnetic susceptibility for Ce$_2$RhSi$_3$ copied from (a)~Ref.~\citenum{leciejewicz1995antiferromagnetic}, (b)~Ref.~\citenum{chevalier1984anew}, (c)~Ref.~\citenum{das1994magnetic}, and (d)~Ref.~\citenum{kase2009antiferromagnetic}. In (d), the applied magnetic field is along either $[100]$ or $[001]$.}
    \label{fig:Ce2_Rh_Si3_experimental_figures_from_literature_magnetic}
\end{figure}

\begin{figure}
    \centering
    \includegraphics[width=1\linewidth]{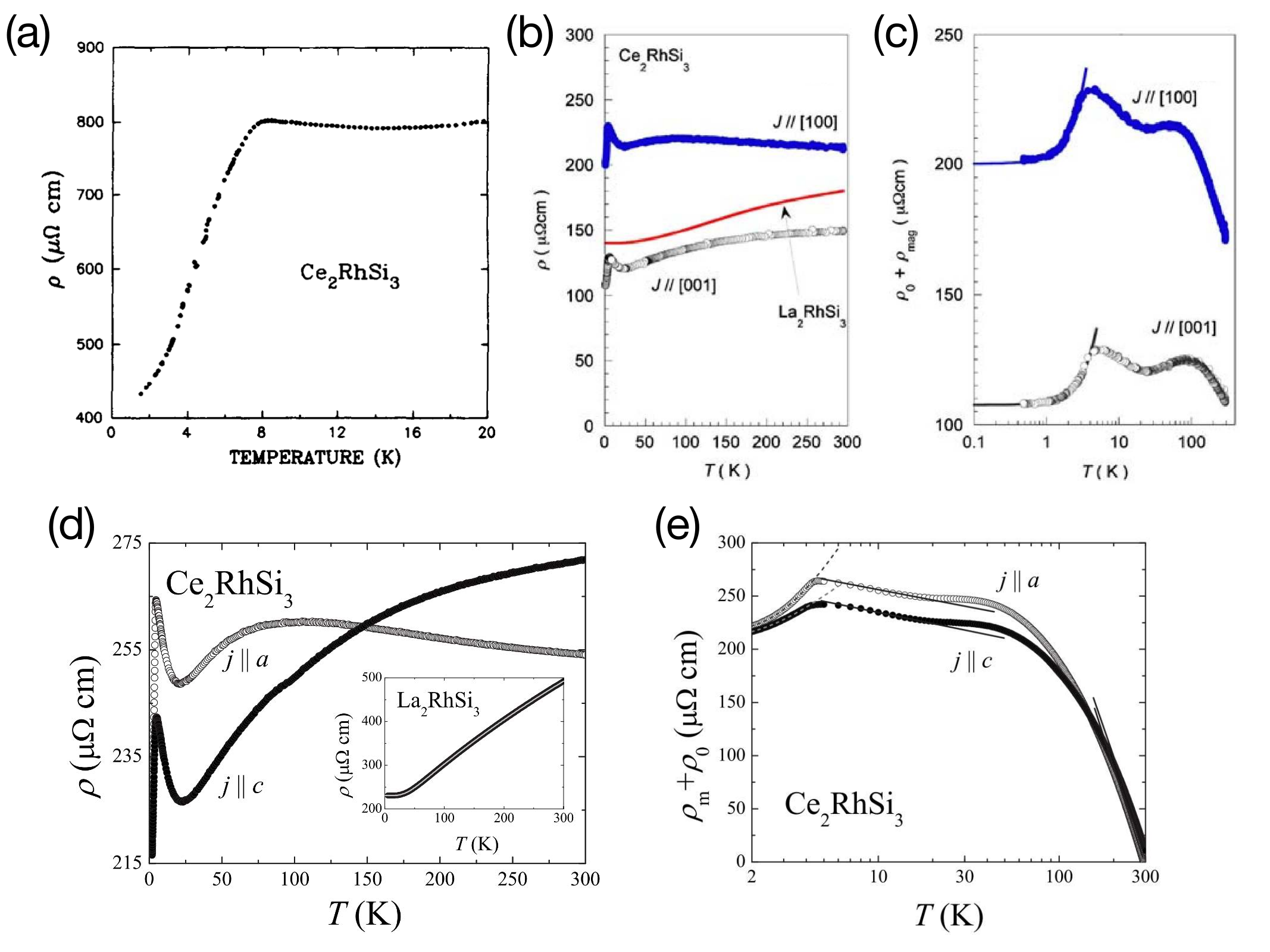}
    \caption{Temperature dependence of the electrical resistivity for Ce$_2$RhSi$_3$ copied from (a)~Ref.~\citenum{das1994magnetic}, (b)~Ref.~\citenum{kase2009antiferromagnetic}, and (d)~Ref.~\citenum{szlawska2009antiferromagnetic}. Temperature dependence of the magnetic contribution to the electrical resistivity for Ce$_2$RhSi$_3$ copied from (c)~Ref.~\citenum{kase2009antiferromagnetic} and (e)~Ref.~\citenum{szlawska2009antiferromagnetic}. In (b), (c), (d), and (e), the current flows along the $a$ ($[100]$) and $c$ ($[001]$) axes.}
    \label{fig:Ce2_Rh_Si3_experimental_figures_from_literature_electrical_resistivity}
\end{figure}

\begin{figure}
    \centering
    \includegraphics[width=1\linewidth]{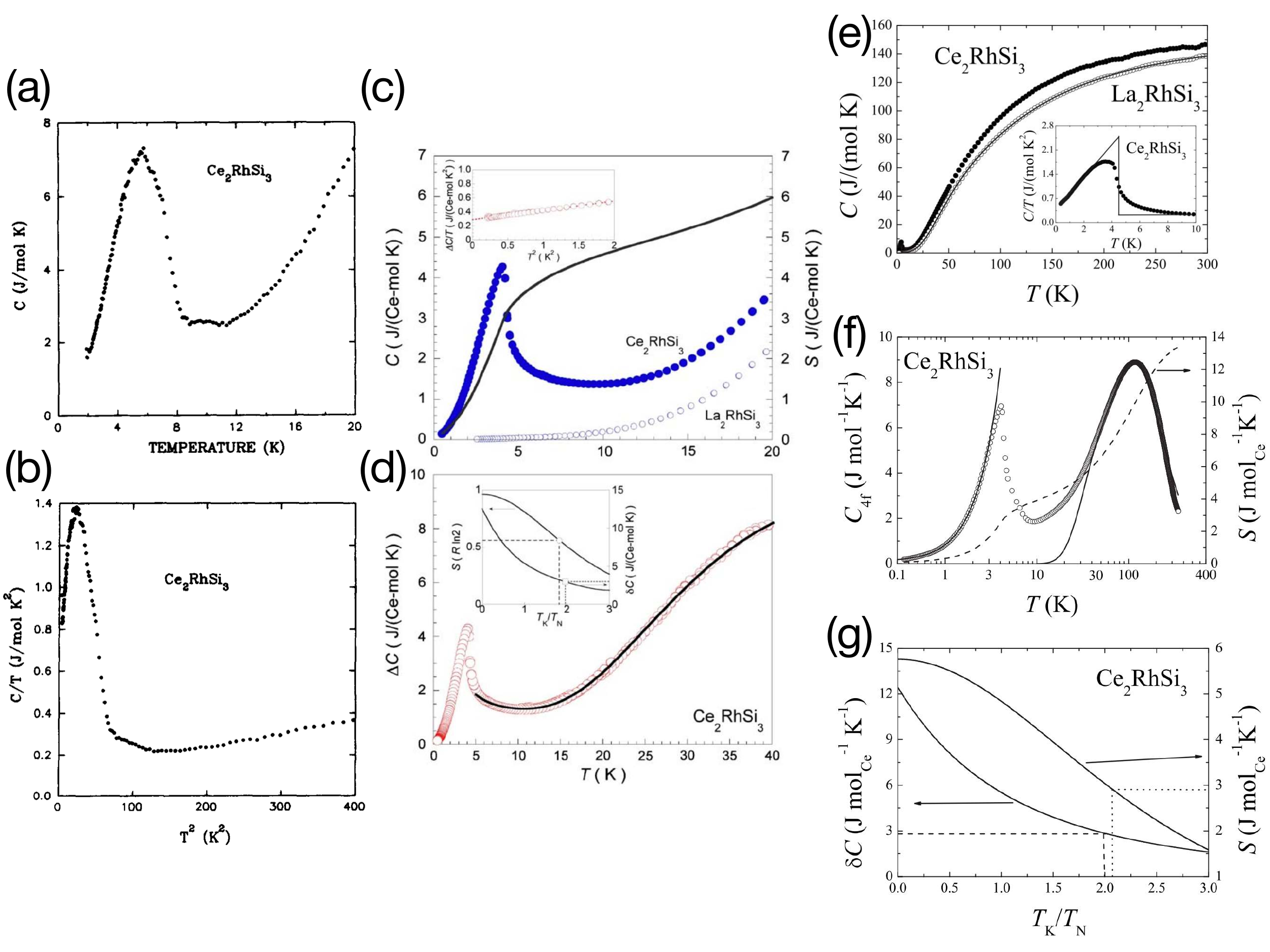}
    \caption{(a)~Specific heat as a function of temperature, and (b)~specific heat divided by temperature ($T$) as a function of $T^2$ for Ce$_2$RhSi$_3$ copied from Ref.~\citenum{das1994magnetic}. (c)~Specific heat and (d)~magnetic contribution to the specific heat $\Delta C$ as a function of temperature for Ce$_2$RhSi$_3$ copied from Ref.~\citenum{kase2009antiferromagnetic}. $\Delta C$ is obtained by subtracting the specific heat of La$_2$RhSi$_3$ from the specific heat of Ce$_2$RhSi$_3$. The inset of (c) is $\Delta C / T$ as a function of $T^2$. (e)~Specific heat and (f)~$C_{4f} \equiv $ specific heat of Ce$_2$RhSi$_3$ minus specific heat of La$_2$RhSi$_3$ as a function of temperature for Ce$_2$RhSi$_3$ copied from Ref.~\citenum{szlawska2009antiferromagnetic}. The inset of (d) and the panel (g) are the procedure used in Ref.~\citenum{kase2009antiferromagnetic} and Ref.~\citenum{szlawska2009antiferromagnetic}, respectively, to estimate the Kondo temperature, based on the magnetic entropy and the jump of specific heat at $T_{\rm N}$.}
    \label{fig:Ce2_Rh_Si3_experimental_figures_from_literature_specific_heat}
\end{figure}

\begin{figure}
    \centering
    \includegraphics[width=1\linewidth]{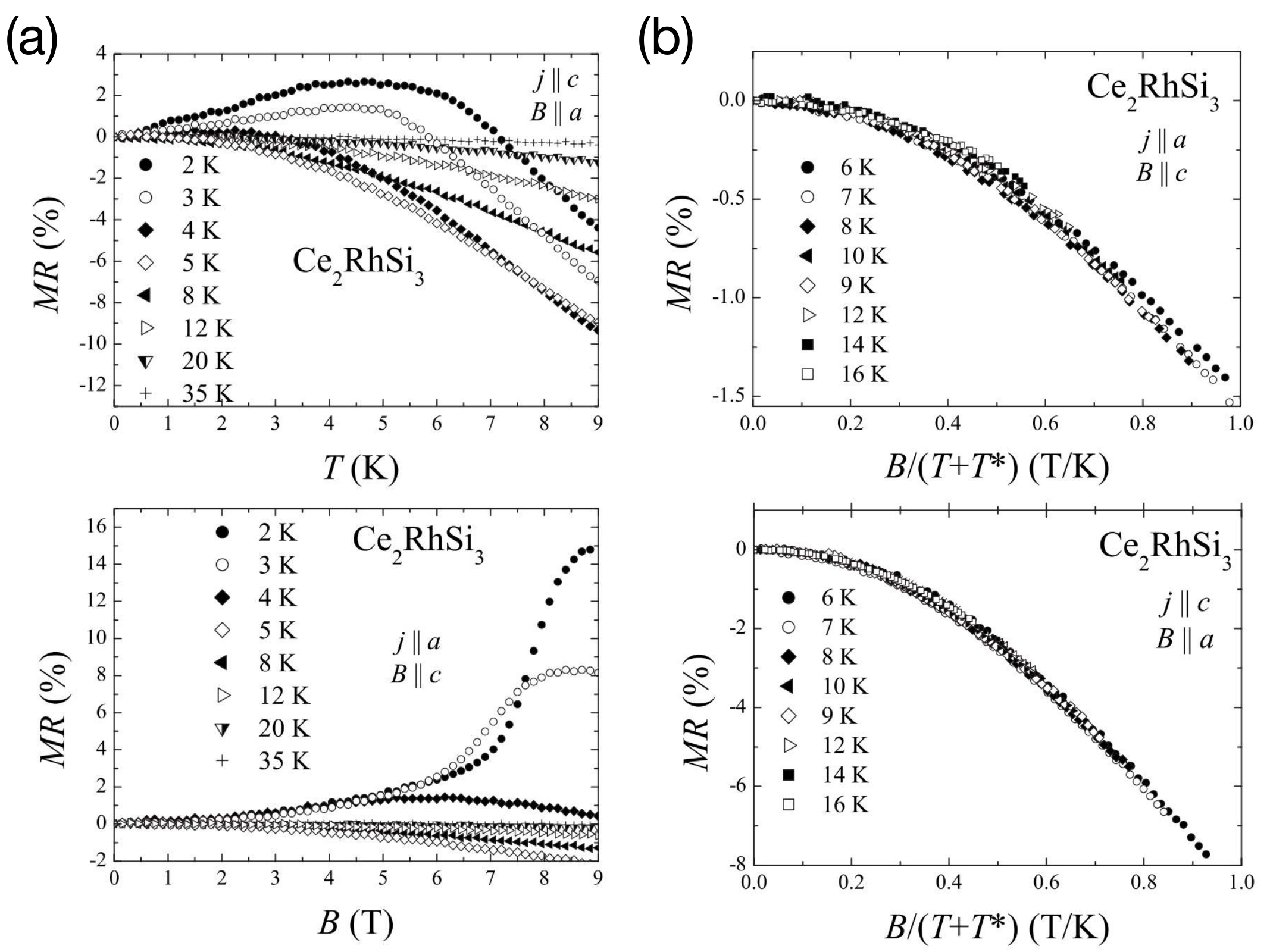}
    \caption{(a)~Transverse magnetoresistivity $\mathrm{MR = \frac{\rho(B)-\rho(0)}{\rho(0)}}$, where $\rho(B)$ is the electrical resistivity as a function of the applied magnetic field $B$, and (b)~Kondo scaling behavior of MR~\cite{schlottmann1983betheansatz,dzsaber2021giant} for Ce$_2$RhSi$_3$ copied from Ref.~\citenum{szlawska2009antiferromagnetic}. $j \parallel a$ and $j \parallel c$ mean the current flows along $a$ and $c$ axes, respectively. $B \parallel a$ and $B \parallel c$ mean the applied magnetic field is along $a$ and $c$ axes, respectively. $T^*$ denotes the characteristic temperature in the scaling behavior.}
    \label{fig:Ce2_Rh_Si3_experimental_figures_from_literature_magnetoresistivity}
\end{figure}

\section{Additional candidate materials}\label{sec:SI_additional_candidate}

In this section, we provide additional candidate materials where their properties are summarized in Supplementary Table~\ref{tab:additional_candidate_materials}.

\begin{table}
    \centering
    \begin{tabular}{c|cc|c|c|c|c|c}
        \multicolumn{8}{c}{Additional candidate material} \\
        \hline
        Material & SG \# & SG & Magnetic transition $T$ & $T_{\rm K}$ & Resistivity & $\gamma$ & Crossing \\
        \hline
        Ce$_6$Ni$_{1.67}$Si$_3$ & $176$ & $P 6_3 / m$ & $3.8(2)$K &  &  & $144$ mJ/K$^2$ Ce-mol & DNL \\
        \hline
        Ce$_5$Ni$_{1.85}$Si$_3$ & $176$ & $P 6_3 / m$ & $T_{\rm N} = 4.8(2)$K &  &  &  & DNL \\
        \hline
        Ce$_5$Ni$_2$Si$_3$ & $176$ & $P 6_3 / m$ & $T_{\rm N} = 7.3$K & $31$K & Semimetallic & $600$ mJ/K$^2$ Ce-mol & DNL \\
        \hline
        Ce$_{20}$Ni$_{42}$As$_{31}$ & $176$ & $P 6_3 / m$ & \xmark &  & Semimetallic &  & DNL \\
        \hline
        Ce$_{20}$Ni$_{42}$P$_{30}$ & $176$ & $P 6_3 / m$ & $32$K & $67$K & Semimetallic &  & DNL 
    \end{tabular}
    \caption{Summary of the physical properties of additional candidate materials Ce$_6$Ni$_{1.67}$Si$_3$~\cite{gaudin2007on}, Ce$_5$Ni$_{1.85}$Si$_3$~\cite{gaudin2007on}, Ce$_5$Ni$_2$Si$_3$~\cite{lee2004magnetic,ryu2005magnetic}, Ce$_{20}$Ni$_{42}$As$_{31}$~\cite{babizhetskyy2002ternary}, and Ce$_{20}$Ni$_{42}$P$_{30}$~\cite{babizhetskyy2001structural,babizhetskyy2004intermediate}. ``SG'' means ``space group''. A ``\xmark'' mark in magnetic transition $T$ means no magnetic transition is reported. ``DNL'' means ``Dirac nodal line''.
    We refer readers to \ref{sec:SI_additional_candidate} for detailed descriptions of the additional candidate materials, such as how different physical properties are determined.}
    \label{tab:additional_candidate_materials}
\end{table}

\setcounter{subsection}{0}
\subsection{Ce$_6$Ni$_{2-\delta}$Si$_3$ ($\delta = 0.33$)}

The chemical formula of Ce$_6$Ni$_{2-\delta}$Si$_3$ we consider in this work is Ce$_6$Ni$_{1.67}$Si$_3$~\cite{gaudin2007on}, which has the achiral space group no.~176 ($P 6_3 / m$) and it could be a potential candidate for the paramagnetic Dirac-Kondo semimetal with fourfold degenerate nodal lines on the BZ boundary~\cite{zhang2018topological}, provided that the temperature is above the magnetic transition temperature. The measurements of the magnetic susceptibility yield the effective magnetic moment $\mu_\mathrm{eff} = 2.45\mu_\mathrm{B}/$Ce and the paramagnetic Curie temperature $\theta_p = -36$K. The negative value of $\theta_p$ indicates anti-ferromagnetic couplings. 
The magnetic measurements further suggest a magnetic ordering below $T = 3.8(2)$K, while further experiments are required to determine the nature of this magnetic ordering.
This means that above $T = 3.8(2)$K Ce$_6$Ni$_{1.67}$Si$_3$ could be in the paramagnetic state.
The Sommerfeld coefficient $\gamma$ is determined as $\gamma = 144$ mJ/K$^2$ Ce-mol using the specific-heat data in the temperature range between $T = 10$K and $22$K, suggesting that Ce$_6$Ni$_{1.67}$Si$_3$ is a moderate heavy-fermion compound.
In addition, the specific-heat data exhibit a peak around $T = 4.1(2)$K, which is consistent with the magnetic ordering temperature $T = 3.8(2)$K from the magnetic measurements.
Using the magnetic entropy extracted from the specific-heat data, Ref.~\citenum{gaudin2007on} also suggests that Ce$_6$Ni$_{1.67}$Si$_3$ may have a moderate Kondo effect.
In addition, as reported in Ref.~\citenum{gaudin2007on}, the colour of Ce$_6$Ni$_{1.67}$Si$_3$ is metallic light grey, which suggests that Ce$_6$Ni$_{1.67}$Si$_3$ might be non-insulating.
We note that in Ce$_6$Ni$_{1.67}$Si$_3$ some Ni elements have partial occupancy, as reported in Ref.~\citenum{gaudin2007on}. Since the element with partial occupancy is not Ce, we may expect that this will not affect the lattice-coherent behavior formed from the Ce elements.

\subsection{Ce$_5$Ni$_{2-\delta}$Si$_3$ ($\delta = 0.15$)}

The chemical formula of Ce$_5$Ni$_{2-\delta}$Si$_3$ we consider in this work is Ce$_5$Ni$_{1.85}$Si$_3$~\cite{gaudin2007on}, which has the achiral space group no.~176 ($P 6_3 / m$) and it could be a potential candidate for the paramagnetic Dirac-Kondo semimetal with fourfold degenerate nodal lines on the BZ boundary~\cite{zhang2018topological}, provided that the temperature is above the magnetic transition temperature. The measurements of the magnetic susceptibility yield the effective magnetic moment $\mu_\mathrm{eff} = 2.51 \mu_\mathrm{B} / $Ce and the paramagnetic Curie temperature $\theta_p = -43$K. The negative value of $\theta_p$ indicates anti-ferromagnetic couplings. 
The magnetic measurements further suggest an anti-ferromagnetic ordering at $T_{\rm N} = 4.8(2)$K as a kink in the temperature dependence of the magnetization divided by the applied field.
This means that above $T_{\rm N} = 4.8(2)$K Ce$_5$Ni$_{1.85}$Si$_3$ could be in the paramagnetic state.
In addition, as reported in Ref.~\citenum{gaudin2007on}, the colour of Ce$_5$Ni$_{1.85}$Si$_3$ is metallic light grey, which suggests that Ce$_5$Ni$_{1.85}$Si$_3$ might be non-insulating.
We note that in Ce$_5$Ni$_{1.85}$Si$_3$ some Ni elements have partial occupancy, as reported in Ref.~\citenum{gaudin2007on}. Since the element with partial occupancy is not Ce, we may expect that this will not affect the lattice-coherent behavior formed from the Ce elements.

\subsection{Ce$_5$Ni$_2$Si$_3$}

Ce$_5$Ni$_2$Si$_3$~\cite{lee2004magnetic,ryu2005magnetic} has the achiral space group no.~176 ($P 6_3 / m$) and it could be a potential candidate for the paramagnetic Dirac-Kondo semimetal with fourfold degenerate nodal lines on the BZ boundary~\cite{zhang2018topological}, provided that the temperature is above the magnetic transition temperature. The measurements of the magnetic susceptibility yield the effective magnetic moment $\mu_\mathrm{eff} = 2.42 \mu_\mathrm{B} / $Ce and the paramagnetic Curie temperature $\theta_p = -61.3$K. The relatively large negative value of $\theta_p$ indicates relatively strong anti-ferromagnetic couplings, suggesting that Ce$_5$Ni$_2$Si$_3$ might be a Kondo compound. The magnetic measurements further suggest an anti-ferromagnetic ordering at $T_{\rm N} = 7.3$K. This means that above $T_{\rm N} = 7.3$K Ce$_5$Ni$_2$Si$_3$ could be in the paramagnetic state. The Sommerfeld coefficient, also known as the electronic specific heat coefficient, is $\gamma = 600$ mJ/K$^2$ Ce-mol~\cite{ryu2005magnetic}, which is a relatively enhanced value suggesting heavy-fermion behavior in a Kondo system. We note here that this value of $\gamma$ is obtained via the data of the magnetic contribution to the specific heat $C_\mathrm{mag}(T)$, which is obtained by subtracting the specific-heat data of La$_5$Ni$_2$Si$_3$ from the specific data of Ce$_5$Ni$_2$Si$_3$~\cite{ryu2005magnetic}, and the expression $\gamma = \frac{C_\mathrm{mag}(T)}{T}$ with $T \to 0$ is used. The electrical resistivity $\rho(T)$ takes value between $245$ $\mu\Omega$cm and $260$ $\mu\Omega$cm in the experimental temperature range, suggesting semimetallic-type resistivity. In addition, the magnetic contribution to the resistivity $\rho_\mathrm{mag}(T)$, which is obtained by subtracting the $\rho(T)$ data of La$_5$Ni$_2$Si$_3$ from the $\rho(T)$ data of Ce$_5$Ni$_2$Si$_3$, shows a $-\ln T$-behavior in the high temperature regime, which may be attributed to the Kondo effect. In Ref.~\citenum{lee2004magnetic} and Ref.~\citenum{ryu2005magnetic}, the Kondo temperature of Ce$_5$Ni$_2$Si$_3$ is estimated as $T_{\rm K} = 31$K.

\subsection{Ce$_{20}$Ni$_{42}$As$_{31}$}

Ce$_{20}$Ni$_{42}$As$_{31}$~\cite{babizhetskyy2002ternary} has the achiral space group no.~176 ($P 6_3 / m$) and it could be a potential candidate for the paramagnetic Dirac-Kondo semimetal with fourfold degenerate nodal lines on the BZ boundary~\cite{zhang2018topological}.
Ce$_{20}$Ni$_{42}$As$_{31}$ remains paramagnetic down to $1.8$K.
Magnetic susceptibility measurements at high temperatures yield $\mu_\mathrm{eff} = 2.55 \mu_B$ and $\theta_p = -28$K.
The electrical resistivity of Ce$_{20}$Ni$_{42}$As$_{31}$ does not vary significantly as a function of temperature, and the value of the electrical resistivity is about 300 $\mu \Omega$cm at room temperature.
Hence, we could classify Ce$_{20}$Ni$_{42}$As$_{31}$ as having a semimetallic-type resistivity.
In addition, the magnetic resistivity has $-\ln{T}$ behaviors in two different temperature regimes, the first one at $T > 100$ K and the second one at $T = 5-15$K.
The observed $-\ln{T}$ behaviors in the magnetic resistivity could suggest the presence of Kondo effect.
The magnetic resistivity further shows a maximum around $53$K.
Finally, we note that the empirical formula of this compound is Ce$_{20}$Ni$_{39}$As$_{30}$.

\subsection{Ce$_{20}$Ni$_{42}$P$_{30}$}

Ce$_{20}$Ni$_{42}$P$_{30}$~\cite{babizhetskyy2001structural,babizhetskyy2004intermediate} has the achiral space group no.~176 ($P 6_3 / m$), and within a certain temperature range it could be a potential candidate for the paramagnetic Dirac-Kondo semimetal with fourfold degenerate nodal lines on the BZ boundary~\cite{zhang2018topological}.
The measurements of the magnetic susceptibility yield the effective magnetic moment $\mu_\mathrm{eff} =1 \mu_\mathrm{B}$ and the paramagnetic Curie temperature $\theta_p = 29$K for $T < 200$K.
On the other hand, $\mu_\mathrm{eff} =2 \mu_\mathrm{B}$ and $\theta_p = -386$K for $T > 200$K.
In addition, a magnetic transition at $32$K is observed, while the determination of the nature of the magnetic order requires further studies. This suggests that above $T=32$K Ce$_{20}$Ni$_{42}$P$_{30}$ is in the paramagnetic state. 
The measurements of the electrical resistivity reveal the potential Kondo physics of Ce$_{20}$Ni$_{42}$P$_{30}$. 
The Kondo temperature of Ce$_{20}$Ni$_{42}$P$_{30}$ is estimated as $T_{\rm K}= 67$K~\cite{babizhetskyy2001structural}.
In addition, the experimental data of $\rho (T)$ does not change significantly as a function of temperature, indicating a semimetallic-type resistivity. The magnetic contribution $\rho_\mathrm{mag}(T)$ to the resistivity $\rho(T)$ is also extracted experimentally by subtracting the $\rho(T)$ data of Sm$_{20}$Ni$_{42}$P$_{30}$ from the $\rho(T)$ data of Ce$_{20}$Ni$_{42}$P$_{30}$. In particular, $\rho_\mathrm{mag}(T)$ exhibits a $T^2$-behavior when $T < 32$K and a $- \ln T$-behavior when $T > 100$K. The $- \ln T$-behavior of $\rho_\mathrm{mag}(T)$ is consistent with the Kondo theory. Although there is a magnetic transition at $T = 32$K, we expect that between the magnetic transition temperature $T = 32$K and the Kondo temperature $T_{\rm K} = 67$K, Ce$_{20}$Ni$_{42}$P$_{30}$ can be a potential candidate of the Kondo-driven paramagnetic state with fourfold degenerate nodal lines on the BZ boundary.

\section{Other candidate materials}\label{sec:SI_other_candidate}

In this section, we provide other candidate materials that have either relatively limited experimental information or experimental information that conflicts with the properties of topological Kondo semimetal phases.
We summarize the properties of these other candidate materials in Supplementary Table~\ref{tab:other_candidate_materials}.

\begin{table}
    \centering
    \begin{tabular}{c|cc|c|c|c|c}
        \multicolumn{7}{c}{Other candidate material} \\
        \hline
        Material & SG \# & SG & $T_{\rm C}$ & $T_{\rm N}$ & Resistivity & $\gamma$ \\
        \hline
        \hline
        \multicolumn{7}{c}{Ce-based material} \\
        \hline
        \hline
        Ce$_3$CuSnSe$_7$ & $173$ & $P 6_3$ &  & $5$K & Non-metallic [\ref{sec:Ce3_Cu_Sn_Se7_our_experiment}] & \\
        \hline
        Ce$_3$Ge$_{1.47}$Se$_7$ & $173$ & $P 6_3$ & \multicolumn{2}{c|}{\xmark} & Possibly insulating & \\
        \hline
        Ce$_3$Mg$_{0.5}$SiS$_7$ & $173$ & $P 6_3$ & \multicolumn{2}{c|}{\xmark} & Possibly insulating & \\
        \hline
        Ce$_{18}$W$_{10}$O$_{57}$ & $190$ & $P \bar{6} 2c$ & \multicolumn{2}{c|}{\xmark} & Possibly insulating & \\
        \hline
        \hline
        \multicolumn{7}{c}{U-based material} \\
        \hline
        \hline
        Cu$_2$U$_3$S$_7$ & $173$ & $P 6_3$ & \multicolumn{2}{c|}{\xmark} &  & \\
        \hline
        Cu$_2$U$_3$Se$_7$ & $173$ & $P 6_3$ &  & $13$K &  & \\
        \hline
        U$_6$Ni$_{20}$P$_{13}$ & $176$ & $P 6_3 / m$ &  & $41 \pm 4$K &  & \\
        \hline
        U$_7$Te$_{12}$ & $176$ & $P 6_3 / m$ & $48 - 73$K &  & Semimetallic & $48$ mJ/K$^2$ mol \\
        \hline
        \hline
        \multicolumn{7}{c}{Yb-based material} \\
        \hline
        \hline
        Ba$_3$YbB$_9$O$_{18}$ & $176$ & $P 6_3 / m$ & \multicolumn{2}{c|}{\xmark} & Possibly insulating & \\
        \hline
        YbBO$_3$ & $176$ & $P 6_3 / m$ & \multicolumn{2}{c|}{\xmark} & Possibly insulating & 
    \end{tabular}
    \caption{Summary of the physical properties of other candidate materials Ce$_3$CuSnSe$_7$~\cite{gulay2005crystalR3CuSnSe7,gulay2005crystalCe3CuSnSe7}, Ce$_3$Ge$_{1.47}$Se$_7$~\cite{daszkiewicz2010crystal}, Ce$_3$Mg$_{0.5}$SiS$_7$~\cite{king2023crystal}, Ce$_{18}$W$_{10}$O$_{57}$~\cite{abeysinghe2017crystal}, Cu$_2$U$_3$S$_7$~\cite{daoudi1996new}, Cu$_2$U$_3$Se$_7$~\cite{daoudi1996new}, U$_6$Ni$_{20}$P$_{13}$~\cite{troc1992crystal,ebel1998preparation}, U$_7$Te$_{12}$~\cite{breeze1971thecrystal,breeze1971aninvestigation,suski1972magnetic,tougait1998characterization,opletal2023ferromagnetic}. ``SG'' means ``space group''. $T_C$ and $T_{\rm N}$ denote the ferromagnetic and antiferromagnetic transition temperatures, respectively. A ``\xmark'' mark in the magnetic transition temperatures mean no magnetic order is reported. We note that for Ce$_3$CuSnSe$_7$ our experimental analysis in \ref{sec:Ce3_Cu_Sn_Se7_our_experiment} verifies the presence of magnetic transition and characterizes Ce$_3$CuSnSe$_7$ as a non-metallic compound. We note that for both U$_6$Ni$_{20}$P$_{13}$ and U$_7$Te$_{12}$, in addition to space group no.~176 ($P 6_3 / m$)~\cite{troc1992crystal,breeze1971thecrystal,breeze1971aninvestigation,suski1972magnetic}, there are also studies reporting group no.~174 ($P \bar{6}$) for U$_6$Ni$_{20}$P$_{13}$~\cite{ebel1998preparation} and U$_7$Te$_{12}$~\cite{tougait1998characterization,opletal2023ferromagnetic}. We have also included in this table the physical properties of U$_6$Ni$_{20}$P$_{13}$ and U$_7$Te$_{12}$ present in these references~\cite{ebel1998preparation,tougait1998characterization,opletal2023ferromagnetic} reporting space group no.~174 ($P \bar{6}$). We refer readers to \ref{sec:SI_other_candidate} for more information on the other candidate materials.}
    \label{tab:other_candidate_materials}
\end{table}

\setcounter{subsection}{0}
\subsection{Ce$_3$CuSnSe$_7$}

Ce$_3$CuSnSe$_7$~\cite{gulay2005crystalR3CuSnSe7,gulay2005crystalCe3CuSnSe7} has the chiral space group no.~173 ($P 6_3$).
From the magnetic susceptibility measurements, the effective magnetic moment $\mu_\mathrm{eff} = 2.46(3) \mu_\mathrm{B}$ and the paramagnetic Curie temperature $\theta_p = -18(1)$K for Ce$_3$CuSnSe$_7$.
The negative value of $\theta_p$ suggests that Ce$_3$CuSnSe$_7$ has anti-ferromagnetic coupling.
The magnetic susceptibility measurements further suggest an anti-ferromagnetic transition at $T_{\rm N} = 5$K.
Due to the lack of experimental information on the specific heat and the Sommerfeld coefficient, we are not able to verify whether Ce$_3$CuSnSe$_7$ is a heavy-fermion compound.
Also, the lack of experimental electrical resistivity in the literature makes it difficult to establish whether Ce$_3$CuSnSe$_7$ is a semimetal.
Finally, the lack of experimental information about the above physical properties also makes it challenging to determine whether Ce$_3$CuSnSe$_7$ exhibits Kondo physics.
Thus motivated, we perform experimental analyses on Ce$_3$CuSnSe$_7$ and report our results in \ref{sec:Ce3_Cu_Sn_Se7_our_experiment}, from which we establish that Ce$_3$CuSnSe$_7$ presents a non-metallic behavior.

\subsection{Ce$_3$Ge$_{1.47}$Se$_7$}

Ce$_3$Ge$_{1.47}$Se$_7$~\cite{daszkiewicz2010crystal} has the chiral space group no.~173 ($P 6_3$).
The measurements of magnetic susceptibility yield the effective magnetic moment $\mu_\mathrm{eff} = 2.40(8) \mu_\mathrm{B}/$Ce and the paramagnetic Curie temperature $\theta_p = -12.9(7)$K.
Although there is no magnetic transition in Ce$_3$Ge$_{1.47}$Se$_7$ down to $1.72$K, it is highly possible that Ce$_3$Ge$_{1.47}$Se$_7$ is an insulator.

\subsection{Ce$_3$Mg$_{0.5}$SiS$_7$}

Ce$_3$Mg$_{0.5}$SiS$_7$~\cite{king2023crystal} has the chiral space group no.~173 ($P 6_3$). 
The measurements of magnetic susceptibility yield the effective magnetic moment $\mu_\mathrm{eff} = 2.33 \mu_\mathrm{B}/$Ce and the paramagnetic Curie temperature $\theta_p = -43.7$K.
Although no antiferromagnetic transition is observed in Ce$_3$Mg$_{0.5}$SiS$_7$ down to $2$K, it is highly possible that Ce$_3$Mg$_{0.5}$SiS$_7$ is an insulator.

\subsection{Ce$_{18}$W$_{10}$O$_{57}$}

Ce$_{18}$W$_{10}$O$_{57}$~\cite{abeysinghe2017crystal} has the achiral space group no.~190 ($P \bar{6} 2c$).
The measurements of magnetic susceptibility yield the effective magnetic moment $\mu_\mathrm{eff} = 10.73 \mu_\mathrm{B}/$formula unit and the paramagnetic Curie temperature $\theta_p = -71.14$K.
Although Ce$_{18}$W$_{10}$O$_{57}$ does not order magnetically, it is highly possible that Ce$_{18}$W$_{10}$O$_{57}$ is an insulator.

\subsection{Cu$_2$U$_3$S$_7$}

Cu$_2$U$_3$S$_7$~\cite{daoudi1996new} has the chiral space group no.~173 ($P 6_3$).
The measurements of magnetic susceptibility yield the effective magnetic moment $\mu_\mathrm{eff} = 2.50 \mu_\mathrm{B}/$U and the paramagnetic Curie temperature $\theta_p = -33$K.
In addition, Cu$_2$U$_3$S$_7$ is paramagnetic down to $5$K.

\subsection{Cu$_2$U$_3$Se$_7$}

Cu$_2$U$_3$Se$_7$~\cite{daoudi1996new} has the chiral space group no.~173 ($P 6_3$).
The measurements of magnetic susceptibility yield the effective magnetic moment $\mu_\mathrm{eff} = 2.64 \mu_\mathrm{B}/$U and the paramagnetic Curie temperature $\theta_p = -28$K.
In addition, Cu$_2$U$_3$Se$_7$ has an antiferromagnetic transition at $T_{\rm N} = 13$K.

\subsection{U$_6$Ni$_{20}$P$_{13}$}

U$_6$Ni$_{20}$P$_{13}$ could crystallize in the achiral space group no.~176 ($P 6_3 / m$)~\cite{troc1992crystal} or the achiral space group no.~174 ($P \bar{6}$)~\cite{ebel1998preparation}. In Ref.~\citenum{ebel1998preparation} which reports space group $P \bar{6}$ for U$_6$Ni$_{20}$P$_{13}$, an antiferromagnetic transition at $T_{\rm N} = 41 \pm 4$K is observed, with $\mu_\mathrm{eff} = 2.1 \pm 0.1 \mu_\mathrm{B}/$U and $\theta_p = -37 \pm 4$K from the magnetic susceptibility measurements.

\subsection{U$_7$Te$_{12}$}

U$_7$Te$_{12}$ could crystallize in the achiral space group no.~176 ($P 6_3 / m$) or the achiral space group no.~174 ($P \bar{6}$). Specifically, the ICSD Collection Codes 16493, 601162, and 653139 report U$_7$Te$_{12}$ as having space group $P 6_3 / m$, in which the corresponding journal references are Ref.~\citenum{breeze1971thecrystal}, Ref.~\citenum{breeze1971aninvestigation}, and Ref.~\citenum{suski1972magnetic}, respectively. On the other hand, Ref.~\citenum{tougait1998characterization} and Ref.~\citenum{opletal2023ferromagnetic} report U$_7$Te$_{12}$ as having space group $P \bar{6}$. In Ref.~\citenum{suski1972magnetic}, a ferromagnetic transition at $T = 73$K is reported for U$_7$Te$_{12}$. Also, in Ref.~\citenum{tougait1998characterization}, U$_7$Te$_{12}$ is reported to have a ferromagnetic transition at $T = 54$K and a semimetallic behavior. Such a semimetallic behavior in Ref.~\citenum{tougait1998characterization} could be deduced from $\rho (T = 77\mathrm{K}) = 3.3 \times 10^{-3}$ $\Omega$cm and $\rho (T = 300\mathrm{K}) = 3.5 \times 10^{-3}$ $\Omega$cm, leading to $\frac{\rho (T = 300\mathrm{K})}{\rho (T = 77\mathrm{K})} \approx 1.06$. Finally, Ref.~\citenum{opletal2023ferromagnetic} reports a ferromagnetic transition at $T = 48$K, a semimetallic behavior, and the Sommerfeld coefficient $\gamma = 48$ mJ/K$^2$ mol for U$_7$Te$_{12}$. The Sommerfeld coefficient in Ref.~\citenum{opletal2023ferromagnetic} is determined using the specific-heat data below $T = 10$K.

\subsection{Ba$_3$YbB$_9$O$_{18}$}

Ba$_3$YbB$_9$O$_{18}$~\cite{khatua2022magnetic} has the achiral space group no.~176 ($P 6_3 / m$).
The measurements of magnetic susceptibility yield the effective magnetic moment $\mu_\mathrm{eff} = 4.73 \mu_\mathrm{B}$ and the paramagnetic Curie temperature $\theta_p = -90$K in the temperature range $100-340$K.
On the other hand, in the temperature range $5-10$K, $\mu_\mathrm{eff} = 2.32 \mu_\mathrm{B}$ and $\theta_p = -0.12 \pm 0.02$K.
Although there is no magnetic order in Ba$_3$YbB$_9$O$_{18}$ down to $1.9$K, it is highly possible that Ba$_3$YbB$_9$O$_{18}$ is an insulator.

\subsection{YbBO$_3$}

YbBO$_3$~\cite{somesh2023absence} has the achiral space group no.~176 ($P 6_3 / m$).
The measurements of magnetic susceptibility yield the effective magnetic moment $\mu_\mathrm{eff} = 4.53 \mu_\mathrm{B}$ and the paramagnetic Curie temperature $\theta_p = -62.6$K in the temperature range above $150$K.
On the other hand, $\mu_\mathrm{eff} = 3.2 \mu_\mathrm{B}$ and $\theta_p = -0.8$K in the temperature range below $50$K.
Although there is no signature of magnetic order observed in YbBO$_3$ down to $0.4$K, it is highly possible that YbBO$_3$ is an insulator.

\section{New experiments on candidate materials}

Motivated by the experimental information on the candidate materials discussed in \ref{sec:SI_prime_candidate}, \ref{sec:SI_additional_candidate}, and \ref{sec:SI_other_candidate}, we carried out crystal growth for selected compounds. We successfully synthesized pure phase $\rm Ce_2NiGe_3$ and $\rm Ce_3CuSnSe_7$, while attempts on $\rm Ce_6Co_{2-\delta}Si_3$ ($\delta = 0.33$) and $\rm Ce_6Rh_{32}P_{17}$ resulted only in samples with mixture of phases. Physical measurements were performed on the successfully synthesized phase-pure crystals.
Specifically, we performed magnetization, specific heat, and electrical resistivity measurements between room temperatures and 1.8\,K and in magnetic fields up to 9\,T, using a  Physical Property Measurement System (PPMS) by Quantum Design.

\subsection{Ce$_2$NiGe$_3$}

Motivated by the experimental information of Ce$_2$NiGe$_3$ from the literature in \ref{sec:Ce2_Ni_Ge3_experimental_information_from_literature}, we synthesized polycrystalline samples of Ce$_2$NiGe$_3$ in order to study its physical properties.
Earlier reports on the crystal structure of Ce$_2$NiGe$_3$ suggested that this compound crystallizes in the AlB$_2$ type structure (space group no. 191, $P6/mmm$)~\cite{huo2001electric}. However, neutron diffraction studies on polycrystalline samples showed a better agreement with the Er$_2$RhSi$_3$ type structure (space group no. 190, $P \bar{6} 2c$)~\cite{kalsi2014neutron}, which is in agreement with the Rietveld refinement analysis of the powder X-ray diffraction (XRD) patterns of our samples, as seen in Supplementary Fig.~\ref{fig:Ce2_Ni_Ge3_powder_XRD_our_experiment}.

Only few studies have explored the physical properties of Ce$_2$NiGe$_3$~\cite{huo2001electric,kalsi2014neutron} and classified it as a Kondo lattice compound with a moderately enhanced Sommerfeld coefficient $\gamma=25$\,mJ/K$^2$\,Ce-mol. These experimental results are consistent with the presence of a spin glass state in Ce$_2$NiGe$_3$, evidenced by a long exponential decay of isothermal magnetization curves, the irreversibility of the zero-field and field cooling magnetic susceptibility curves below the anomaly at 3.5\,K, and the presence of a broad peak in the specific heat data around 5\,K~\cite{huo2001electric}. Additionally, neutron diffraction data show no evidence for long-range order down to 2.8\,K~\cite{kalsi2014neutron}.
Our results are in overall good agreement with these characteristics [Supplementary Fig.~\ref{fig:Ce2_Ni_Ge3_magnetic_susceptibility_our_experiment} and Supplementary Fig.~\ref{fig:Ce2_Ni_Ge3_physical_properties_characterizations_our_experiment}(a-c)].

In Supplementary Fig.~\ref{fig:Ce2_Ni_Ge3_magnetic_susceptibility_our_experiment}, we show the measured magnetic susceptibility of Ce$_2$NiGe$_3$, which confirms that there is a magnetic transition around $3.5$K, consistent with Ref.~\citenum{kalsi2014neutron}.
The magnetic susceptibility data are consistent with the Curie-Weiss behavior with a negative paramagnetic Curie temperature of $\theta_p = -7.8$\,K, suggesting antiferromagnetic correlations and an effective moment $\mu_{\rm eff} = 2.6 \mu_{\rm B}$ close to the full Ce$^{3+}$ moment.
We further perform more characterizations of Ce$_2$NiGe$_3$ and present our results in Supplementary Fig.~\ref{fig:Ce2_Ni_Ge3_physical_properties_characterizations_our_experiment}.
Our measurements on the electrical resistivity [Supplementary Fig.~\ref{fig:Ce2_Ni_Ge3_physical_properties_characterizations_our_experiment}(a)] is consistent with Ref.~\citenum{huo2001electric}, verifying that Ce$_2$NiGe$_3$ has a semimetallic-type resistivity.
Our results on the magnetization [Supplementary Fig.~\ref{fig:Ce2_Ni_Ge3_physical_properties_characterizations_our_experiment}(b)] and specific heat [Supplementary Fig.~\ref{fig:Ce2_Ni_Ge3_physical_properties_characterizations_our_experiment}(c)] of Ce$_2$NiGe$_3$ are also consistent with the literature~\cite{huo2001electric,kalsi2014neutron}.

We now describe new experimental results for Ce$_2$NiGe$_3$.
We begin with the magnetoresistivity, which is the longitudinal resistivity in the presence of an applied magnetic field. As shown in Supplementary Fig.~\ref{fig:Ce2_Ni_Ge3_physical_properties_characterizations_our_experiment}(d), the magnetoresistivity of Ce$_2$NiGe$_3$ exhibits the qualitative behavior of a Kondo system~\cite{schlottmann1983betheansatz,dzsaber2021giant}.

The Hall resistivity of Ce$_2$NiGe$_3$ is linear-in-field at low fields, but tends to saturate at larger fields [Supplementary Fig.~\ref{fig:Ce2_Ni_Ge3_physical_properties_characterizations_our_experiment}(e)]. Similar behavior is observed for the magnetization [Supplementary Fig.~\ref{fig:Ce2_Ni_Ge3_physical_properties_characterizations_our_experiment}(b)]. 
From Hall resistivity, we estimate a carrier density of $n=1/(eR_0)=2.16 \times 10^{20}\,{\rm cm}^{-3}$ for Ce$_2$NiGe$_3$ at 2\,K using a single-band model. The corresponding carrier densities at different temperatures are shown in Supplementary Fig.~\ref{fig:Ce2_Ni_Ge3_physical_properties_characterizations_our_experiment}(f).
The low carrier density extracted from Hall measurement also confirms its semimetallicity.

\begin{figure}
    \centering
    \includegraphics[width=0.75\linewidth]{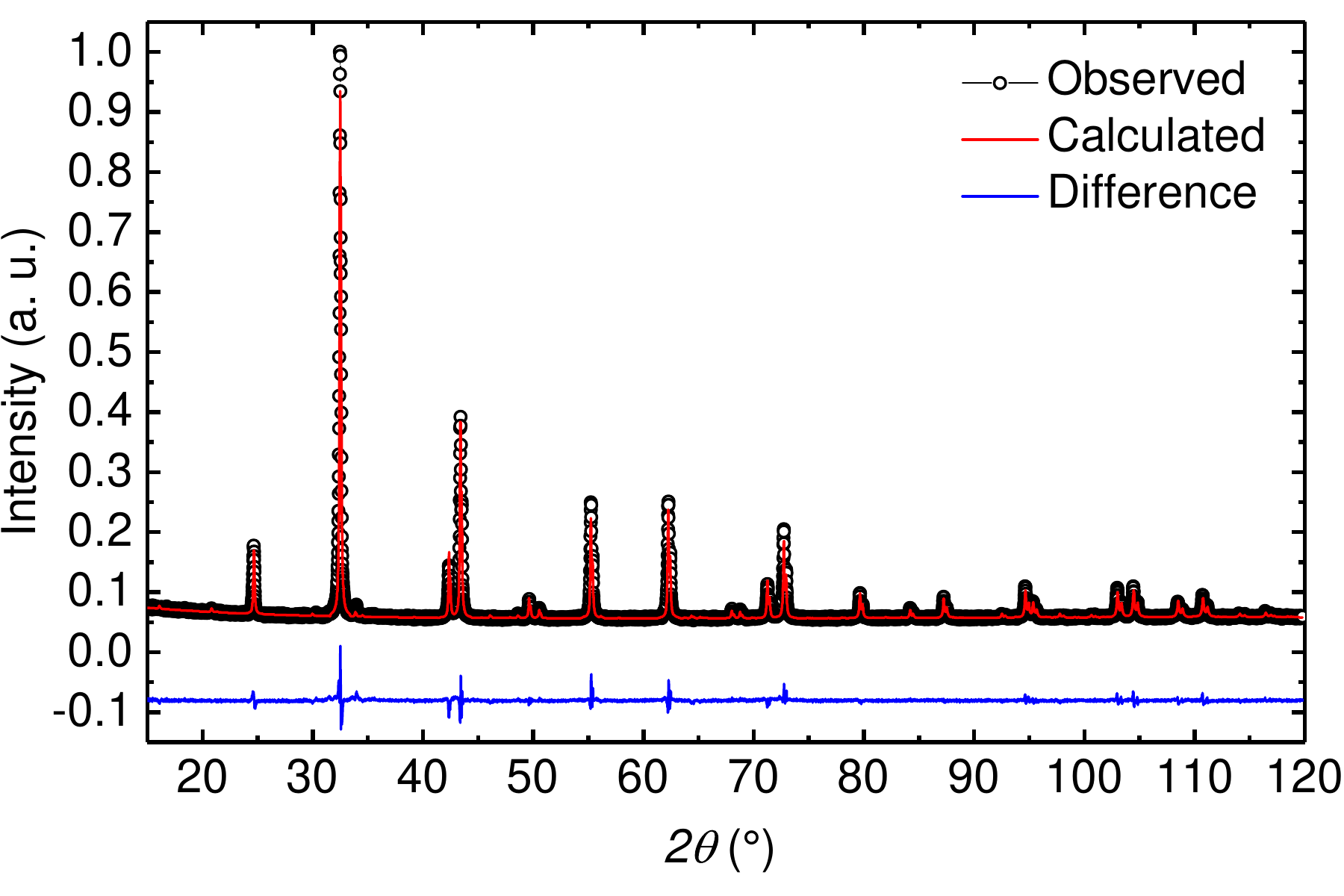}
    \caption{Powder XRD pattern for a Ce$_2$NiGe$_3$ sample. The open circles are the experimental data, the red curve is obtained from the Rietveld refinement for the Er$_2$RhSi$_3$ type structure (space group no.~190, $P \bar{6} 2c$), and the blue curve is the difference between the experimental data and the Rietveld refinement.}
    \label{fig:Ce2_Ni_Ge3_powder_XRD_our_experiment}
\end{figure}

\begin{figure}
    \centering
    \includegraphics[width=\linewidth]{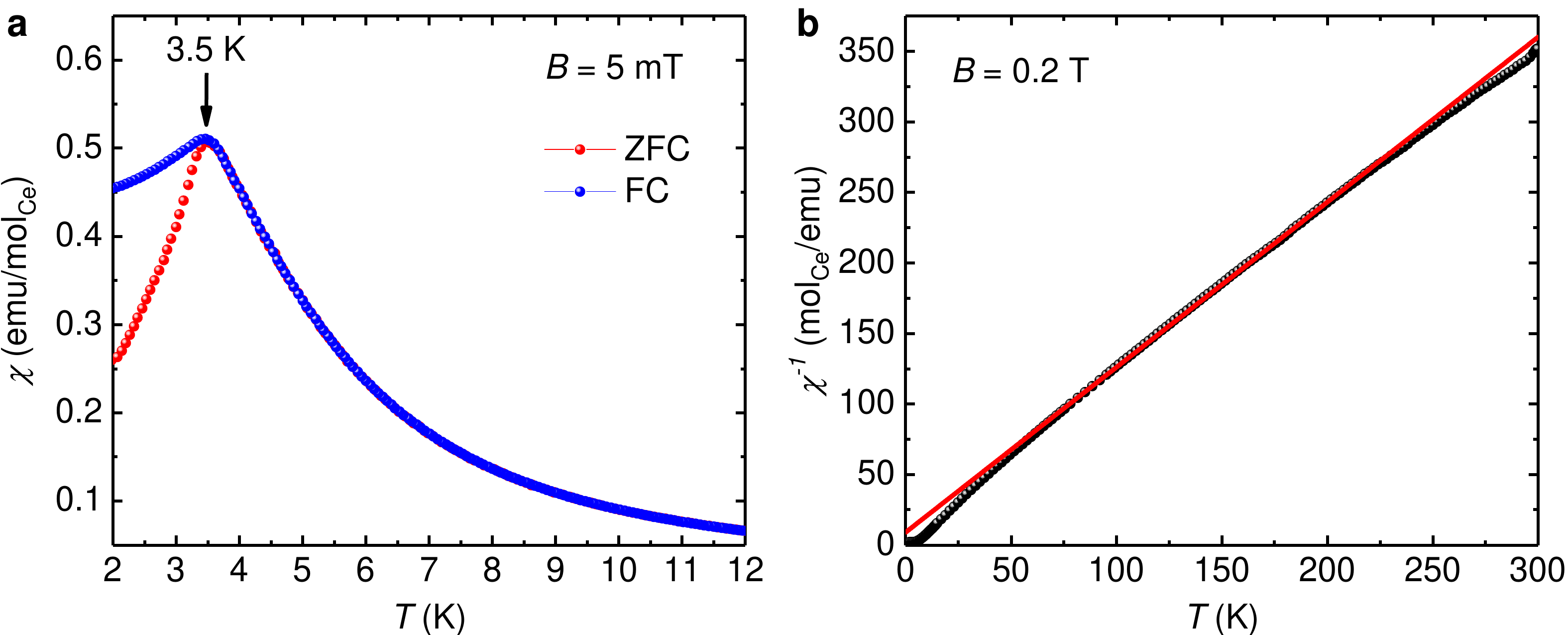}
    \caption{(a) Zero-field cooling (ZFC) and field cooling (FC) magnetic susceptibility curves of Ce$_2$NiGe$_3$ measured in a field of 5\,mT. (b) Temperature dependence of $\chi^{-1}$ measured at 0.2\,T and Curie-Weiss fits (red line).}
    \label{fig:Ce2_Ni_Ge3_magnetic_susceptibility_our_experiment}
\end{figure}

\begin{figure}
    \centering
    \includegraphics[width=1\linewidth]{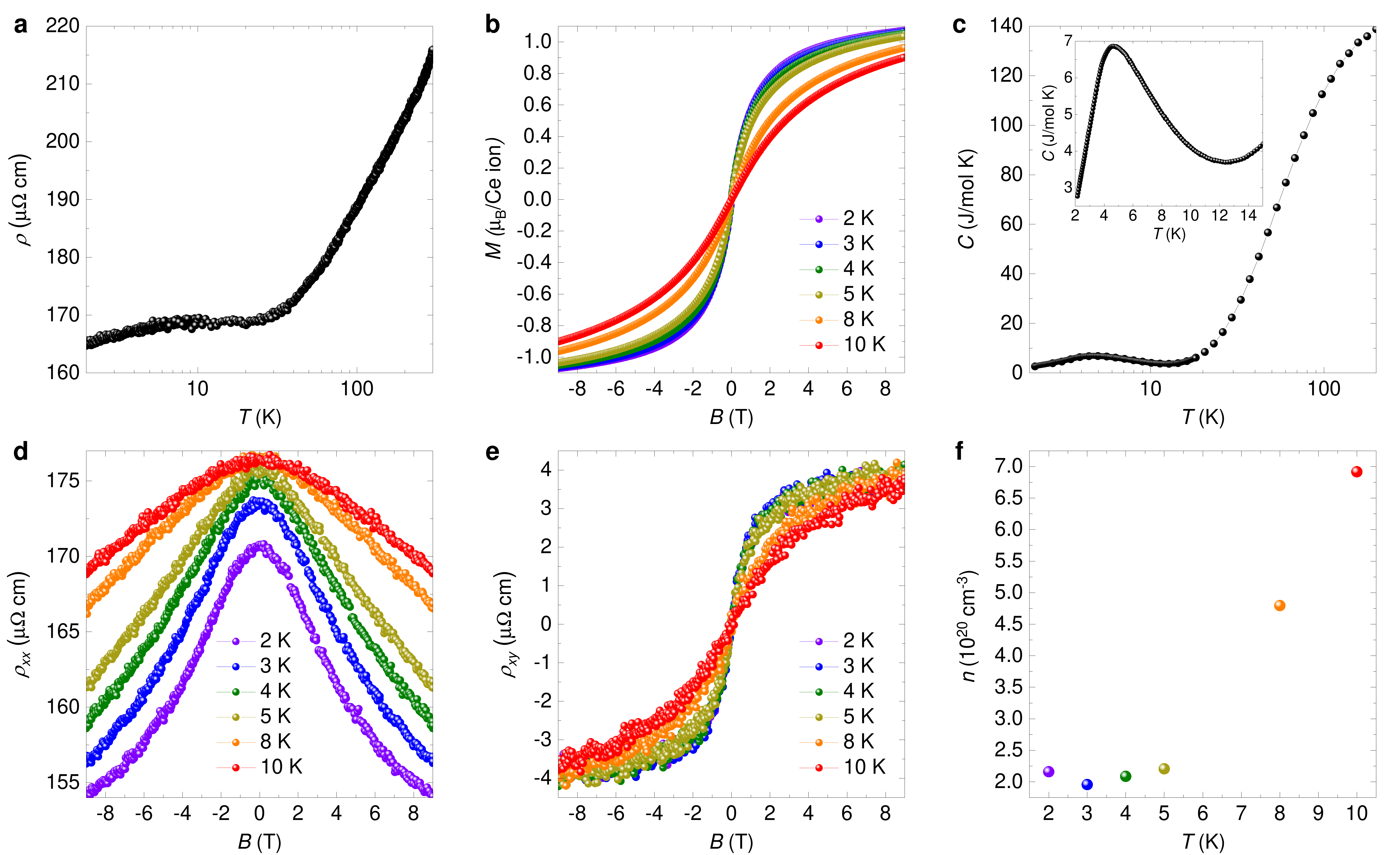}
    \caption{Physical properties of Ce$_2$NiGe$_3$. (a) Temperature dependence of electrical resistivity. (b), (d), (e): Applied magnetic field dependence of magnetization, longitudinal resistivity, and Hall resistivity at different temperatures. (c)~Temperature dependence of specific heat. (f)~Temperature dependence of carrier density estimated from the linear portion of the curves in panel~(e).}
    \label{fig:Ce2_Ni_Ge3_physical_properties_characterizations_our_experiment}
\end{figure}

\subsection{Ce$_3$CuSnSe$_7$}
\label{sec:Ce3_Cu_Sn_Se7_our_experiment}

We synthesized polycrystalline samples of Ce$_3$CuSnSe$_7$ in order to study its physical properties.
In Supplementary Fig.~\ref{fig:Ce3_Cu_Sn_Se7_powder_XRD_our_experiment}, we show our powder XRD analysis on Ce$_3$CuSnSe$_7$, which confirms that the crystal structure of our Ce$_3$CuSnSe$_7$ sample belongs to the chiral space group no.~173 ($P 6_3$).

Previous magnetic susceptibility studies of Ce$_3$CuSnSe$_7$ revealed a sharp anomaly around 5\,K and an upturn at lower temperatures, which was interpreted as evidence for antiferromagnetic order at 5\,K and a more complex form of magnetism at lower temperatures~\cite{gulay2005crystalCe3CuSnSe7}.
Our magnetic susceptibility data on Ce$_3$CuSnSe$_7$ also show an anomaly around 5\,K and an upturn at lower temperatures [Supplementary Fig.~\ref{fig:Ce3_Cu_Sn_Se7_magnetic_susceptibility_our_experiment}], consistent with the above. At high temperatures, the data exhibit Curie-Weiss behavior, with a negative paramagnetic Curie temperature of $\theta_p = -45$~K suggestive of antiferromagnetic correlations and an effective moment of $\mu_{\rm eff} = 2.8 \mu_{\rm B}$, which is not too different from a full Ce$^{3+}$ moment.

The electrical resistivity of Ce$_3$CuSnSe$_7$ increases with decreasing temperature [Supplementary Fig.~\ref{fig:Ce3_Cu_Sn_Se7_electrical_resistivity_our_experiment}], evidencing a non-metallic state. Between 300\,K and 150\,K, the data are consistent with the Arrhenius law $\rho(T)=\rho_0 \exp(E_{\rm G}/k_{\rm B} T)$, with a band gap of $E_{\rm G} = 0.57$\,eV. A band gap of the same order, $E_{\rm G}=0.49$\,eV, was found for a related $P6_3$ space group quaternary compound, La$_3$Fe$_{0.61}$SnSe$_7$ \cite{Assoud2014}. For temperatures below 90\,K, the resistivity is well described by the Mott variable range hopping (VRH) model, which is suitable for describing disordered semiconductors \cite{PRAMANIK1981,NGUYEN2003,Essaleh2017}.

As we aim to find hexagonal Weyl-Kondo semimetal candidate materials, further investigations of the non-metallic compound Ce$_3$CuSnSe$_7$ are beyond the scope of this work. In general, future studies should address the complex magnetic behavior at lower temperatures and its relation to the VRH behavior of the electrical resistivity. Several reports of quaternary $R_3MM'X_7$ ($R=$ rare earth, $M$ and $M'=$ metal or semimetal, and $X=$ S, Se) compounds of space group $P6_3$ discuss partial occupancy of the $M$ and $M'$ sites, with implications for magnetic and electric properties \cite{Assoud2014,Daszkiewicz2009,Iyer2016,Iyer2017,Guo2009,Daszkiewicz2007,Akopov2021}.

\begin{figure}
    \centering
    \includegraphics[width=0.75\linewidth]{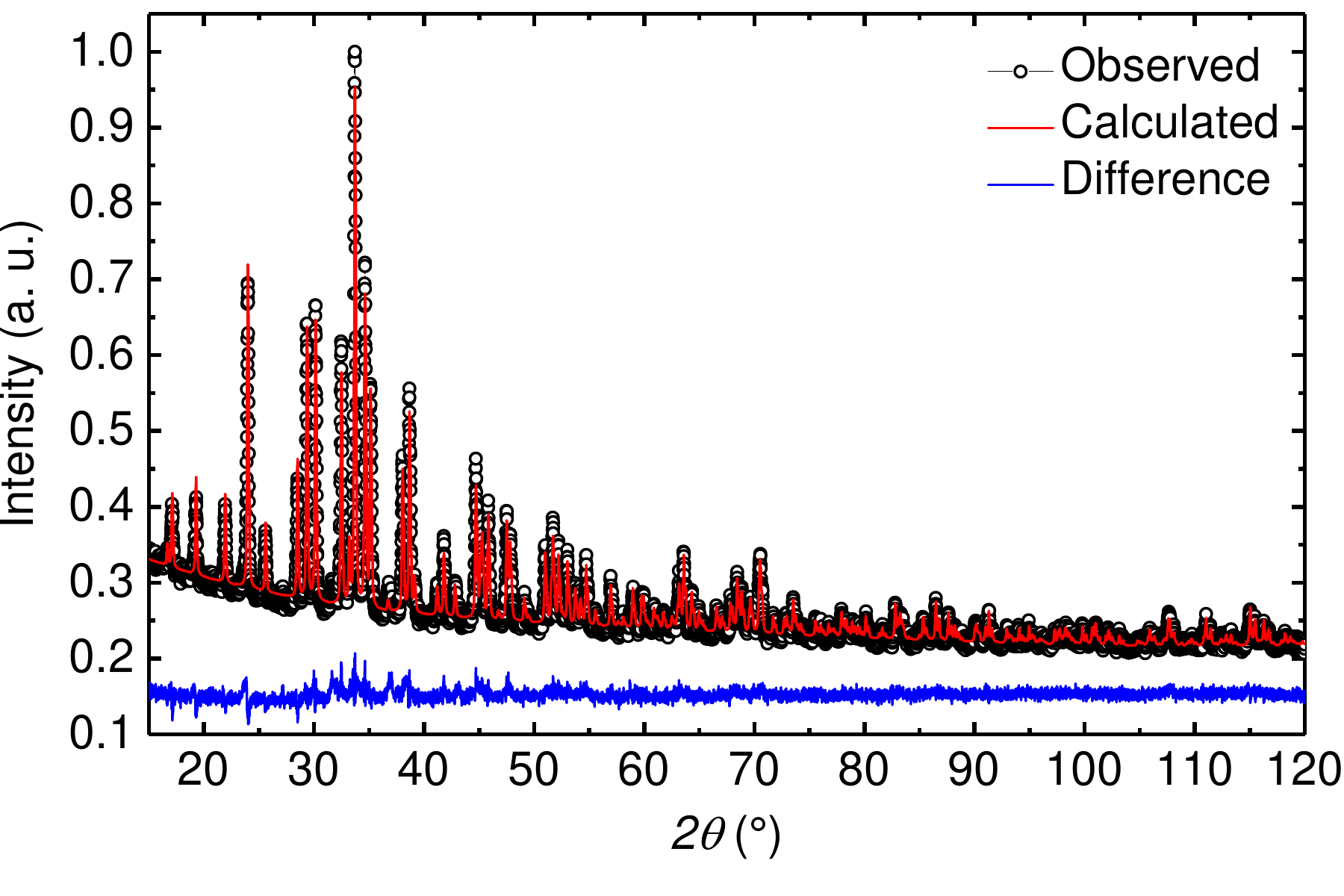}
    \caption{Powder XRD pattern for a Ce$_3$CuSnSe$_7$ sample. The open circles are the experimental data, the red curve is obtained from the Rietveld refinement, and the blue curve is the difference between the experimental data and the Rietveld refinement.}
    \label{fig:Ce3_Cu_Sn_Se7_powder_XRD_our_experiment}
\end{figure}

\begin{figure}
    \centering
    \includegraphics[width=\linewidth]{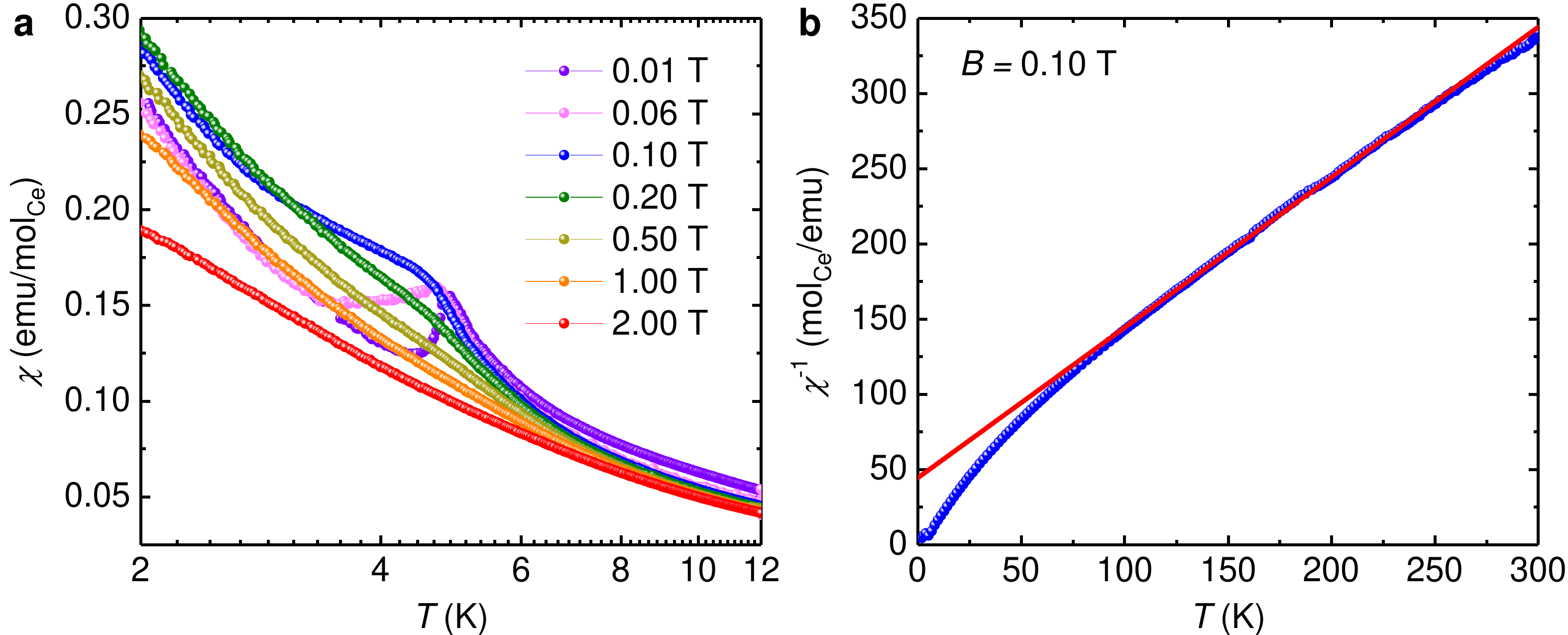}
    \caption{(a) Zero-field cooling (ZFC) magnetic susceptibility curves of Ce$_3$CuSnSe$_7$ measured in different magnetic fields. (b) Temperature dependence of $\chi^{-1}$ measured at 0.1\,T and a Curie-Weiss fit (red line).}
    \label{fig:Ce3_Cu_Sn_Se7_magnetic_susceptibility_our_experiment}
\end{figure}

\begin{figure}
    \centering
    \includegraphics[width=0.75\linewidth]{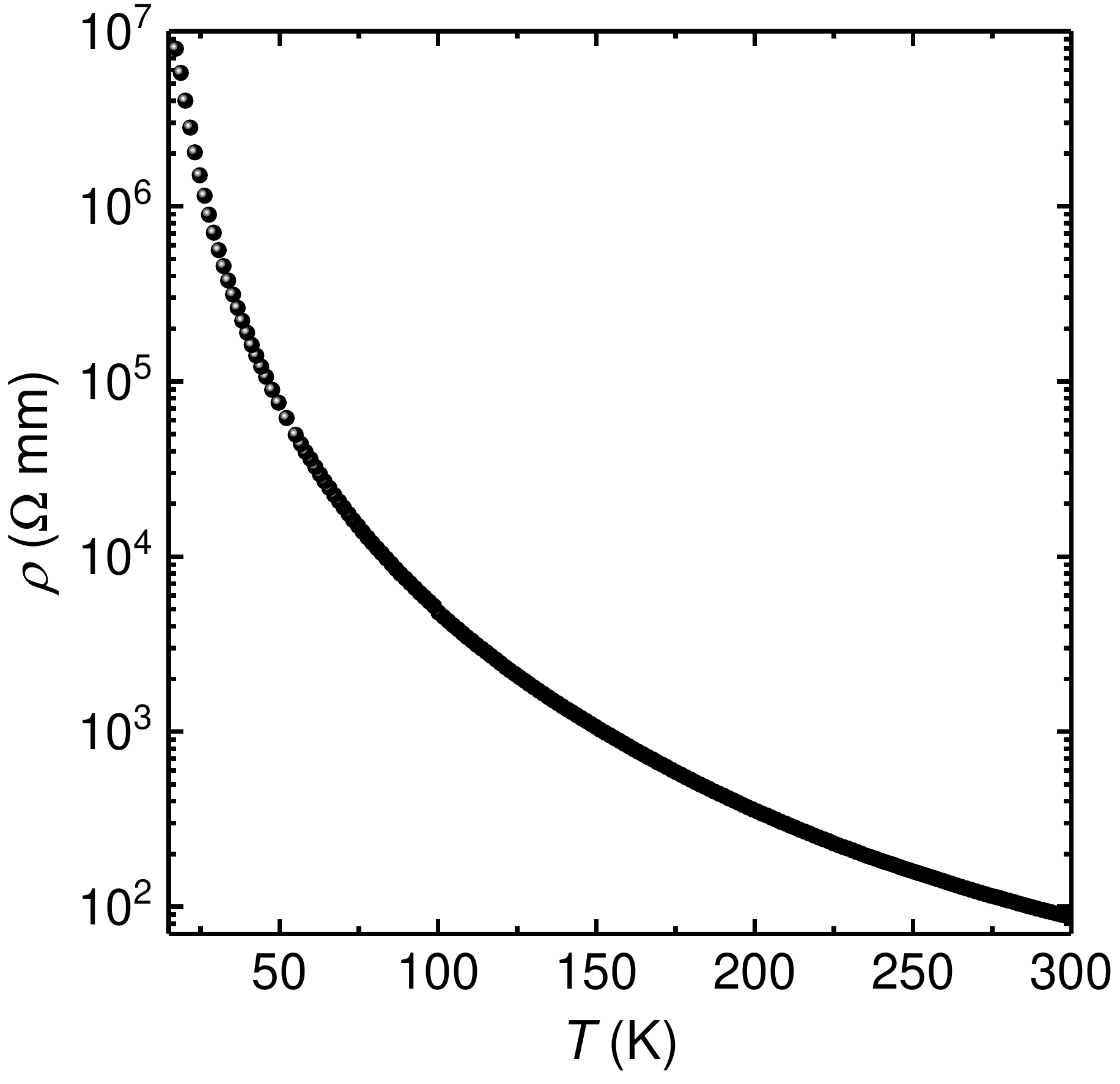}
    \caption{Temperature dependence of the electrical resistivity of Ce$_3$CuSnSe$_7$.}
    \label{fig:Ce3_Cu_Sn_Se7_electrical_resistivity_our_experiment}
\end{figure}

\section{Geometric frustration in hexagonal crystals}

We now explore the geometric frustration in our candidate materials. For example, in the lattice structure, the presence of triangular plaquettes promotes geometric frustration when the interactions are antiferromagnetic.

For CePt$_2$B, as shown in Supplementary Fig.~\ref{fig:frustration}(a,b), its Ce ions form a three-dimensional kagome-like lattice structure, where the abundance of triangular plaquettes indicates geometric frustration.

For Ce$_2$NiGe$_3$, as shown in Supplementary Fig.~\ref{fig:frustration}(c,d), its Ce ions form a nearly-ideal kagome lattice structure, suggesting the presence of geometric frustration.

For Ce$_6$Co$_{2-\delta}$Si$_3$ with $\delta = 0.33$, namely Ce$_6$Co$_{1.67}$Si$_3$, as shown in Supplementary Fig.~\ref{fig:frustration}(e,f), its Ce ions form a lattice structure with abundant triangular plaquettes, suggesting geometric frustration.

\begin{figure}
    \centering
    \includegraphics[width=0.8\linewidth]{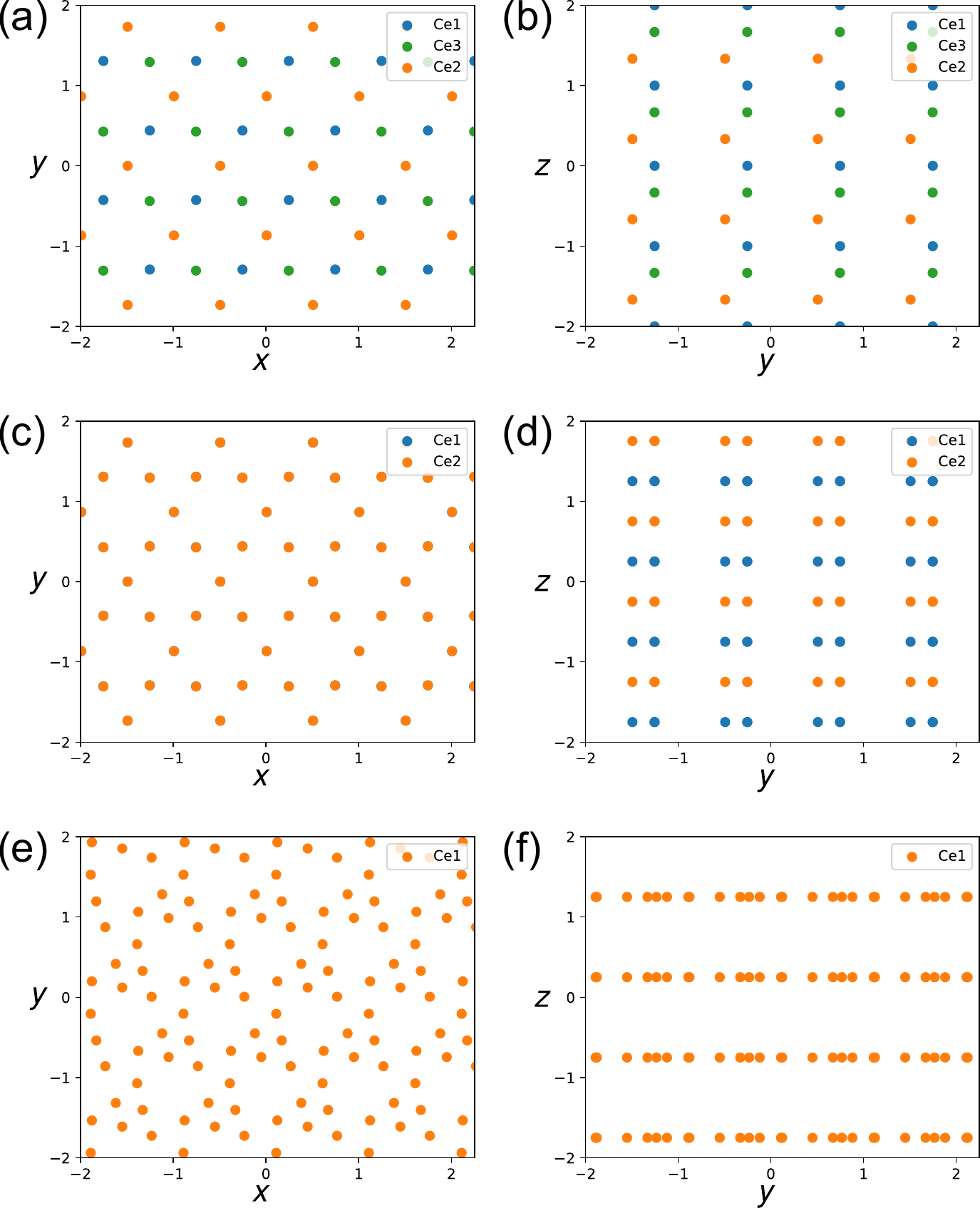}
    \caption{(a) Top view (along $c$-axis) of Ce sites in $\rm CePt_2 B$. 
    (b) Side view (along $a$-axis) of Ce sites in $\rm CePt_2 B$. 
    (c) Top view (along $c$-axis) of Ce sites in $\rm Ce_2 Ni Ge_3$. 
    (d) Side view (along $a$-axis) of Ce sites in $\rm Ce_2 Ni Ge_3$. 
    (e) Top view (along $c$-axis) of Ce sites in $\rm Ce_6 Co_{2-\delta} Si_3$ with $\delta =0.33$. 
    (f) Side view (along $a$-axis) of Ce sites in $\rm Ce_6 Co_{2-\delta} Si_3$ with $\delta =0.33$. 
    In these plots color represents different $z$-coordinates in the unit cell.
    }
    \label{fig:frustration}
\end{figure}

\end{document}